\documentclass{lmcs} %%% last changed 2014-08-20
\pdfoutput=1
\usepackage[utf8]{inputenc}

\usepackage{lastpage}
\lmcsdoi{22}{1}{13}
\lmcsheading{}{\pageref{LastPage}}{}{}%
{Sep.~08,~2021}{Feb.~17,~2026}{}


\usepackage{hyperref}

\usepackage{xspace}
\usepackage{tikz}
\usetikzlibrary{positioning,arrows}

\usepackage{xcolor}
\usepackage{hhline}
\usepackage{multirow}

\usepackage{mathtools}
\usepackage{amsmath,amsfonts,amssymb,amsthm}

\usepackage{microtype}
\usepackage{multicol}

\tikzstyle{every node} =
[draw = none, fill = white, thin]
\tikzstyle{noall} =
[draw = none, fill = none]
\tikzstyle{nodraw} =
[draw = none, fill = white]
\tikzstyle{nofill} =
[draw = black, fill = black]

\tikzstyle{cnode} =
[circle, draw = black, thick]

\newcommand \tpl[1]{\langle #1 \rangle}
\newcommand \details[1]{}

\newcommand{\Prop}{{\mathit{AP}}}
\newcommand{\Agents}{{\textit{Ag}}}
\newcommand{\agent}{{\textit{a}}}
\newcommand{\Actions}{{\textit{Ac}}}
\newcommand{\Decisions}{{\textit{Dc}}}
\newcommand{\decision}{{\textit{d}}}
\newcommand{\States}{{\textit{S}}}
\newcommand{\Trans}{{\tau}}
\newcommand{\Lab}{{\textit{Lab}}}
\newcommand{\Range}{{\textit{Ran}}}

\newcommand{\Undef}{{\dashv}}%branching degree
\newcommand{\bottom}{{\gamma_0}}

\newcommand{\ATLStar}{\text{\sffamily ATL$^{*}$}}
\newcommand{\ATLPlus}{\text{\sffamily ATL$^{+}$}}
\newcommand{\CTLStar}{\text{\sffamily CTL$^{*}$}}
\newcommand{\ATL}{\text{\sffamily ATL}}

\newcommand{\ACG}{\text{\sffamily ACG}}
\newcommand{\CGS}{\text{\sffamily CGS}}
\newcommand{\PMS}{\text{\sffamily PMS}}
\newcommand{\CTL}{\text{\sffamily CTL}}
\newcommand{\CGT}{\text{\sffamily CGT}}
\newcommand{\NPTA}{\text{\sffamily NPTA}}

\newcommand{\GS}{{\mathcal{G}}}
\newcommand{\PS}{{\mathcal{S}}}
\newcommand{\Au}{{\mathcal{A}}}

\newcommand{\Pu}{{\mathcal{P}}}

\newcommand{\dir}{{\textit{dir}}}

\newcommand{\Bu}{{\mathcal{B}}}

\newcommand{\NTA}{\text{\sffamily NTA}}

\newcommand{\LTL}{\text{\sffamily LTL}}

\newcommand{\until}{\textsf{U}}
\newcommand{\release}{\textsf{R}}
\newcommand{\Next}{\textsf{X}}
\newcommand{\A}{\textsf{A}}
\newcommand{\E}{\textsf{E}}
\newcommand{\Always}{\textsf{G}}
\newcommand{\Eventually}{\textsf{F}}
\newcommand{\Exists}[1]{\langle\hspace{-0.05cm}\langle #1 \rangle\hspace{-0.05cm}\rangle}

\newcommand{\NLOGSPACE}{{\sc NLogspace}\xspace}
\newcommand{\FOUREXPTIME}{{\sc 4Exptime}\xspace}
\newcommand{\THREEXPTIME}{{\sc 3Exptime}\xspace}
\newcommand{\TWOEXPTIME}{{\sc 2Exptime}\xspace}
\newcommand{\THREEXPSPACE}{{\sc 3Expspace}\xspace}

\newcommand{\EXPTIME}{{\sc Exptime}\xspace}
\newcommand{\PTIME}{{\sc Ptime}\xspace}
\newcommand{\PSPACE}{{\sc Pspace}\xspace}

\newcommand{\MainP}{\textit{Main}}

\newcommand{\outcome}{{\textit{out}}}
\newcommand{\Unw}{{\textit{Unw}}}

\newcommand{\Con}{{\textit{Cons}}}

\newcommand{\env}{{\textit{env}}}

\newcommand{\sys}{{\textit{sys}}}
\newcommand{\exec}{{\textit{exec}}}
\newcommand{\Ann}{{\textit{Ann}}}
\newcommand{\an}{{\textit{an}}}
\newcommand{\Dom}{{\textit{Dom}}}
\newcommand{\Cod}{{\textit{Cod}}}
\newcommand{\Atoms}{{\textit{Atoms}}}
\newcommand{\atom}{{\textit{atom}}}
\newcommand{\Tau}{{\mathcal{T}}}

\newcommand{\Lang}{{\mathcal{L}}}
\newcommand{\Tower}{\mathit{Tower}}

\newcommand{\ReqOne}{\textsf{R1}}
\newcommand{\ReqTwo}{\textsf{R2}}
\newcommand{\ReqThree}{\textsf{R3}}
\newcommand{\ReqFour}{\textsf{R4}}
\newcommand{\ReqFive}{\textsf{R5}}

\newcommand{\CondOne}{\textsf{C1}}
\newcommand{\CondTwo}{\textsf{C2}}
\newcommand{\CondThree}{\textsf{C3}}

\newcommand{\lst}{{\textit{lst}}}
\def\B{{\mathbb{B}}}
\def\Nat{{\mathbb{N}}}
\newcommand{\true}{\texttt{true}}

\newcommand{\Tag}{{\textit{tag}}}

\newcommand{\BCT}{{\textit{CBT}}}

\newcommand{\der}[1]{\ensuremath{\;\;{\mathop{{ %
            \longrightarrow}}\limits^{{#1}}}\!}\;\;} %
\newcommand{\Stack}{{\textsf{Stack}}}
\newcommand{\dire}{{\textit{dir}}}
\newcommand{\Succ}{{\textit{next}}}
\newcommand{\first}{{\textit{first}}}
\newcommand{\last}{{\textit{last}}}
\newcommand{\fair}{{\textit{fair}}}
\newcommand{\conf}{{\textit{conf}}}
\newcommand{\init}{{\textit{init}}}
\newcommand{\good}{{\textit{good}}}
\newcommand{\push}{{\textsf{push}}}
\newcommand{\pop}{{\textsf{pop}}}
\newcommand{\inc}{{\textit{inc}}}
\newcommand{\Start}[1]{\textsf{s}_{#1}}
\newcommand{\End}[1]{\textsf{e}_{#1}}
\newcommand{\bl}{\textit{bl}}
\newcommand{\bla}{\textit{bl}\kern .1em '}
\newcommand{\CheckB}[1]{{\textit{check}_{#1}}}
\newcommand{\CheckMark}[1]{{\widehat{\textit{check}_{#1}}}}
\begin{document}

\title[Module checking of pushdown multi-agent systems]{Module checking of pushdown multi-agent systems\rsuper*}
\titlecomment{{\lsuper*}The paper is an extended version completed with proofs of the conference paper \cite{BozzelliMP20}.}

\author[L.~Bozzelli]{Laura Bozzelli\lmcsorcid{0009-0004-8555-8229}}[a]	
\author[A.~Murano]{Aniello Murano\lmcsorcid{0000-0003-4876-3448}}[b]
\author[A.~Peron]{Adriano Peron\lmcsorcid{0000-0002-7111-3171}}[c]	

\address{University of Napoli ``Federico II'', Napoli, Italy}
\email{laura.bozzelli@unina.it}

\address{University of Napoli ``Federico II'', Napoli, Italy}	
\email{aniello.murano@unina.it}

\address{University of Napoli ``Federico II'', Napoli, Italy}	
\email{adrperon@unina.it}

\begin{abstract}
In this paper, we investigate the module-checking problem of pushdown multi-agent systems (\PMS)
against \ATL\ and \ATLStar\ specifications. We establish that for \ATL, module checking of \PMS\ is \TWOEXPTIME-complete, which is the
same complexity as pushdown module-checking   for \CTL. On the other hand, we show that  \ATLStar\ module-checking of \PMS\  turns out to be \FOUREXPTIME-complete, hence exponentially harder than  both \CTLStar\ pushdown module-checking and \ATLStar\ model-checking of \PMS.
Our result for \ATLStar\ provides a rare example of a natural decision problem that is elementary yet
but with a complexity that is higher than triply exponential-time.
\end{abstract}

\maketitle

%-------------------------------------------------------------------------------
\section{Introduction}\label{sec:intro}
%-------------------------------------------------------------------------------

\emph{Model checking} is a well-established formal-method technique to automatically check for global correctness of   systems~\cite{Clarke81ctl,Queille81verification}. Early use of model checking mainly considered
\emph{finite-state closed systems}, modelled as
labelled state-transition graphs (Kripke structures) equipped with some internal degree of nondeterminism,  and
specifications given in terms of standard   temporal logics   such as the linear-time temporal logic $\LTL$~\cite{Pnueli77} and the branching-time temporal logics $\CTL$ and $\CTLStar$~\cite{EmersonH86}.
In the last two decades, model-checking techniques have been extended to the analysis
 of reactive and distributed component-based systems, where the behavior of a component depends on assumptions on its
environment (the other components).   One of the first approaches  to model check \emph{finite-state open systems} is \emph{module checking}~\cite{KV96},
a framework for handling the interaction between a system and an external unpredictable environment. In this setting, the system
is modeled as a \emph{module} that is a finite-state Kripke structure whose
 states are partitioned into those controlled by the system and those controlled by the environment. The latter ones intrinsically carry an additional source of nondeterminism describing the possibility that the computation, from these states, can continue with any subset of its possible successor states. This means that while in model checking, we have only one computation tree representing the possible evolution of the system, in module checking we have an infinite number of trees to handle, one for each possible behavior of the environment. Deciding whether a system satisfies a property amounts to check that all such trees satisfy the property.
This makes %the
module checking %problem
harder to deal with. Classically, module checking has been investigated with respect to $\CTL$ and $\CTLStar$~\cite{KV96,Kupferman97revisited,Basu07local} specifications and for $\mu$-calculus specifications~\cite{Ferrante08enriched}.  An extension of module checking has been also used to reason about three-valued abstractions in~\cite{Alfaro04three,Godefroid03open}.
Other approaches to the verification of multi-component finite-state systems (\emph{multi-agent systems}) are based on the game paradigm:
the system is modeled by a multi-player finite-state concurrent game, where at each step, the next state is
determined by considering the “intersection” between the choices made
simultaneously and independently by all the players  (the agents).
 In this setting, properties are specified in logics for strategic reasoning
such as the alternating-time temporal logics \ATL\ and \ATLStar~\cite{AlurHK02},  the latter ones being
well-known extensions of \CTL\ and \CTLStar, respectively, which allow
to express cooperation and competition among agents in order to achieve certain goals. In particular, they can express selective quantification over those paths that are the result of the infinite game between a given coalition and the rest of the agents.

\begin{table}[tp]
\begin{center}
%\small
\begin{tabular}{|c||c|c|c|c|}
\hline &
        \small Model Checking  &  \small Model Checking   &  \small Module Checking
&   \small Module Checking \\
& &  \small (fixed formula)  & &   \small (fixed formula)  \\
\hline \hline
\small  \CTL  & \small \PTIME  & \small  \NLOGSPACE & \small \EXPTIME & \small  \PTIME \\
& \small \cite{EmersonH86} & \small \cite{BernholtzVW94} & \small \cite{KV96} & \small \cite{KV96} \\
\hline \small \CTLStar  & \small  \PSPACE  & \small \NLOGSPACE  & \small \TWOEXPTIME \ & \small \PTIME  \\
& \small \cite{EmersonH86} & \cite{BernholtzVW94} & \small \cite{KV96}  & \small \cite{KV96}  \\
\hline \small \ATL  & \small  \PTIME  & \small \PTIME  & \small \EXPTIME \ & \small \PTIME  \\
& \small \cite{AlurHK02} & \cite{AlurHK02} & \small \cite{BozzelliM2017}  & \small \cite{BozzelliM2017}  \\
\hline \small \ATLStar  & \small  \TWOEXPTIME  & \small \PTIME  & \small \THREEXPTIME \ & \small \PTIME  \\
& \small \cite{AlurHK02} & \cite{AlurHK02} & \small \cite{BozzelliM2017}  & \small \cite{BozzelliM2017}  \\
\hline \hline
\end{tabular}
\normalsize
\end{center}
\caption{Complexity results on finite-state model checking and
finite-state module checking } \label{tableFiniteState}
\end{table}

For a long time, there has been a common believe that module checking of \CTL/\CTLStar\ is a special case of model checking of \ATL/\ATLStar.
The belief has been refuted in~\cite{JamrogaM14} where it is proved that module checking includes two features inherently absent in the semantics of \ATL/\ATLStar, namely irrevocability and nondeterminism of strategies. On the other hand, branching-time
temporal logics like \CTL\ and \CTLStar\ do not accommodate strategic reasoning.  These facts have  motivated the extension of module checking to a finite-state multi-agent setting for handling specifications in \ATLStar\ \cite{JamrogaM15,BozzelliM2017},   which turns out to be more expressive than both $\CTLStar$ module checking and $\ATLStar$ model checking~\cite{JamrogaM14,JamrogaM15}. Table~\ref{tableFiniteState} summarizes known results about the complexity of finite-state model checking and finite-state module checking.  All the complexities in Table~\ref{tableFiniteState} denote tight bounds.

\paragraph{Verification of pushdown systems} An active field of research is model checking of pushdown systems. These represent an infinite-state formalism suitable to capture the control flow
of procedure calls and returns in programs. Model checking of (closed) pushdown systems against standard regular temporal logics (such as $\LTL$, $\CTL$, $\CTLStar$,
and the modal $\mu$-calculus) is decidable and it has been intensively studied leading to efficient verification algorithms and tools (see ~\cite{Wal96,BEM97,BR00,AminofKM12,AminofMM14}).
The verification of open pushdown systems in a two-player turn-based setting has been investigated in many works (e.g. see~\cite{LodingMS04,HagueO09}).  Open pushdown systems along with the module-checking paradigm have been considered in~\cite{Bozzelli10pushdown}. As in the case of finite-state systems, for the logic  $\CTL$ (resp., $\CTLStar$), pushdown module-checking is singly exponentially harder than pushdown model-checking, being precisely
\TWOEXPTIME-complete (resp., \THREEXPTIME-complete), although with the same program complexity as pushdown model-checking (that is \EXPTIME-complete). Pushdown module-checking  has been investigated under  several restrictions~\cite{Aminof13pushdown-jc,Bozzelli11newpushown,Murano08hierarchical}, including the imperfect-information setting case, where the latter variant is in general undecidable~\cite{Aminof13pushdown-jc}. In~\cite{MuranoP15,ChenSW16}, the verification of open pushdown systems has been extended to a concurrent game setting (\emph{pushdown multi-agent  systems}) by  considering %multi-agent
specifications in  $\ATLStar$ and the alternating-time modal $\mu$-calculus. In particular, model checking of \PMS\ against \ATLStar\ has the same complexity as pushdown module-checking against \CTLStar~\cite{ChenSW16}.   %. The work in \cite{} extends the approach in \cite{} by considering \emph{global} model checking of \PMS\ against \ATLStar\ and the alternating-time modal $\mu$-calculus.

\begin{table}[tp]
\begin{center}
%\small
\begin{tabular}{|c||c|c|c|c|}
\hline &
        \small Pushdown  &  \small Pushdown   &  \small Pushdown
&   \small Pushdown \\
& \small Model Checking &  \small Model Checking  & \small Module Checking &   \small Module Checking  \\
& &  \small (fixed formula)  & &   \small (fixed formula)  \\
\hline \hline
\small  \CTL  & \small \EXPTIME  & \small  \EXPTIME & \small \TWOEXPTIME & \small  \EXPTIME \\
& \small \cite{Walukiewicz00} & \small \cite{Bozzelli06} & \small \cite{Bozzelli10pushdown} & \small \cite{Bozzelli10pushdown} \\
\hline \small \CTLStar  & \small  \TWOEXPTIME  & \small \EXPTIME  & \small \THREEXPTIME \ & \small \EXPTIME  \\
& \small \cite{Bozzelli06} & \cite{Bozzelli06} & \small \cite{Bozzelli10pushdown}  & \small \cite{Bozzelli10pushdown}  \\
\hline \small \ATL  & \small  \EXPTIME  & \small { \EXPTIME}  & \small \color{red}\TWOEXPTIME \ & \small \color{red} \EXPTIME  \\
& \small \cite{ChenSW16} & \cite{ChenSW16} &  \color{red} Corollary \ref{cor:UpperBounds}  & \color{red} Corollary \ref{cor:UpperBounds}  \\
\hline \small \ATLStar  & \small  \THREEXPTIME  & \small \EXPTIME  & \small \color{red} \FOUREXPTIME \ & \small \color{red} \EXPTIME  \\
& \small \cite{ChenSW16} & \cite{ChenSW16} &  \color{red} Cor.~\ref{cor:UpperBounds} \& Theorem \ref{theorem:lowerbound}  & \color{red} Corollary \ref{cor:UpperBounds} \\
\hline \hline
\end{tabular}
\normalsize
\end{center}
\caption{Complexity results on pushdown model checking and
pushdown module checking } \label{tablePushdown}
\end{table}

\paragraph{Our contribution} In this paper, we extend the module-checking framework to the verification of multi-agent pushdown systems (\PMS)
by addressing the module-checking problem of  \PMS\ against \ATL\ and \ATLStar\ specifications.
By~\cite{JamrogaM14}, the considered setting for $\ATL$ (reps., $\ATLStar$) is strictly more expressive than both pushdown module checking
for \CTL\ (resp., $\CTLStar$) and  $\ATL$ (reps., $\ATLStar$) model-checking of $\PMS$.
We establish
that  \ATL\ module-checking for \PMS\  has the same complexity as  pushdown module-checking  for \CTL, that is \TWOEXPTIME-complete. On the other hand,
we show that $\ATLStar$ module-checking of \PMS\ has a very high complexity: it turns out to be exponentially harder than \ATLStar\ model-checking of
\PMS\ and pushdown module-checking for \CTLStar, being,  precisely, \FOUREXPTIME-complete with an \EXPTIME-complete complexity for a fixed-size formula. The upper bounds are obtained by %applying
an automata-theoretic approach.
The matching lower bound for \ATLStar\ is shown by a technically non-trivial reduction from the acceptance problem for \THREEXPSPACE-bounded  alternating
Turing Machines. Our result for \ATLStar\ provides a rare example of a natural decision problem that is elementary yet
but with a complexity that is higher than triply exponential-time. To the best of our knowledge, the unique known characterization of the class
\FOUREXPTIME\ concerns   validity of \CTLStar\ on alternating automata %enriched
with bounded cooperative concurrency~\cite{HarelRV90}.

Our results confirm that pushdown module checking is exponentially harder than finite-state module checking. Indeed, like the logics \CTL\ and \CTLStar,
pushdown module checking against \ATL\ (resp., \ATLStar) turns out to be exponentially harder that finite-state module checking against 
\ATL\ (resp., \ATLStar) even for a fixed formula. This is illustrated in 
Tables~\ref{tableFiniteState} and~\ref{tablePushdown}, where all the complexities denote tight bounds.

The rest of the paper is organized as follows. In Section~\ref{sec:Preliminaries}, we recall the  concurrent game setting, the class of multi-agent pushdown systems (\PMS), and the logics $\ATL$ and $\ATLStar$.
Moreover, we introduce the $\PMS$ module-checking framework for $\ATL$ and $\ATLStar$ specifications. In Section~\ref{sec:DecisionProcedures}, we describe the proposed automata-theoretic approach for solving the module-checking problem of $\PMS$ against $\ATL$ and $\ATLStar$, and in Section~\ref{sec:LowerBound}, we show that for the logic $\ATLStar$, the considered problem is \FOUREXPTIME-hard. Finally  Section~\ref{sec:conc} provides an assessment of the work done, and outlines future research directions.

\section{Preliminaries}\label{sec:Preliminaries}

We fix the following notations. Let $\Prop$ be a finite nonempty set of atomic propositions,
$\Agents$ be a finite nonempty set of agents, and $\Actions$ be a finite nonempty set of actions  that can be made by agents.
For a set  $A\subseteq \Agents$ of agents, an $A$-decision $\decision_A$  is an element in $\Actions^{A}$ assigning to each agent $\agent\in A$ an action $\decision_A(\agent)$. For $A,A'\subseteq \Agents$ with $A\cap A'=\emptyset$, an $A$-decision  $\decision_A$ and $A'$-decision $\decision_{A'}$, $\decision_A\cup \decision_{A'}$ denotes the $(A\cup A')$-decision   defined in the obvious way.
Let $\Decisions = \Actions^{\Agents}$ be the set of \emph{full decisions} of all the agents in $\Agents$.

Let $\Nat$ be the set of natural numbers. %For all  $i,j\in\Nat $, with $i\leq j$, $[i,j]$ denotes the set of natural numbers $h$ such that $i\leq h\leq j$.
 For an infinite word $w$ over an alphabet $\Sigma$ and $i\geq 0$, $w(i)$ denotes the $(i+1)^{th}$ letter of $w$ and $w_{\geq i}$ the suffix of $w$ starting from the $(i+1)^{th}$ letter of $w$, i.e., the infinite word $w(i)w(i+1)\ldots$. For a finite word $w$ over
 $\Sigma$, $|w|$ is the length of $w$.

Given a set $\Upsilon$ of directions, an (\emph{infinite}) 
\emph{$\Upsilon$-tree} $T$ is a non-empty   prefix closed subset of $\Upsilon^{*}$ such that
for all $\nu\in T$, $\nu\cdot \gamma\in T$ for some $\gamma\in \Upsilon$. 
Elements of $T$ are called nodes and $\varepsilon$ is the root of $T$. For $\nu\in T$, a child of $\nu$ in $T$ is a
node  of the form $\nu\cdot \gamma$ for some $\gamma\in \Upsilon$.  An (infinite) path of $T$ is an infinite sequence $\pi$ of nodes such that  $\pi(i+1)$ is a child in $T$ of $\pi(i)$
 for all $i\geq 0$.  For an alphabet $\Sigma$, a $\Sigma$-labeled  $\Upsilon$-tree is a pair $\tpl{T, \Lab}$ consisting of a  $\Upsilon$-tree
 and a labelling $\Lab:T \mapsto \Sigma$  assigning to each node in $T$ a symbol in $\Sigma$.
We extend the labeling $\Lab$ to  paths $\pi$ in the obvious way, i.e.
$\Lab(\pi)$ is the infinite word over $\Sigma$ given by $\Lab(\pi(0))\Lab(\pi(1))\ldots$.
The labeled tree $\tpl{T, \Lab}$ is \emph{complete} if $T=\Upsilon^{*}$. Given $k\in\Nat\setminus\{0\}$, a
 \emph{$k$-ary  tree} is a $\{1,\ldots,k\}$-tree.

\paragraph{Concurrent game structures (\CGS)} \CGS~\cite{AlurHK02} extend Kripke structures
 to a setting involving multiple agents.
 They can be viewed as multi-player games in which players perform concurrent actions, chosen strategically as a function of the history of the game.

\begin{defi}[\CGS]\label{def:CGS} A \CGS\ (over $\Prop$, $\Agents$, and $\Actions$) is a tuple $\GS =\tpl{\States,s_0,\Lab,\Trans}$, where $\States$ is a  set of states, $s_0\in \States$ is the initial state, $\Lab: \States \mapsto 2^{\Prop}$ maps each state to a set of atomic propositions, and
$\Trans: \States\times \Decisions \mapsto \States \cup \{\Undef\}$ is a transition function
 that maps a state and a full decision either to a state or to the special symbol $\Undef$ ($\Undef$ is for `undefined') such that for all states $s$, there exists  $\decision\in\Decisions$
 so that $\Trans(s,\decision)\neq \Undef$.
 Given a set  $A\subseteq \Agents$ of agents, an $A$-decision $\decision_A$, and a state $s$, we say that  \emph{$\decision_A$ is available at state $s$} if there exists an $(\Agents\setminus A)$-decision $\decision_{\Agents\setminus A}$ such that $\Trans(s,\decision_A\cup \decision_{\Agents\setminus A})\in \States$.

For a state $s$ and an agent $\agent$, \emph{state $s$ is controlled by $\agent$} if there is a unique $(\Agents\setminus\{a\})$-decision available at state $s$.  Agent \emph{$\agent$ is passive in  $s$} if there is a unique $\{a\}$-decision available at state $s$. A \emph{multi-agent turn-based game} is a \CGS\ where each state is controlled by an agent.
\end{defi}

Note that in modelling independent agents, usually one assumes that at each state $s$, each agent $\agent$ has a set $\Actions_{\agent,s}\subseteq \Actions$ of actions which are enabled at the state $s$. This is reflected in the transition function $\Trans$ by requiring that the set of full decisions $\decision$ such that
$\Trans(s,\decision)\neq \Undef$ corresponds to $(\Actions_{\agent,s})_{\agent\in \Agents}$.

We now recall the notion of strategy
in a \CGS\ $\GS =\tpl{\States,s_0,\Lab,\Trans}$. Here, we consider \emph{perfect recall} strategies where an agent decides the next action by using 
all the available information up to the current round.
 A \emph{play} is an infinite sequence of states $s_1 s_2 \ldots $
 such that for all $i\geq 1$, $s_{i+1}$ is a \emph{successor} of $s_i$, i.e. $s_{i+1}= \Trans(s_i,\decision)$ for some full decision $\decision$. A \emph{track} (or \emph{history}) $\nu$ is a nonempty prefix of some play.
  Given a set  $A\subseteq \Agents$ of agents, a \emph{strategy for $A$} is a mapping $f_A$
  assigning to each track $\nu$ (representing the history the agents saw so far)  an $A$-decision available at the last state, denoted $\lst(\nu)$, of $\nu$.
  The \emph{outcome} function $\outcome(s,f_A)$ for a state $s$ and the strategy $f_A$ returns the set of all the plays starting at state $s$ that can occur when agents $A$ execute strategy $f_A$ from state $s$ on. Formally, $\outcome(s,f_A)$ is the set of plays $\pi=s_1 s_2\ldots $
  such that $s_1=s$ and for all $i\geq 1$, there is $d\in \Actions^{\Agents\setminus A}$ so that $s_{i+1}=\Trans(s_i,f_A(s_1\ldots s_i)\cup d)$.

\begin{defi} For a set $\Upsilon$ of directions, a \emph{Concurrent Game $\Upsilon$-Tree} ($\Upsilon$-\CGT) is a \CGS\ $\tpl{T,\varepsilon,\Lab,\Trans}$, where $\tpl{T,\Lab}$ is a $2^{\Prop}$-labeled $\Upsilon$-tree, and for each node $x\in T$, the %set of
successors of $x$ correspond  to the %set of
children of $x$ in $T$. Every $\CGS$ $\GS =\tpl{\States,s_0,\Lab,\Trans}$ induces  a
$\States$-\CGT, denoted by $\Unw(\GS)$,
obtained by unwinding $\GS$ from the initial state in the usual way.
  Formally,
  $\Unw(\GS)= \tpl{T,\varepsilon,\Lab',\Trans'}$, where  $\nu\in T$ %is the set of elements $\nu$ in $S^{*}$ such that
   iff $s_0\cdot \nu$ is a track of $\GS$, and for all $\nu\in T$ and   $\decision \in \Decisions$, $\Lab'(\nu) = \Lab(\lst(\nu))$ and $\Trans'(\nu,\decision) = \nu\cdot \Trans(\lst(\nu),\decision)$, with $\lst(\varepsilon)=s_0$.
\end{defi}

\paragraph{Pushdown multi-agent systems (\PMS)} $\PMS$, introduced in~\cite{MuranoP15},
generalize standard pushdown systems to a concurrent multi-player setting.

\begin{defi}%[\PMS]
\label{def:PMS} A \PMS\ (over $\Prop$, $\Agents$, and $\Actions$) is a tuple $\PS =\tpl{Q,\Gamma\cup \{\bottom\},q_0,\Lab,\Delta}$, where $Q$ is a finite set of (control) states, $\Gamma\cup \{\bottom\}$ is a finite stack alphabet ($\bottom\notin\Gamma$ is the special
\emph{stack bottom symbol}),   $q_0\in Q$ is the initial state, $\Lab: Q \mapsto 2^{\Prop}$ maps each state to a set of atomic propositions, and
$\Delta: Q\times (\Gamma\cup \{\gamma_0\}) \times\Decisions \mapsto (Q \times \Gamma^{*}) \cup \{\Undef\}$ is a transition function
 ($\Undef$ is for `undefined') such that for all pairs
 $(q,\gamma)\in Q\times (\Gamma\cup \{\gamma_0\})$, there is  $\decision\in\Decisions$
 so that $\Delta(q,\gamma,\decision)\neq\Undef$.
\end{defi}

The size $|\Delta|$ of the transition function $\Delta$ is given by $|\Delta|=\sum_{ (q',\beta) \in\Range(\Delta)}|\beta|$, where 
$\Range(\Delta)$ is the set of pairs $(q',\beta)\in Q \times \Gamma^{*}$ such that 
$(q',\beta)=\Delta(q,\gamma,d)$ for some $(q,\gamma,d)\in Q\times (\Gamma\cup \{\gamma_0\}) \times\Decisions$. 
A  \emph{configuration} of the $\PMS$ $\PS$ is a pair $(q,\beta)$ where $q$ is a (control) state and $\beta\in  \Gamma^{*}\cdot \bottom$ is a stack content.
Intuitively, when the  $\PMS$ $\PS$ is in state $q$, the stack top symbol is $\gamma$ and the agents take a full decision $d$ available at the current configuration, i.e.~such that $\Delta(q,\gamma,d)=(q',\beta)$ for some $(q',\beta)\in Q\times \Gamma^{*}$, then $\PS$ moves to the configuration with state $q'$ and stack content obtained by removing $\gamma$ and pushing $\beta$ (if $\gamma=\bottom$ then $\gamma$ is not removed).
Formally, the $\PMS$ $\PS =\tpl{Q,\Gamma\cup \{\bottom\},q_0,\Lab,\Delta}$
 induces the infinite-state $\CGS$ $\GS(\PS)=\tpl{\States,s_0,\Lab',\Trans}$, where
$\States$ is the set of configurations of $\PS$, $s_0=(q_0,\bottom)$ (initially, the stack contains just the bottom symbol $\bottom$), $\Lab'((q,\beta))=\Lab(q)$ for each configuration $(q,\beta)$, and the transition function $\Trans$ is defined as follows for all $((q,\gamma\cdot \beta),d)\in \States \times \Decisions$, where $\gamma\in\Gamma\cup\{\bottom\}$:
\begin{itemize}
  \item  \emph{either} $\Delta(q,\gamma,d)=\Undef$ and $\Trans((q,\gamma\cdot \beta),d)=\Undef$,
 \item  \emph{or} $\gamma\in \Gamma$, $\Delta(q,\gamma,d)=(q',\beta')$, and  $\Trans((q,\gamma\cdot \beta),d)=(q',\beta'\cdot \beta)$,
 \item  \emph{or} $\gamma=\bottom$ (hence, $\beta=\varepsilon$), $\Delta(q,\gamma,d)=(q',\beta')$, and $\Trans((q,\gamma\cdot \beta),d)=(q',\beta'\cdot \bottom)$.
\end{itemize}

\subsection{The logics   ATL* and  ATL}\label{sec:LogicsATL}

We recall the alternating-temporal logics $\ATLStar$ and $\ATL$ proposed by Alur et al.~\cite{AlurHK02} as extensions of the standard branching-time temporal logics $\CTLStar$ and
$\CTL$ (respectively)~\cite{EmersonH86}, where the path quantifiers are replaced by more general parameterized quantifiers  which allow for reasoning about the strategic capability of groups of agents.
For the given sets $\Prop$ and $\Agents$ of atomic propositions and agents, $\ATLStar$ formulas $\varphi$ are defined by the following grammar:
\[
\varphi ::=  \true \ | \ \ p \ | \ \neg \varphi   \ | \ \varphi \vee \varphi \ | \ \Next \varphi\ | \ \varphi \,\until\, \varphi\ | \  \Exists{A}  \varphi
\]
where $p\in \Prop$, $A\subseteq \Agents$, $\Next$ and  $\until$ are the standard
``next'' and ``until'' temporal modalities,   and $\Exists{A}$ is the
 ``existential strategic quantifier" parameterized by a set $A$ of agents. Formula $\Exists{A}\varphi$ expresses the property that the group of agents $A$ has a collective strategy
    to enforce  property $\varphi$. In addition, we  use standard shorthands:
  the ``eventually"  temporal modality
  $\Eventually \varphi := \true\,\until\, \varphi$, the ``release" temporal modality $\varphi_1\,\release\, \varphi_2 := \neg (\neg \varphi_1\,\until\, \neg \varphi_2)$,  and the ``always" temporal modality
    $\Always \varphi:= \neg\true\, \release\, \varphi$.
    \newline
    A \emph{state formula} is a formula where each temporal modality is in the scope of a strategic quantifier.
 The logic $\ATL$ is the  fragment of $\ATLStar$ where each temporal modality is immediately preceded by a strategic quantifier. Formally, the set of $\ATL$ formulas are defined by the 
 following grammar:
\[
\varphi ::=  \true \ | \ \ p \ | \ \neg \varphi   \ | \ \varphi \vee \varphi  \ | \  \Exists{A} \Next \varphi
\ | \  \Exists{A}(\varphi \, \until\, \varphi)  \ | \  \Exists{A}(\varphi\, \release\, \varphi)
\]
  Note that \CTLStar\ (resp., \CTL) corresponds
 to the fragment of $\ATLStar$ (resp., $\ATL$), where only the strategic modalities
 $\Exists{\Agents}$  and $\Exists{\emptyset}$ (equivalent to  the existential and universal path quantifiers  $\E$ and $\A$, respectively) are allowed.

Given a \CGS\ $\GS$ with labeling $\Lab$ and a play $\pi$ of $\GS$, the
satisfaction relation $\GS,\pi  \models \varphi$ for
$\ATLStar$  is defined as follows (Boolean connectives are treated as usual):\vspace{-0.05cm}
\[ \begin{array}{ll}
\GS,\pi  \models p  &  \Leftrightarrow  p \in \Lab(\pi(0))\\
\GS,\pi  \models \Next \varphi  & \Leftrightarrow   \GS,\pi_{\geq 1} \models \varphi \\
\GS,\pi  \models \varphi_1\,\until\, \varphi_2  &
  \Leftrightarrow  \text{there is }\,j\geq 0: \GS,\pi_{\geq j}
  \models \varphi_2
  \text{ and }  \GS,\pi_{\geq k} \models  \varphi_1 \text{ for all }0\leq k<j\\
\GS,\pi \models \Exists{A} \varphi  & \Leftrightarrow \text{for some strategy } f_A \text{ for }A,\, \GS,\pi' \models \varphi \text{ for all }\pi'\in \outcome(\pi(0),f_A). %\vspace{-0.1cm}
\end{array} \]
For a state $s$ of $\GS$, $\GS,s  \models \varphi$ if there is a play $\pi$ starting from $s$ such that
$\GS,\pi  \models \varphi$. Note that if $\varphi$ is a state formula, then for all plays $\pi$ and $\pi'$ from $s$,
$\GS,\pi  \models \varphi$  iff $\GS,\pi'  \models \varphi$.
$\GS$ is a model of $\varphi$, denoted $\GS\models \varphi$, if for the initial state $s_0$,
$\GS,s_0  \models \varphi$. Note that $\GS\models \varphi$ iff $\Unw(\GS)\models \varphi$.

\subsection{ATL* and ATL  Pushdown Module-checking}

The module-checking framework was proposed in~\cite{KV96} for the verification of finite open systems,
that is systems that interact with an environment whose behavior cannot be determined in advance.
In such a framework, the system is modeled by a \emph{module} corresponding to a two-player turn-based game
between the system and the environment. Thus, in a module, the set of states is partitioned into a set of system
states (controlled by the system) and a set of environment states (controlled by the environment).

The module-checking problem takes  two inputs: a module $M$ and a branching-time temporal formula $\psi$. The idea is
that the open system should satisfy the specification $\psi$ no matter how the environment behaves.
Let us consider the unwinding   $\Unw(M)$ of $M$ into an infinite tree. Checking whether $\Unw(M)$ satisfies
$\psi$ is the usual model-checking problem. On the other hand, for an open system,
$\Unw(M)$ describes the interaction of the system with a maximal environment, i.e. an environment
that enables all the external nondeterministic choices. In order to take into account all the
possible behaviors of the environment, we have to consider all the trees $T$ obtained from $\Unw(M)$
by pruning subtrees whose root is a successor of an environment state (pruning these subtrees
corresponds to disabling possible environment choices). Therefore, a module $M$ satisfies
$\psi$ if all these trees $T$ satisfy $\psi$.

It has been proved in~\cite{JamrogaM14}  that module checking of \CTL/\CTLStar\ includes two features inherently absent in the semantics of \ATL/\ATLStar, namely irrevocability of strategies and nondeterminism of strategies. Intuitively, unlike the standard \ATLStar\ semantics, in module checking, a formula is evaluated over a restricted behaviour of the full execution tree which corresponds
to the strategy tree induced by a possible environment behaviour (\emph{irrevocability of the environment's strategies}). On the other hand, %branching-time
temporal logics like \CTL\ and \CTLStar\ do not accommodate strategic reasoning.  These facts have  motivated the extension of module checking to a multi-agent setting for handling specifications in \ATLStar\ \cite{JamrogaM15},   which turns out to be more expressive than both $\CTLStar$ module checking and $\ATLStar$ model checking~\cite{JamrogaM14,JamrogaM15}.

In this section, we  first recall the $\ATLStar$ module-checking framework.
% which turns out to be more expressive than both $\CTLStar$ module-checking and $\ATLStar$ %model-checking~\cite{JamrogaM14,JamrogaM15}.
Then, we %further
generalize this setting to pushdown multi-agent  systems.
In the multi-agent module-checking setting, one considers $\CGS$ with a distinguished agent (the \emph{environment}).

\begin{defi}[Open \CGS] An open \CGS\  is a $\CGS$ $\GS =\tpl{\States,s_0,\Lab,\Trans}$ containing a special agent called ``the environment" ($\env\in\Agents$). Moreover,
for every state $s$, either $s$ is controlled by the environment (\emph{environment state}) or the environment is passive in $s$ (\emph{system state}).
\end{defi}

For an open $\CGS$ $\GS=\tpl{\States,s_0,\Lab,\Trans}$, the set of \emph{environment}   \emph{strategy trees of $\GS$}, denoted $\exec(\GS)$, is the set of $\States$-\CGT\ obtained from $\Unw(\GS)$ by possibly pruning some environment transitions.  Formally, $\exec(\GS)$ is the set of $\States$-\CGT\ $\Tau =\tpl{T,\varepsilon,\Lab',\Trans'}$ such that $T$ is a prefix closed subset of the set of $\Unw(\GS)$-nodes  and for all $\nu\in T$  and   $\decision \in \Decisions$, $\Lab'(\nu) = \Lab(\lst(\nu))$, and $\Trans'(\nu,\decision) = \nu\cdot\Trans(\lst(\nu),\decision)$ if $\nu\cdot \Trans(\lst(\nu),\decision)\in T$, and $\Trans'(\nu,\decision)=\Undef$ otherwise,  where $\lst(\varepsilon)=s_0$. Moreover, for all $\nu \in T$, the following holds:
\begin{itemize}
\item if $\lst(\nu)$ is a system state, then for each successor $s$ of $\lst(\nu)$ in $\GS$, $\nu\cdot s \in T$;
\item if $\lst(\nu)$ is an environment state, then there is a nonempty subset $\{s_1,\ldots,s_n\}$ of the set of $\lst(\nu)$-successors such that the set of children of $\nu$ in
$T$ is $\{\nu\cdot s_1,\ldots,\nu\cdot s_n\}$.
\end{itemize}

Intuitively, when $\GS$ is in a system state $s$, then all the transitions from $s$ are enabled. When
 $\GS$ is instead in an environment state, the set of enabled transitions from $s$ depend on the current environment.
 Since the behavior of the environment is nondeterministic, we have to consider
  all the possible subsets of the set of $s$-successors.
The only constraint, since we consider environments that cannot block the system, is that
not all the transitions from $s$ can be disabled. Note that $\Unw(\GS)\in\exec(\GS)$ ($\Unw(\GS)$ corresponds to the maximal environment that never restricts the set of its next states).

It is worth noting that the choices made by the environment along an environment strategy tree describe a strategy of the environment which is nondeterministic.
This is in contrast with the given notion of strategy for a coalition $A$ of agents which is instead deterministic (at each round, the coalition $A$ selects exactly one
$A$-decision available at the current state).

For an open $\CGS$ $\GS$  and an \ATLStar\ formula $\varphi$, $\GS$ \emph{reactively satisfies} $\varphi$, denoted $\GS\models^{r} \varphi$, if
for all environment strategy trees $\Tau\in \exec(\GS)$, $\Tau\models \varphi$.
Note that $\GS\models^{r} \varphi$ implies $\GS\models \varphi$ (since
$\Unw(\GS)\in\exec(\GS)$), but the converse in general does not hold. Moreover,  $\GS\not\models^{r} \varphi$ is not equivalent to
$\GS \models^{r} \neg\varphi$. Indeed, $\GS\not\models^{r} \varphi$ just states that there is some $\Tau\in \exec(\GS)$
satisfying $\neg\varphi$.\vspace{0.2cm}

\begin{figure}[t]
    \centering
  \begin{center}
		{\begin{tikzpicture}
			[scale=1.0, bend angle = 15,  every node/.style={scale=0.7}]
	 
            %%%%% NODES
			\node [cnode,draw = none]
			(SI)
			{ };
			
			\node [cnode, node distance = 4em]
			(Choice)
			[below of = SI]
			{\normalsize $\textsf{choice}$};

            \node [cnode, draw = none, node distance = 8em]
		 	(SII)
		 	[below left of = Choice]
		 	{ };	

            \node [cnode, node distance = 10em]
			(ReqB)
			[left of = SII]
			{\normalsize$\textsf{reqb}$};

            \node [cnode, draw = none, node distance = 8em]
			(SIII)
			[below right of = Choice]
			{ };
			
			\node [cnode, node distance = 10em]
			(ReqW)
			[right of = SIII]
			{\normalsize$\textsf{reqw}$};

            \node [cnode, node distance = 12em]
			(Black)
			[right of = ReqB]
			{\normalsize$\textsf{black}$};

            \node [cnode, node distance = 12em]
			(White)
			[left of = ReqW]
			{\normalsize$\textsf{white}$};

            \node [cnode, draw = none, node distance = 6em]
			(SIV)
			[left of = ReqW]
			{ };

            \node [cnode, node distance = 2.5em]
			(Milk)
			[below of = SIV]
			{\normalsize$\textsf{milk}$};

            \node [cnode, draw = none, node distance = 12em]
			(SV)
			[left of = Choice]
			{ };

			\node [cnode, node distance = 3em, scale = 0.8,inner sep=0pt,minimum size=1pt]
			(RejB)
			[above of = SV]
			{\Large \begin{tabular}{c} \textsf{reqb} \\ \textsf{rej}  \end{tabular}};

            \node [cnode, draw = none, node distance = 12em]
			(SVI)
			[right of = Choice]
			{ };

            \node [cnode, node distance = 3em, scale = 0.8,inner sep=0pt,minimum size=1pt]
			(RejW)
			[above of = SVI]
			{\Large \begin{tabular}{c} \textsf{reqw} \\ \textsf{rej}  \end{tabular}};

            %%%%%%% EDGES

            \path[black, thick,->] (SI) edge (Choice);									
			
            \path[black, thick,->] (Choice) edge [bend right = 15] node[fill=white, anchor=center,  align=left, midway, sloped, font=\normalsize]
                { $b_+,\,\push(\gamma)$ }(ReqB);

            \path[black, thick,->] (Choice) edge node[fill=white, anchor=center,  align=left, midway, sloped, font=\normalsize]
                {  $b_-,\,\pop(\gamma)$ }(ReqB);
			
            \path[black, thick,->] (Choice) edge [bend left = 15] node[fill=white, anchor=center,  align=left, midway, sloped, font=\normalsize]
                { $b$ }(ReqB);

            \path[black, thick,->] (Choice) edge [bend left = 15]
			node[fill=white, anchor=center, align=left, midway, sloped, font=\normalsize] {
				$w_+,\,\push(\gamma)$ } (ReqW);

            \path[black, thick,->] (Choice) edge
			node[fill=white, anchor=center, align=left, midway, sloped, font=\normalsize] {
				 $w_-,\,\pop(\gamma)$ } (ReqW);

            \path[black, thick,->] (Choice) edge [bend right = 15]
			node[fill=white, anchor=center, align=left, midway, sloped, font=\normalsize] {
				$w$  } (ReqW);

            \path[black, thick,->] (ReqB) edge [bend right]
			node[fill=white, anchor=center, align=left, midway, sloped, font=\normalsize] {
				$pour$  } (Black);

            \path[black, thick,->] (Black) edge  (Choice);

            \path[black, thick,->] (White) edge  (Choice);

            \path[black, thick,->] (ReqW) edge
			node[above, align=left, midway, sloped, font=\normalsize] {
				$pour$  } (Milk);

            \path[black, thick,->] (Milk) edge
			node[above, align=center, midway, sloped, font=\normalsize] {
				$milk$  } (White);

           \path[black, thick,->] (Milk) edge [bend left = 20]
			node[fill=white, anchor=center, align=center, midway, sloped, font=\normalsize] {
				$ign$  } (Black);

            \path[black, thick,->] (ReqB) edge [bend left = 30]
			node[fill=white, anchor=center, align=left, midway, sloped, font=\normalsize] {
				$ign$  } (Choice);

            \path[black, thick,->] (ReqW) edge [bend right = 30]
			node[fill=white, anchor=center, align=left, midway, sloped, font=\normalsize] {
				$ign$  } (Choice);

             \path[black, thick,->] (Choice) edge  [bend right = 20]
			node[fill=white, anchor=center, align=left, midway, sloped, font=\normalsize] {
				$b_-,\, \pop(\bottom)$  } (RejB);

           \path[black, thick,->] (Choice) edge [bend left = 20]
			node[fill=white, anchor=center, align=left, midway, sloped, font=\normalsize] {
				$w_-,\, \pop(\bottom)$  } (RejW);

            \path[black, thick,->] (RejB) edge [bend left = 50] (Choice);

            \path[black, thick,->] (RejW) edge [bend right = 50] (Choice);		
			\end{tikzpicture} }
		\end{center}
    \caption{Multi-agent pushdown coffee machine $\PS_{cof}$}
    \label{fig:ModuleChecking}
\end{figure}

\paragraph{Pushdown Module-checking} An \emph{open $\PMS$} is a $\PMS$ $\PS$ such that the induced $\CGS$
$\GS(\PS)$ is open. Note that for an open $\PMS$, the property of a configuration of being an environment or system configuration  depends only on the control state and the symbol on the top of the stack.
The \emph{pushdown module-checking problem against} \ATL\ (resp., \ATLStar) is checking for a given
open \PMS\ $\PS$ and an \ATL\ formula (resp., \ATLStar\ state formula)  $\varphi$ whether $\GS(\PS)\models^{r}\varphi$.

\begin{exa} Consider a coffee machine that allows customers (acting the role of the environment) to choose between the following actions:
\begin{itemize}	
\item
ordering and paying a black or white coffee (actions $b$ or $w$);
\item the same as in the previous point but, additionally, paying a ``suspended"
coffee (a prepaid coffe) for the benefit of any unknown needy customer claiming it in the future (actions $b_+$ or $w_+$);
\item  asking for an available prepaid (black or white) coffee
(actions $b_-$ or $w_-$).
\end{itemize}

The coffee machine
 is modeled by a turn-based open $\PMS$ $\PS_{cof}$ with three agents:
the environment, the brewer  $br$ whose function is to pour coffee into  the cup (action $pour$), and the milk provider who can
add milk (action $milk$). The two system agents can be faulty and ignore the request from the environment (action $ign$).
The stack is exploited for keeping track of the number of prepaid coffees:  a request for a prepaid coffee can be accepted only if the stack is not empty. After the completion of a request, the machine waits for further selections. The $\PMS$ $\PS_{cof}$ is represented as a graph in Figure~\ref{fig:ModuleChecking} where each node (control state) is labeled by the propositions
holding at it: the state labeled by $\textsf{choice}$ is controlled by the environment, the  states labeled by $\textsf{reqb}$ or $\textsf{reqw}$ are controlled by the brewer $br$, while the state labeled by
$\textsf{milk}$ is controlled by the milk provider. The notation $\push(\gamma)$ denotes a push stack operation (pushing the symbol $\gamma\neq \bottom$), while
$\pop(\gamma)$ (resp., $\pop(\bottom)$) denotes a pop operation onto a non-empty (resp., empty) stack. The set of propositions is $\{\textsf{reqw},\textsf{reqb},\textsf{rej}, \textsf{black}, \textsf{white}\}$.

In module checking, we can condition the property to be achieved on the behaviour of the environment. For instance, users who never order white coffee and whose request is never rejected can be served by the brewer alone:  $\GS(\PS_{cof})\models^{r} \A\Always(\neg \textsf{reqw} \wedge \neg \textsf{rej}) \rightarrow  \Exists{br}\Eventually\, \textsf{black}$. In model checking, the same formula does not express any interesting property since $\GS(\PS_{cof})\not\models \A\Always(\neg \textsf{reqw} \wedge \neg \textsf{rej})$.
Likewise $\GS(\PS_{cof})\models  \A\Always \neg \textsf{reqw}  \rightarrow  \Exists{br}\Eventually\, \textsf{black}$, whereas module checking gives a different and more intuitive answer:
  $\GS(\PS_{cof})\not \models^{r} \A\Always \neg \textsf{reqw}  \rightarrow  \Exists{br}\Eventually\, \textsf{black}$ (there are environments where requests for a prepaid coffee are always rejected).
\end{exa}

\section{Decision procedures}\label{sec:DecisionProcedures}

In this section, we provide an automata-theoretic framework
for solving the pushdown  module-checking problem against \ATL\ and \ATLStar\ which is based on the  use of
 \emph{parity alternating automata for \CGS} (parity \ACG)~\cite{ScheweF06}  and \emph{parity Nondeterministic Pushdown Tree Automata}
 (parity \NPTA)~\cite{KPV02}.
The proposed approach (which is proved to be asymptotically optimal in Section~\ref{sec:LowerBound}) consists of two steps.
For the given open $\PMS$   $\PS$ and  $\ATL$ formula (resp., $\ATLStar$ state formula)  $\varphi$,
by exploiting known results,
we first build in linear-time (resp., double exponential time) a parity \ACG\ $\Au_{\neg\varphi}$ accepting the set of $\CGT$ which satisfy $\neg\varphi$.
Then in the second step,  we show how to construct %in time singly exponential  in the size of $\Au_{\neg\varphi}$ and polynomial in the size of  $\PS$,
a parity $\NPTA$ $\Pu$ accepting suitable encodings of the environment strategy trees of $\GS(\PS)$ accepted by $\Au_{\neg\varphi}$.
Hence, $\GS(\PS)\models^r\varphi$ iff the language accepted by $\Pu$ is empty.

In the following, we first recall the frameworks of parity \NPTA\ and parity \ACG, and the known translations  of  $\ATLStar$   and $\ATL$ formulas into equivalent parity \ACG. Then, in Subsection~\ref{sec:UpperBoundsATL}, by exploiting parity \NPTA,  we show that given an open $\PMS$   $\PS$ and a parity $\ACG$ $\Au$, checking that no environment strategy tree of
$\GS(\PS)$ is accepted by $\Au$ can be done in time double exponential in the size of $\Au$ and singly exponential in the size of $\PS$.

\paragraph{Parity \NPTA~\cite{KPV02}}
 Here, we describe parity \NPTA\  (without
$\varepsilon$-transitions)  over labeled complete $k$-ary  trees for a given $k\geq 1$, which
are tuples $\Pu=\tpl{\Sigma,Q,\Gamma\cup \{\bottom\},q_0,\rho,\Omega}$, where
$\Sigma$ is a finite input alphabet, $Q$ is a finite set of (control) states, $\Gamma\cup \{\bottom\}$ is a finite stack
alphabet ($\bottom\notin\Gamma$ is the special bottom symbol),  $q_0\in Q$ is
an initial state,  $\rho:Q\times \Sigma \times (\Gamma\cup
\{\bottom\})\rightarrow 2^{(Q\times \Gamma^*)^k}$ is a transition
function, and $\Omega: Q \mapsto \Nat$ is a \emph{parity acceptance condition} over $Q$ assigning to each state a natural number called \emph{color}.
The \emph{index} of $\Pu$ is the number of colors in $\Omega$, i.e., the cardinality of $\Omega(Q)$.

Intuitively, when the automaton is in state $q$, reading an input
node $x$ labeled by $\sigma\in\Sigma$, and the stack contains a
word $\gamma\cdot \beta$ in $\Gamma^*. \bottom$, then the automaton
chooses a tuple
$\tpl{(q_1,\beta_1),\ldots,(q_k,\beta_k)}\in\rho(q,\sigma,\gamma)$ and
splits in $k$ copies such that for each $1\leq i\leq k$, a copy in
state $q_i$, and stack content obtained by removing $\gamma$ and
pushing $\beta_i$, is sent to   the  node $x \cdot i$ in the input
tree.

Formally, a run of the \NPTA\ $\Pu$  on a
$\Sigma$-labeled complete  $k$-ary tree $\tpl{T,\Lab}$ (with
$T=\{1,\ldots,k\}^*$) is a $( Q\times \Gamma^*.\bottom)$-labeled
tree $r=\tpl{T,\Lab_r}$ such that $\Lab_r(\varepsilon)=(q_0,\bottom)$ (initially, the stack contains just the bottom symbol $\bottom$) and for
each $x\in T$ with $\Lab_r(x)=(q,\gamma\cdot\beta)$, there is
$\tpl{(q_1,\beta_1),\ldots,(q_k,\beta_k)}\in\rho(q,\Lab(x),\gamma)$ such
that for all $1\leq i\leq k$, $\Lab_r(x\cdot i)=(q_i,\beta_i\cdot
    \beta)$ if $\gamma\neq \bottom$, and $\Lab_r(x\cdot i)=(q_i,\beta_i\cdot
    \bottom)$ otherwise (note that in this case $\beta=\varepsilon$).
The run $r=\tpl{T,\Lab_r}$ is accepting  if for all infinite paths $\pi$ starting from the root, the highest color $\Omega(q)$ of the states $q$ appearing infinitely often along $\Lab_r(\pi)$ is even. The language $\Lang(\Pu)$ accepted by $\Pu$ consists of the $\Sigma$-labeled complete  $k$-ary trees $\tpl{T,\Lab}$ such that there is an accepting run of $\Pu$
over $\tpl{T,\Lab}$.

For complexity analysis, we consider the following two parameters: 
\begin{itemize}
  \item the size $|\rho|$ of $\rho$ given by
\[
|\rho|:=\sum_{(q,\sigma,\gamma)\in Q\times \Sigma \times (\Gamma\cup
\{\bottom\})}\sum_{\tpl{(q_1,\beta_1),\ldots,(q_k,\beta_k)}\in\rho(q,\sigma,\gamma)}
|\beta_1|+\ldots + |\beta_k|,
\] 
  \item  and the smaller parameter $||\rho||$  given by $||\rho||:=\sum_{\beta\in\rho_0}|\beta|$ where $\rho_0$
is the set of words $\beta\in  \Gamma^*.\bottom$ occurring in $\rho$.
\end{itemize}
It is well-known~\cite{KPV02} that emptiness of parity $\NPTA$ can be solved in single exponential time by a polynomial time reduction to emptiness
of standard two-way alternating tree automata~\cite{Var98}. In particular, the following holds (see~\cite{KPV02,Bozzelli10pushdown}).

\begin{propC}[\cite{KPV02,Bozzelli10pushdown}] \label{prop:EmptinessForNPTA}
The emptiness problem for a parity \NPTA\ of index $m$  with $n$
states and transition function $\rho$ can be solved in time
$O(|\rho|\cdot 2^{O(||\rho||^2\cdot n^2\cdot m^2 \log m)})$.
\end{propC}

\paragraph{Parity alternating automata for \CGS\ (parity \ACG)~\cite{ScheweF06}} \ACG\ generalize  alternating automata  by branching universally or existentially
 over all successors that result from the decisions of agents. Formally, for a  set $X$, let
$\B^{+}(X)$ be the set
of \emph{positive} Boolean formulas over $X$, i.e. Boolean formulas  built from elements in $X$
using $\vee$ and $\wedge$. A subset $Y$ of $X$ is a \emph{model} of $\theta \in \B^{+}(X)$ if the truth assignment that assigns true (resp., false) to
the elements in $Y$ (resp., $X \setminus Y$) satisfies $\theta$.

A parity \ACG\ over $2^{\Prop}$ and \Agents\ is a tuple
 $\Au=\tpl{Q,q_0,\delta,\Omega}$, where $Q$, $q_0$, and
 $\Omega$ are defined as for \NPTA, while  $\delta$  is a transition function of the form
  $\delta:Q\times 2^{\Prop}\rightarrow \B^{+}(Q\times \{\Box,\Diamond\}\times 2^{\Agents})$.
The transition function $\delta$ maps a state and an input letter to a positive Boolean combination of universal atoms $(q,\Box,A)$  and existential atoms $(q,\Diamond,A)$.
Intuitively, a universal atom $(q,\Box,A)$ prescribes that for some $A$-decision $d_A$ available at the current state $s$ of the input \CGS,
copies of the automaton in state $q$ are sent to \emph{all} the successors of $s$ which are consistent with $d_A$.
Dually, an existential atom $(q,\Diamond,A)$ prescribes that for all $A$-decisions $d_A$ available at the current state $s$ of the input \CGS,
a copy of the automaton in state $q$ is sent to \emph{some} successor  of $s$ which is consistent with $d_A$.

 The size $|\Au|$ of $\Au$ is $|Q|+ |\Atoms(\Au)|$, where
$\Atoms(\Au)$ is the set of atoms  of $\Au$, i.e. the set of tuples in $Q\times \{\Box,\Diamond\}\times 2^{\Agents}$ occurring in the transition function $\delta$.

 We interpret the parity \ACG\ $\Au$ over $\CGT$. Note that since the set of full decisions of all agents is finite, 
 a $\CGT$, which is a special $\CGS$, is \emph{finitely-branching}, i.e., each state has a finite number of successors.
 Thus, given a $\CGT$ $\Tau =\tpl{T,\varepsilon,\Lab,\Trans}$ over $\Prop$ and $\Agents$, a possible behaviour of 
 $\Au$ over the input $\Tau$ can be formalized by  an $\Nat$-tree (i.e., a tree with the \emph{countable} set of directions $\Nat$) whose nodes are labeled
 by pairs $(q,\nu)\in Q\times T$ describing a copy of the automaton that is in the state $q$ and reads the node $\nu$ of
$T$. Formally,  a run of $\Au$ over the input $\Tau$ is a   $(Q\times T)$-labeled  $\Nat$-tree
$r=\tpl{T_r,\Lab_r}$ such  that
\begin{itemize}
  \item  $\Lab_r(\varepsilon)=(q_0,\varepsilon)$ (initially, the automaton is in state $q_0$ reading the root node of  $\Tau$),
  \item for each  $y\in T_r$ with $\Lab_r(y)=(q,\nu)$, there is a set $H\subseteq Q\times \{\Box,\Diamond\}\times 2^{\Agents}$ such that $H$ is a model of
$\delta(q,\Lab(\nu))$ and the set $L$ of labels associated with the  children of $y$ in $T_r$ %minimally
satisfies the following conditions:
\begin{itemize}
  \item for all universal atoms $(q',\Box,A)\in H$, there is an available $A$-decision $d_A$ in the node $\nu$ of $\Tau$ such that for all the children $\nu'$ of $\nu$
  which are consistent with  $d_A$, $(q',\nu')\in L$;
  \item for all existential atoms $(q',\Diamond,A)\in H$ and for all available $A$-decisions $d_A$ in the node $\nu$ of $\Tau$, there is some child $\nu'$ of $\nu$
  which is consistent with $d_A$ such that $(q',\nu')\in L$.
\end{itemize}
\end{itemize}
The run $r$ is accepting if for all infinite paths $\pi$ starting from the root, the highest color of the states appearing infinitely often along $\Lab_r(\pi)$ is even. The language $\Lang(\Au)$ accepted by $\Au$ consists of the $\CGT$ $\Tau$ on $\Prop$ and $\Agents$ such that there is an accepting run of $\Au$
over $\Tau$.

\paragraph{From  \ATLStar\ and \ATL\ to parity \ACG} In the following we shall exploit a known translation of $\ATLStar$ state formulas (resp., $\ATL$ formulas)
 into equivalent parity $\ACG$ which has been provided in~\cite{BozzelliM2017}.  To this end we recall that, for a finite set $B$ disjunct from $\Prop$ and a  $\CGT$ $\Tau =\tpl{T,\varepsilon,\Lab,\Trans}$ over $\Prop$, a
 \emph{$B$-labeling extension of $\Tau$} is a \CGT\ over $\Prop\cup B$ of the form  $\tpl{T,\varepsilon,\Lab',\Trans}$, where $\Lab'(\nu)\cap \Prop=\Lab(\nu)$ for all $\nu\in T$. A \emph{basic formula} of $\ATLStar$ is a state formula of $\ATLStar$ having  the form  $\Exists{A}  \varphi$.
 The result exploited in the following, which corresponds to Theorem~1 in~\cite{BozzelliM2017},   is summarized as follows.

 \begin{thmC}[\cite{BozzelliM2017}] \label{theo:TranslationATLStar} For an \ATLStar\ state formula (resp., \ATL\ formula) $\varphi$ over $\Prop$, one can construct in doubly exponential time (resp., linear time) a parity  $\ACG$ $\Au_\varphi$  over $2^{\Prop\cup B_\varphi}$, where $B_\varphi$ is the set of basic subformulas of $\varphi$, such that for all
$\CGT$ $\Tau$ over $\Prop$, $\Tau$ is a model of $\varphi$ iff there exists a $B_\varphi$-labeling extension of $\Tau$ which is accepted by
 $\Au_\varphi$.  Moreover, $\Au_\varphi$ has
size $O( 2^{2^{O(|\varphi|\cdot \log(|\varphi|))}})$ and  index $2^{O(|\varphi|)}$ (resp., size $O(|\varphi|)$ and index $2$).
\end{thmC}

Note that for the proof of Theorem~\ref{theo:TranslationATLStar}, one exploits the fact that for a given 
$\CGT$ $\Tau$ and \ATLStar\ state formula $\varphi$, there exists a \emph{unique} $B_\varphi$-labeling extension of $\Tau$ which is \emph{well-formed}, i.e., such that 
for each node $\nu$ of $\Tau$, the $B_\varphi$-labeling of $\nu$ coincides with the set of basic subformulas of $\varphi$ which hold at node $\nu$
of $\Tau$. Thus, the automaton  $\ACG$ $\Au_\varphi$ of Theorem~\ref{theo:TranslationATLStar}, by crucially exploiting alternation, recursively checks that 
the given $B_\varphi$-labeling extension of $\Tau$ is well-formed. Hence, $\Au_\varphi$ 
accepts a $B_\varphi$-labeling extension of $\Tau$ \emph{only if}  it is well-formed. 

It is worth noting that while the well-known translation
of $\CTLStar$ formulas into  alternating automata involves just a single exponential blow-up, by Theorem~\ref{theo:TranslationATLStar}, the translation of $\ATLStar$ formulas in alternating automata for $\CGS$  entails a double exponential blow-up.  This seems in contrast with the automata-theoretic approach used in~\cite{Schewe08}
for solving satisfiability of \ATLStar\ (recall that 
   \ATLStar\ satisfiability has the same complexity as \CTLStar\ satisfiability, i.e., it is \TWOEXPTIME-complete~\cite{Schewe08}). In particular,
given an \ATLStar\ state formula $\varphi$, one can construct in singly exponential time a parity \ACG\ accepting the set of \CGT\ satisfying some special requirements %(depending on $\varphi$)
which provide a necessary and sufficient condition for ensuring the existence of some model of $\varphi$~\cite{Schewe08}. These requirements are based on an equivalent representation
 of the models of a formula obtained by a sort of widening operation. However, when applied to the environment strategy trees of a   \CGS, such an encoding is not regular
 since one has to require that for all nodes in the encoding which are copies of the same environment node in the given environment strategy tree, the associated subtrees are isomorphic. Hence, the approach used in~\cite{Schewe08} cannot be applied to the module-checking setting.

\subsection{Upper bounds for ATL  and  ATL* pushdown module-checking}\label{sec:UpperBoundsATL}

Let $\PS$ be an open $\PMS$, $\varphi$   an $\ATLStar$ (resp., $\ATL$) formula, and $\Au_{\neg\varphi}$  the parity
$\ACG$ over $2^{\Prop\cup B_\varphi}$ ($B_\varphi$ is the set of basic subformulas of $\varphi$) of Theorem~\ref{theo:TranslationATLStar}
associated with the negation of $\varphi$. By Theorem~\ref{theo:TranslationATLStar}, checking that $\GS(\PS)\models^r\varphi$ reduces
to checking that there are no $B_\varphi$-labeling extensions of the environment strategy trees of $\GS(\PS)$ accepted by $\Au_{\neg\varphi}$.
In this section, we provide an algorithm for checking this last condition.
In particular, we establish the following result.

 \begin{thm}\label{theo:EmptinessACG} Given an open $\PMS$ $\PS$ over $\Prop$, a finite set $B$ disjoint from $\Prop$, and a parity \ACG\ $\Au$ over $2^{\Prop\cup B}$, checking that there are no $B$-labeling extensions of the environment strategy trees of $\GS(\PS)$ accepted by $\Au$ can be done in time doubly exponential in the size of $\Au$ and singly exponential in the size of $\PS$.
\end{thm}

Thus, by Theorem~\ref{theo:TranslationATLStar} and Theorem~\ref{theo:EmptinessACG}, and since the pushdown module-checking problem against \CTL\ is already \TWOEXPTIME-complete, and \EXPTIME-complete for a fixed \CTL\ formula~\cite{Bozzelli10pushdown},  we obtain the following corollary.

\begin{cor}\label{cor:UpperBounds} Pushdown module-checking for $\ATLStar$  is in \FOUREXPTIME\ while pushdown module-checking for $\ATL$   is \TWOEXPTIME-complete. Moreover, for a fixed
$\ATLStar$ state formula (resp., $\ATL$ formula), the pushdown module-checking problem is \EXPTIME-complete.
\end{cor}

In Section~\ref{sec:LowerBound}, we provide a lower bound for $\ATLStar$ %module-checking problem
matching the upper bound in the corollary above. We present now the proof of Theorem~\ref{theo:EmptinessACG} which is based on a reduction to the emptiness problem of parity \NPTA.  Given an open \PMS\  $\PS$ over $\Prop$ and a parity $\ACG$ $\Au$ over $2^{\Prop\cup B }$, we construct in single exponential time  a parity $\NPTA$ $\Pu$ over $2^{\Prop\cup B }$  accepting the $B$-labeling extensions of suitable encodings of the environment strategy trees of $\GS(\PS)$ which are accepted by $\Au$.
Since the set $B$ just occurs in the input alphabet $2^{\Prop\cup B }$ and the behaviour of $\GS(\PS)$ does not depend  on $B$, for simplicity and without loss of
generality,  we assume that the set $B$ in the statement of Theorem~\ref{theo:EmptinessACG} is empty. % (the general case where $B\neq \emptyset$ is similar).

\paragraph{Encoding of environment strategy trees of open $\PMS$}
Let us fix an open \PMS\  $\PS =\tpl{Q,\Gamma\cup \{\bottom\},q_0,\Lab,\Delta}$ over $\Prop$, and let  $\GS(\PS) =\tpl{\States,s_0,\Lab_{\States},\Trans}$.
For all pairs $(q,\gamma)\in Q\times (\Gamma\cup \{\bottom\})$, we denote by $\Succ_\PS(q,\gamma)$  the finite set
of pairs $(q',\beta)\in Q\times \Gamma^{*}$ such that there is a full decision $d$ so that $\Delta(q,\gamma,d)=(q',\beta)$.
We fix an ordering on the  set $\Succ_\PS(q,\gamma)$ which induces an ordering on the finite set of successors of all
the configurations  of the form $(q,\gamma\cdot \alpha)$. Moreover, we consider the parameter $k_\PS= \max \{|\Succ_\PS(q,\gamma)|\mid (q,\gamma)\in Q\times (\Gamma\cup \{\bottom\})\}$ which represents the finite branching degree of $\Unw(\GS(\PS))$. Thus, we can encode  each track $\nu=s_0,s_1,\ldots,s_n$ of $\GS(\PS)$ starting from the initial state,  by the finite word $i_1,\ldots,i_n$ over $\{1,\ldots,k_{\PS}\}$ of length $n$ where for all $1\leq h\leq n$, $i_h$ represents the index of state
 $s_h$ in the ordered set of successors of state $s_{h-1}$.
Now, we observe that  the transition function $\Trans'$ of an environment strategy tree $\Tau =\tpl{T,\varepsilon,\Lab',\Trans'}$ of $\GS(\PS)$ is completely determined by $T$ and the transition function $\Trans$ of $\GS(\PS)$. Hence, for the fixed open $\CGS$ $\GS(\PS)$, $\Tau$ can be simply specified by the underlying $2^{\Prop}$-labeled $\States$-tree $\tpl{T,\Lab'}$.

We consider an equivalent representation of $\tpl{T,\Lab'}$ by a   $(2^{\Prop}\cup\{\bot\})$-labeled \emph{complete} $k_\PS$-tree  $\tpl{\{1,\ldots,k_\PS\}^{*},\Lab_\bot}$, called the
\emph{$\bot$-completion encoding} of $\Tau$ ($\bot$ is a fresh proposition), where the labeling $\Lab_\bot$ is defined as follows for each node $x\in \{1,\ldots,k_\PS\}^{*}$:
\begin{itemize}
  \item if $x$ encodes a track $s_0\cdot \nu$ such that $\nu$ is a node of $T$, then $\Lab_\bot(x)=\Lab'(\nu)$ (\emph{concrete nodes});
  \item otherwise, $\Lab_\bot(x)=\{\bot\}$ (\emph{completion nodes}).
\end{itemize}
In this way, all the labeled trees encoding environment strategy trees $\Tau$ of $\GS(\PS)$ have the same structure (they all coincide with
$\{1,\ldots,k_\PS\}^{*}$), and they differ only in their labeling. Thus, the proposition $\bot$ is used to denote both
``completion" nodes and nodes in $\Unw(\GS(\PS))$ which are absent in $\Tau$  (corresponding to possible disabling of environment choices).

\paragraph{Proof of Theorem~\ref{theo:EmptinessACG}} We now prove  Theorem~\ref{theo:EmptinessACG} for the case $B=\emptyset$. We establish by Theorem~\ref{Theor:FromACGToNPTA} below 
that given an open \PMS\  $\PS$ and a parity $\ACG$ $\Au=\tpl{Q_\Au,q_\Au^{0},\delta,\Omega}$
over $2^{\Prop}$, one can build  a parity $\NPTA$ $\Pu$  accepting the  $\bot$-completion encodings of the environment strategy trees of $\GS(\PS)$ which are accepted by $\Au$.
Thus,  checking that there are no environment strategy trees of $\GS(\PS)$ accepted by $\Au$ reduces to emptiness of the language accepted by
$\Pu$. Moreover, the size of $\Pu$ is polynomial in the size of  $\PS$ and singly exponential in the size of $\Au$. Hence,
by Proposition~\ref{prop:EmptinessForNPTA},  Theorem~\ref{theo:EmptinessACG}  for the case $B=\emptyset$  directly follows.

\begin{thm}\label{Theor:FromACGToNPTA} Given an open \PMS\  $\PS =\tpl{Q,\Gamma\cup \{\bottom\},q_0,\Lab,\Delta}$ over $\Prop$ and a parity $\ACG$ $\Au=\tpl{Q_\Au,q_\Au^{0},\delta,\Omega}$
over $2^{\Prop}$ with index $h$, one can build in single exponential time, a parity $\NPTA$ $\Pu$  accepting the set of $2^{\Prop}\cup \{\bot\}$-labeled complete $k_{\PS}$-trees which are the $\bot$-completion encodings of the environment strategy trees of $\GS(\PS)$
which are accepted by $\Au$. Moreover, $\Pu$ has %stack alphabet $\Gamma\cup \{\bottom\}$,
index $O(h|\Au|^{2})$,   number of states  $O(|Q| \cdot  (h|\Au|^{2})^{O(h|\Au|^{2})})$, and transition function $\rho$ such that $||\rho|| = O(|\Delta| \cdot  (h|\Au|^{2})^{O(h|\Au|^{2})})$.
\end{thm}
\begin{proof}
First, we observe that for the given parity $\ACG$ $\Au$ and an input $\CGT$ $\Tau$, we can associate in a standard
way  to $\Au$ and $\Tau$ an infinite-state two player parity game, where player 0 plays for
acceptance, while player 1 plays for rejection. Winning strategies of player 0
correspond to accepting runs of $\Au$ over $\Tau$.
Thus, since the existence of a winning strategy in parity games
implies the existence of a memoryless one, we can restrict ourselves
to consider only memoryless runs of $\Au$, i.e. runs $r=\tpl{T_r,\Lab_r}$ where the behavior of $\Au$ along $r$ depends only
on the current input node and current state. Formally, $r$ is memoryless if for all nodes $y$ and $y'$ of $r$ having the same label, the subtrees rooted at the nodes $y$
and $y'$ of $r$ are isomorphic. We now provide a representation of the memoryless runs of $\Au$ over the environment strategy trees of the open $\CGS$ $\GS(\PS)$
induced by the given open $\PMS$ $\PS$.

Fix an environment strategy tree $\Tau = \tpl{T,\varepsilon,\Lab_\Tau,\Trans}$ of
$\GS(\PS)$ and let $\tpl{\{1,\ldots,k_\PS\}^{*},\Lab_\bot}$ be the $\bot$-completion encoding  of
$\Tau$. Recall that  $\Atoms(\Au)$ is the set of atoms  of $\Au$, i.e., the set of tuples in $Q_\Au\times \{\Box,\Diamond\}\times 2^{\Agents}$ occurring in the transition function $\delta$ of $\Au$. 

Let $\Ann := 2^{Q_\Au\times \Atoms(\Au)}$ be the finite set of \emph{annotations} and $\Sigma := (2^{\Prop}\times \Ann \times \Ann)\cup \{\bot\}$.
In the following, a \emph{move} is an element in $Q_\Au\times \Atoms(\Au)$. 
For an annotation $\an\in \Ann$, we define the following finite sets:
\begin{itemize}
  \item  $\Dom(\an)$ is the set of $\Au$-states $q$ such that $(q,\atom)\in \an$ for some  $\atom\in\Atoms(\Au)$;
  \item $\Cod(\an)$ is the set of $\Au$-states occurring in the atoms of $\an$;
  \item For each state $q\in Q_\Au$, $\Atoms(q,\an)$ the set of atoms $\atom$ such that $(q,\atom)\in \an$.
\end{itemize}
For example, if $\an = \{(q_1,(q'_1,\Diamond,A_1)),(q_2,(q'_2,\Box,A_2))\}$, then
$\Dom(\an)= \{q_1,q_2\}$, $\Cod(\an)=\{q'_1,q'_2\}$, and $\Atoms(q_1,\an)=\{(q'_1,\Diamond,A_1)\}$.

We represent memoryless runs $r$ of $\Au$ over $\Tau$ as \emph{annotated extensions} of the
$\bot$-completion encoding  $\tpl{\{1,\ldots,k_\PS\}^{*},\Lab_\bot}$ of $\Tau$, i.e.,
$\Sigma$-labeled complete $k_\PS$-trees
$\tpl{\{1,\ldots,k_\PS\}^{*},\Lab_\Sigma}$, where:
\begin{enumerate}[align=left]
  \item[($\ReqOne$)]   for every concrete node $x\in \{1,\ldots,k_\PS\}^{*}$ encoding a node $\nu_x$ of $\Tau$,
$\Lab_\Sigma(x)$ is of the form $(\Lab_\bot(x),\an,\an')$ (recall that $\Lab_\bot(x)=\Lab_\Tau(\nu_x)$), and for every completion node $x$,
$\Lab_\Sigma(x)=\Lab_\bot(x)=\{\bot\}$. 
\end{enumerate}
 Intuitively, the meaning of the first annotation $\an$ and the second annotation $\an'$ in the label of a concrete
node  $x$ is as follows:
\begin{itemize}
  \item  $\Dom(\an)$ represents the set of  $\Au$-states $q$ associated with the copies of $\Au$ in the run $r$ which read the input node $\nu_x$ of $\Tau$, while for each $q\in\Dom(\an)$, $\Atoms(q,\an)$  represents the model of $\delta(q,\Lab_\Tau(\nu_x))$ selected by $\Au$ in $r$ on reading node $\nu_x$ in state $q$. Note that $\Cod(\an)$ represents the set of target states of the moves in $\an$.
  \item Additionally, the second annotation $\an'$ in the labeling of node $x$ keeps tracks, in case $x$ is not the root,  of the subset of the moves in the first annotation of the parent $\nu'$ of $\nu_x$ in $T$ for which, starting from $\nu'$, a copy of $\Au$ is sent to the current node $\nu_x$ along $r$.
\end{itemize}
\noindent Formally, for the concrete node $x$ with label $(\Lab_\bot(x),\an,\an')$, we require  that the following requirements hold, where 
$x_1,\ldots,x_N$ denote the concrete children of node $x$ in  $\tpl{\{1,\ldots,k_\PS\}^{*},\Lab_\bot}$ encoding the children
$\nu_{1},\ldots \nu_{N}$ of node $\nu_x$ in $\Tau$, and $\Lab_\Sigma(x_i)=(\Lab_\bot(x_i),\an_i,\an'_i)$ for each $1\leq i\leq N$: 
\begin{enumerate}[align=left]
  \item[($\ReqTwo$)]   for each $q\in\Dom(\an)$, $\Atoms(q,\an)$  is a  model of $\delta(q,\Lab_\Tau(\nu_x))$;
  \item[($\ReqThree$)]   the two annotations $\an$ and $\an'$ are consistent, i.e., $\an'=\emptyset$ if $x$ is the root and $\Cod(\an')=\Dom(\an)$ otherwise; 
  moreover, if $x$ is the root, then $\Dom(\an)=\{q_\Au^{0}\}$ ($q_\Au^{0}$ is the initial state of $\Au$);
  \item[($\ReqFour$)]   for all $(q,(q',m,A))\in \an$, the following holds:
  \begin{itemize}
  \item case $m=\Box$:  there is an available $A$-decision $d_A$ in the node $\nu_x$ of $\Tau$ such that for all $i\in \{1,\ldots,N\}$ so that  $\nu_i$ is a child of $\nu_x$ in $\Tau$
 consistent with the $A$-decision  $d_A$, it holds that  $(q,(q',m,A)) \in \an'_i$;
  \item case $m=\Diamond$:   for all available $A$-decisions $d_A$ in the node $\nu_x$ of $\Tau$, there is some $i\in \{1,\ldots,N\}$ such  that the child $\nu_i$ of $\nu_x$ in 
  $\Tau$ is consistent with the $A$-decision $d_A$ and $(q,(q',m,A)) \in \an'_i$.
\end{itemize}
   \item[($\ReqFive$)] $\an = \displaystyle{\bigcup_{i=1}^{i=N}} \an'_i$. 
\end{enumerate}
\noindent    
An  \emph{annotated extension} $\tpl{\{1,\ldots,k_\PS\}^{*},\Lab_\Sigma}$ of $\tpl{\{1,\ldots,k_\PS\}^{*},\Lab_\bot}$ is \emph{well-formed} if it satisfies requirements $\ReqOne$--$\ReqFive$ expressed above. We deduce the following result. \vspace{0.2cm}

\paragraph{Claim 1} One can construct in singly exponential time a parity $\NPTA$ $\Pu_{\textit{wf}}$ over $\Sigma$-labeled complete $k_\PS$-trees
accepting the set of  \emph{well-formed} annotated extensions of
the $\bot$-completion encodings of the environment strategy trees of $\GS(\PS)$. Moreover, $\Pu_{\textit{wf}}$ has %stack alphabet $\Gamma\cup \{\bottom\}$,
number of  states $O(|Q| \cdot  2^{O(|Q_\Au |\cdot |\Atoms(\Au)|)})$, index $1$, and transition function $\rho$ such that $||\rho||= O(|\Delta|)$.\vspace{0.2cm}

The proof of Claim~1 is postponed at the end of the proof of Theorem~\ref{Theor:FromACGToNPTA}.

Note that the well-formedness requirement just ensures that the
annotated extension $\tpl{\{1,\ldots,k_\PS\}^{*},\Lab_\Sigma}$ of $\tpl{\{1,\ldots,k_\PS\}^{*},\Lab_\bot}$ encodes 
a memoryless run $r$ of the $\ACG$ $\Au$ over the input $\Tau$. In order to ensure  that 
$\tpl{\{1,\ldots,k_\PS\}^{*},\Lab_\Sigma}$ encodes a run $r$ which is also accepting, we need to enforce additional \emph{global} 
requirements  on the annotated extension $\tpl{\{1,\ldots,k_\PS\}^{*},\Lab_\Sigma}$.

Let $\pi$ be an infinite path of $\tpl{\{1,\ldots,k_\PS\}^{*},\Lab_\Sigma}$ from the root which does not visit $\bot$-labeled nodes. Then,  $\Lab_\Sigma(\pi)$  ``collects" all the infinite sequences $\nu$ of states in $Q_\Au$  along the run $r$ associated with the  input path of the environment strategy tree $\Tau$ encoded by $\pi$.
In order to check the acceptance condition on the  individual parallel paths $\nu$, the infinite sequence of annotations  $\Lab_\Sigma(\pi)$ must allow to distinguish the 
individual infinite paths over $Q_\Au$ grouped by $\Lab_\Sigma(\pi)$. This is because we exploit the second annotation $\an'$ in the labeling $(\Lab_\bot(x),\an,\an')$ of a concrete node $x$. In particular, the individual paths over $Q_\Au$ grouped by $\Lab_\Sigma(\pi)$ correspond to the so-called $\emph{$Q_\Au$-paths}$ of $\Lab_\Sigma(\pi)$   which are defined as follows.

For all $i\geq 0$, let  $\Lab_\Sigma(\pi(i))=(\sigma_i,\an_i,\an'_i)$. Then, a \emph{$Q_\Au$-path of $\Lab_\Sigma(\pi)$} is an infinite sequence $q_0 q_1\ldots$ of $Q_\Au$-states such that for all $i\geq 0$, $q_i\in \Dom(\an_i)$ and $(q_i,(q_{i+1},m,A))\in \an_i\cap \an'_{i+1}$ for some $m\in \{\Box,\Diamond\}$ and set $A$ of agents.

 We need to check that all these $Q_\Au$-paths satisfy the acceptance parity condition of $\Au$. To this end, we construct 
 a standard parity nondeterministic tree automaton (parity \NTA)  $\Au_{\textit{acc}}$ over $\Sigma$-labeled complete $k_\PS$-trees which
 accepts an input tree  if all the $Q_\Au$-paths associated with the infinite paths of the input tree starting at the root satisfy the acceptance parity condition
 of $\Au$. In order to construct $\Au_{\textit{acc}}$,  we proceed as follows.
 
We first  easily construct a co-parity nondeterministic word automaton   $\Bu$ over $\Sigma$
 with $O(|Q_\Au|\cdot |\Atoms(\Au)|)$ states and index $h$ (the index of $\Au$) which accepts an infinite word over $\Sigma$ iff it contains a $Q_\Au$-path that does not satisfy the parity acceptance condition of $\Au$. Recall that a co-parity condition over a set $Q'$ of states is defined as a parity condition over $Q'$, i.e., a mapping of the form $Q'\mapsto \Nat$, but for acceptance of an infinite sequence of states
 $\rho$, we require that the highest color of the states appearing infinitely often along $\rho$ is \emph{odd}. Formally, the co-parity word automaton
 $\Bu$ with input alphabet $\Sigma$ is given by $\Au=\tpl{Q_\Bu,q_\Bu^{0},\delta_\Bu,\Omega_\Bu}$, where
 \begin{itemize}
   \item $Q_\Bu= \{q_\Bu^{0}\}\cup (Q_\Au \times \Atoms(\Au))$.
   \item The transition function $\delta_\Bu: Q_{\Bu}\times \Sigma \mapsto 2^{Q_{\Bu}}$ is defined as follows for all $(\sigma,\an,\an')\in \Sigma$:
   \begin{itemize}
     \item $\delta_{\Bu}(q_\Bu^{0},(\sigma,\an,\an'))=\an$;
     \item for all $(q,(q',m,A))\in Q_\Au \times \Atoms(\Au)$, we have that $\delta_{\Bu}((q,(q',m,A)),(\sigma,\an,\an'))=\emptyset$ if $(q,(q',m,A))\notin \an'$,
     and  $\delta_{\Bu}((q,(q',m,A)),(\sigma,\an,\an'))=\an$ otherwise.
   \end{itemize}
   \item $\Omega_\Bu(q_\Bu^{0})=0$ and $\Omega_\Bu((q,(q',m,A)))=\Omega(q)$ for all $(q,(q',m,A))\in Q_\Au \times \Atoms(\Au)$.  
 \end{itemize}
 By construction, given an input $w\in\Sigma^\omega$ with $w=(\sigma_0,\an_0,\an'_0)(\sigma_1,\an_1,\an'_1)\ldots$, $\Bu$ accepts $w$ iff
   there exists a run of $\Bu$ of the form
   \[
   q_\Bu^{0}\cdot (q_0,(q'_0,m_0,A_0))\cdot (q_1,(q'_1,m_1,A_1))\ldots
   \]
   such that $(q_i,(q'_i,m_i,A_i))\in \an_i\cap \an'_{i+1}$ for all $i\geq 0$, and the infinite sequence of $\Au$-states $q_0q_1\ldots$
   does not satisfy the parity acceptance condition of $\Au$. Hence, $\Bu$ accepts $w$
   iff $w$  contains a $Q_\Au$-path that does not satisfy the parity acceptance condition of $\Au$.
 We now co-determinize the co-parity nondeterministic word automaton $\Bu$, i.e., determinize it and complement it in a singly-exponential construction~\cite{Safra88} to obtain a \emph{deterministic} parity word automaton $\Bu'$ that rejects violating $Q_\Au$-paths. By~\cite{Safra88},  $\Bu'$ has $(nh)^{O(nh)}$ states and index $O(nh)$, where $n= |Q_\Au|\cdot |\Atoms(\Au)|$. 
 Then the parity \NTA\ $\Au_{\textit{acc}}$ is obtained from $\Bu'$ by simply running $\Bu'$ in parallel over all the branches of the input which do not visit a $\bot$-labeled node. Note that $\Au_{\textit{acc}}$ has $(nh)^{O(nh)}$ states and index $O(nh)$.

% From $\Bu'$, we construct a standard parity nondeterministic tree automaton (parity \NTA)  $\Au_{\textit{acc}}$ over $\Sigma$-labeled complete $k_\PS$-trees having %$(nh)^{O(nh)}$ states and index $O(nh)$
% obtained by simply running $\Bu'$ in parallel over all the branches of the input which do not visit a $\bot$-labeled node.
 Finally, the parity $\NPTA$ $\Pu$ satisfying %the statement of
 Theorem~\ref{Theor:FromACGToNPTA} is obtained by projecting out the annotation components of the input trees accepted by the intersection of the %parity
 $\NPTA$ $\Pu_{\textit{wf}}$ of index $1$
 in Claim~1 with the parity \NTA\ $\Au_{\textit{acc}}$ (recall that parity $\NPTA$ are effectively and polynomial-time closed under projection and intersection with nondeterministic tree automata~\cite{KPV02}). 
 
 In order to conclude the proof of Theorem~\ref{Theor:FromACGToNPTA} it remains to prove Claim~1.

 \paragraph{Proof of Claim 1} In order to define the \NPTA\ $\Pu_{\textit{wf}}$ satisfying Claim 1, we need additional definitions
 which allow to express requirements $\ReqFour$ and $\ReqFive$ in terms of the transition function 
 of the \PMS\  $\PS =\tpl{Q,\Gamma\cup \{\bottom\},q_0,\Lab,\Delta}$.
 
%Recall that for an annotation $\an\in \Ann$,   $\Dom(\an)$ denotes the set of $\Au$-states $q$ such that $(q,\atom)\in \an$ for some atom $\atom\in\Atoms(\Au)$, while $\Cod(\an)$ denotes the set of states occurring in %the atoms of $\an$. Moreover, for each  state $q\in Q_\Au$, we denote by $\Atoms(q,\an)$ the set of atoms $\atom$ such that $(q,\atom)\in \an$.

\noindent  Let $(p,\gamma)\in Q\times (\Gamma\cup \{\bottom\})$ with $\Succ_\PS(p,\gamma)=\{(p_1,\beta_1),\ldots,(p_k,\beta_k)\}$ for some $1\leq k\leq k_\PS$.
\noindent For an annotation $\an$ and a tuple $\tpl{\an_1,\ldots,\an_k}$ of $k$ annotations, we say that
$\tpl{\an_1,\ldots,\an_k}$ \emph{is consistent with  annotation $\an$} and the pair $(p,\gamma)$   if the following conditions are fulfilled:
\begin{itemize}
  \item $\an=\bigcup_{i=1}^{i=k}\an_i$;
  \item for each move $\eta=(q,(q',m,A)) \in\an$, let  $X_\eta$ be the subset  of $\{(p_1,\beta_1),\ldots,(q_k,\beta_k)\}$ consisting of the pairs $(p_i,\beta_i)$ such that $\eta\in \an_i$. Then, $X_\eta\neq \emptyset$ and the following holds:
  \begin{itemize}
  \item case $m=\Box$:  there is an  $A$-decision $d_A$ available in $(p,\gamma)$ (i.e.,  $\Delta(p,\gamma, d)\neq \Undef$ for some full decision $d$  consistent with $d_A$)
   such that for each full decision $d$  consistent with $d_A$, either $\Delta(p,\gamma, d)=\Undef$ or  $\Delta(p,\gamma, d)\in X_\eta$;
  \item case $m=\Diamond$:   for all available $A$-decisions $d_A$ available in $(p,\gamma)$, there is some   full decision $d$  consistent with $d_A$
  so that $\Delta(p,\gamma, d)\in X_\eta$.
\end{itemize}    
\end{itemize}
%For a move $\eta=(p,(p',m,A))\in Q_\Au\times \Atoms(\Au)$ and a non-empty subset $X$ of $\Succ_\PS(q,\gamma)$, we say that
%\emph{$X$ is consistent with the move $\eta$} if the following holds:
%\begin{itemize}
%  \item case $m=\Box$: there is an $A$-decision $d_A$ such that $X$ coincides with the set of pairs $\Delta(q,\gamma, d)$ where $d$ is a full decision  consistent with $d_A$;
%  \item case $m=\Diamond$: there is a surjective function $f:\Actions^{A}\mapsto X$ such that for each $A$-decision $d_A$, $f(d_A)=\Delta(q,\gamma,d)$ for
%  some full decision $d$ consistent with $d_A$.
%\end{itemize}
We denote by $\Con(p,\gamma,\an)$ the set of tuples $\tpl{\an_1,\ldots,\an_k}$ of $k$ annotations which are consistent with the annotation $\an$
and the pair $(p,\gamma)$.

Intuitively, the previous definition allows to express requirements $\ReqFour$ and $\ReqFive$ in case all the possible choices are enabled. 
More precisely, in case the concrete node $\nu_x$ of the environment strategy tree $\Tau$ in conditions $\ReqFour$ and $\ReqFive$ is associated with a configuration of the form
$(p,\gamma\cdot\beta)$ and all the children of $\nu_x$ in $\Unw(\GS(\PS))$ are also children of $\nu_x$ in $\Tau$, then $\ReqFour$ and $\ReqFive$ are equivalent to the following condition: $N=k$ and
  $\tpl{\an'_1,\ldots,\an'_k}\in \Con(p,\gamma,\an)$.
  
Now assume that not all the children of node $\nu_x$ in  $\Unw(\GS(\PS))$ are also children of $\nu_x$ in $\Tau$. This entails that node $\nu_x$ is associated with an environment configuration $(p,\gamma\cdot\beta)$.
Since $(p,\gamma\cdot\beta)$ is controlled by the environment, we observe that   $\tpl{\an_1,\ldots,\an_k}\in \Con(p,\gamma,\an)$   iff the following conditions hold:
 \begin{itemize}
  \item $\an=\bigcup_{i=1}^{i=k}\an_i$;
  \item for each move $\eta=(q,(q',m,A)) \in\an$, the following holds:
  \begin{itemize}
  \item if \emph{either} $m=\Diamond$ and $\env\in A$, \emph{or} $m=\Box$ and  $\env\notin A$, then   $\eta\in \an_i$ for some $i\in \{1,\ldots,k\}$; 
  \item otherwise,  $\eta\in \an_i$ for each $i\in \{1,\ldots,k\}$.
\end{itemize}    
\end{itemize} 
This justifies the following definition: an annotation $\an$ is \emph{obligation-free} if it does not contain moves of the form $(q,(q',m,A))$ such that \emph{either} $m=\Diamond$ and $\env\in A$, \emph{or} $m=\Box$ and  $\env\notin A$. Then, in case the concrete node $\nu_x$ in $\Tau$ is associated with an environment configuration $(p,\gamma\cdot\beta)$,  
requirements $\ReqFour$ and $\ReqFive$ are equivalent to the following condition.
  \begin{itemize}
  \item  $N\leq k$ and there are distinct indexes $i_1,\ldots,i_N\in  \{1,\ldots,k\}$ such that the following holds for some $\tpl{\an''_1,\ldots,\an''_k}\in \Con(p,\gamma,\an)$: $\an'_{j}= \an''_{i_j}$
  for all $1\leq j\leq N$, and (ii) $\an''_\ell$ is obligation-free for all $\ell\in \{1,\ldots,k\}\setminus \{i_1,\ldots,i_n\}$.
\end{itemize}    
By using the previous definitions and observations, we now define the parity \NPTA\ $\Pu_{\textit{wf}}$ of index $1$ satisfying Claim~1. Essentially, given a $\Sigma$-labeled complete $k_\PS$-tree
$\tpl{\{1,\ldots,k_\PS\}^{*},\Lab_\Sigma}$, the automaton $\Pu_{\textit{wf}}$, by simulating the behaviour of the open $\PMS$ $\PS$ and by exploiting the transition function of the parity $\ACG$ $\Au$, checks that the input is a \emph{well-formed} annotated extension of
the $\bot$-completion encoding of some environment strategy tree of $\GS(\PS)$.
Formally, the \NPTA\ $\Pu_{\textit{wf}}=  \tpl{\Sigma,P,\Gamma\cup \{\bottom\},p_0,\rho,\Omega: p\in P \mapsto \{0\}}$
is defined as follows.

The set $P$ of  states  consists  of the triples $(q,\an,m)$ where $q\in Q$ is a  state
of the \PMS\ $\PS$, $\an\in\Ann$ is an annotation, and $m\in \{\bot,\top,\vdash\}$ is a state marker such that $\an = \emptyset$ if $m=\bot$.
Intuitively, whenever the current input node $x$ is a concrete node, then $q$ represents the state of $\PS$ associated with node $x$. The meaning of the state marker $m$
is as follows. When the state marker $m$ is $\bot$, the $\NPTA$ $\Pu_{\textit{wf}}$  can read
    only the letter $\bot$, while when the state marker is $\top$, $\Pu_{\textit{wf}}$ can read only letters
  in $\Sigma\setminus \{\bot\}$. Finally, when $\Pu_{\textit{wf}}$ is in states of the form
    $(q,\an,\vdash)$, then it  can read   both letters in $\Sigma\setminus \{\bot\}$ and the letter
    $\bot$. In this  case, it is left to the environment to
    decide whether the transition to a configuration of the simulated $\PMS$ $\PS$ of  the form
    $(q,\beta)$ is enabled. Intuitively, the three types of
    states are used to ensure that the environment enables all
    transitions from enabled system configurations, enables at
    least one transition from each enabled
    environment configuration, and disables transitions from
    disabled configurations. Moreover, the annotation $\an$ in a control state $(q,\an,m)$
    of $\Pu_{\textit{wf}}$ represents the guessed  subset of the moves in the first annotation of the parent $x'$ (if any) of the current \emph{concrete} input node for which, starting from $x'$, a copy of $\Au$ is sent to the current input node:  in the transition function, we require that in case  the current input symbol $\sigma$ is not $\bot$,   $\an$ coincides with the second annotation of $\sigma$.

The initial state $p_0$ is given by $(q_0,\emptyset,\top)$. Finally, the transition function $\rho:P\times \Sigma \times (\Gamma\cup
\{\bottom\})\rightarrow 2^{(P\times \Gamma^*)^{k_\PS}}$ is defined as follows.
According to the definition of
$P$, the automaton $\Pu_{\textit{wf}}$ can be in a state of the form
	$(q,\emptyset,\bot)$, $(q,\an,\top)$, or $(q,\an,\vdash)$.
	Both in the first and the third cases, $\Pu_{\textit{wf}}$ can read $\bot$,
	which means that the automaton is reading a disabled or a completion node.
	Thus, independently from the fact that the actual configuration of the automaton is associated with an environment or a system configuration
of the open $\PMS$ $\PS$,
	$\rho$ propagates states of the form $(q,\emptyset,\bot)$ to all children of the reading node. Note that for states
of the form $(q,\an,\vdash)$ and in case $\Pu_{\textit{wf}}$  reads $\bot$, we require that $\an$ is obligation-free.
	If instead the automaton is in a state of the form $(q,\an,\top)$ or $(q,\an,\vdash)$ and reads a label different from $\bot$,
 the possible successor states further depend on the particular kind of the configuration
	in which the automaton is.
	If $\Pu_{\textit{wf}}$ is in a system configuration of $\PS$,
	then all the children of the reading node associated with the successors of such a configuration in the $\CGS$ $\GS(\PS)$  must not be disabled and so,
	$\rho$ sends to all of them states with marker $\top$.
	If $\Pu_{\textit{wf}}$ is in an environment configuration of $\PS$, then
all the children of the reading node, but one, associated with the successors of such a configuration in the $\CGS$ $\GS(\PS)$
 may be disabled and so,
	$\rho$ sends to all of them states with marker $\vdash$, except one, to which $\rho$ sends a state with marker $\top$.

\noindent Formally, let $(q,\an,m)\in P$, $\sigma\in\Sigma$, and $\gamma\in\Gamma\cup\{\bottom\}$
 with $\Succ_{\PS}(q,\gamma)=\tpl{(q_1,\beta_1),\ldots, (q_k,\beta_k)}$
    ($1\!\leq\! k\!\leq\! k_\PS$). Then,
    $\rho((q,\an,m),\sigma,\gamma)$ is defined as follows:

 \begin{itemize}
        \item Case $m \in \{\bot,\vdash\}$, $\sigma=\bot$, and $\an$ is obligation-free:
        \begin{displaymath}
          \rho((q,\an,m),\bot,\gamma)=\{\tpl{\ \underbrace{((q,\emptyset,\bot),\varepsilon),\ldots,((q,\emptyset,\bot),\varepsilon)}_{k_{\PS} \ pairs} \ }\}
        \end{displaymath}
        That is, $\rho((q,\an,m),\bot,\gamma)$ contains exactly one $k_\PS$-tuple.
      In this case all the successors of the current $\PS$-configuration are disabled.
      \item Case $m \in \{\top,\vdash\}$, $(q,\gamma)$ is associated with \emph{system} $\PS$-configurations, $\sigma=(\Lab(q),\an',\an)$ for some annotation $\an'$ such that $\Cod(\an)=\Dom(\an')$, and for each $q_\Au\in \Dom(\an')$, $\Atoms(q_\Au,\an')$ is a  model of $\delta(q_\Au,\Lab (q))$:
      \[
     \begin{array}{ll}
     \rho((q,\an,m),\sigma,\gamma) = &  \displaystyle{\bigcup_{\tpl{\an_1,\ldots,\an_k}\in \Con(q,\gamma,\an')}}  \{\tpl{((q_1,\an_1,\top),\beta_1),\ldots,((q_k,\an_k,\top),\beta_k), \vspace{0.2cm}\\
      & \phantom{\bigcup_{\tpl{\an_1,\ldots,\an_k}\in \Con(q,\gamma,\an')}}\,\underbrace{((q,\emptyset,\bot),\varepsilon),\ldots((q,\emptyset,\bot),\varepsilon)}_{k_{\PS}-k \ pairs} \ } \}
     \end{array}
     \]
   In this case, all the $k$ successors of the current system $\PS$-configuration are enabled. Moreover,  the automaton
   guesses a tuple  $\tpl{\an_1,\ldots,\an_k}$ of $k$ annotations which are consistent with the first annotation $\an'$ of the input node
and the pair $(q,\gamma)$, and sends state $(q_i,\an_i,\top)$ to the $i$th child of the current input node for all $1\leq i\leq k$.
   \item Case $m \in \{\top,\vdash\}$, $(q,\gamma)$ is associated with \emph{environment} $\PS$-configurations,  $\sigma=(\Lab(q),\an',$ $\an)$ for some annotation $\an'$ such that $\Cod(\an)=\Dom(\an')$, and for each $q_\Au\in \Dom(\an')$, $\Atoms(q_\Au,\an')$ is a  model of $\delta(q_\Au,\Lab (q))$. In this case $\rho((q,\an,m),\sigma,\gamma)$ is defined
      as follows:
      \[
     \begin{array}{l}
      \hspace{0.2cm}  \displaystyle{\bigcup_{\tpl{\an_1,\ldots,\an_k}\in \Con(q,\gamma,\an')}} \,\bigl\{ \vspace{0.2cm}\\
      \hspace{0.2cm}    \tpl{((q_1,\an_1,\top),\beta_1),((q_2,\an_2,\vdash),\beta_2),\ldots,((q_k,\an_k,\vdash),\beta_k),((q,\emptyset,\bot),\varepsilon),\ldots((q,\emptyset,\bot),\varepsilon)}, \\
      \hspace{0.2cm}   \tpl{((q_1,\an_1,\vdash),\beta_1),((q_2,\an_2,\top),\beta_2),\ldots,((q_k,\an_k,\vdash),\beta_k), ((q,\emptyset,\bot),\varepsilon),\ldots((q,\emptyset,\bot),\varepsilon)}, \\
       \hspace{5cm} \vdots \\
      \hspace{0.2cm}   \tpl{((q_1,\an_1,\vdash),\beta_1),((q_2,\an_2,\vdash),\beta_2),\ldots,((q_k,\an_k,\top),\beta_k), ((q,\emptyset,\bot),\varepsilon),\ldots((q,\emptyset,\bot),\varepsilon) } \bigr\}
     \end{array}
     \]
   Thus, the automaton guesses a tuple  $\tpl{\an_1,\ldots,\an_k}$ of $k$ annotations which is consistent with the first annotation $\an'$ of the input node
and the pair $(q,\gamma)$ and, additionally, guesses an index $1\leq i\leq k$. With these choices,  the automaton
   sends state $(q_i,\an_i,\top)$ to the $i$th child of the current input node and, additionally,
   ensures that the $i$th successor  of the current environment $\PS$-configuration is enabled while  all the other successors may be disabled.
   \item \emph{All the other cases:} $\rho((q,\an,m),\sigma,\gamma)=\emptyset$.
    \end{itemize}

Note that  $\Pu_{\textit{wf}}$ has  $O(|Q| \cdot  2^{O(|Q_\Au |\cdot |\Atoms(\Au)|)})$ states,  $||\rho||= O(|\Delta|)$, and
$|\rho|=O(|\Delta|\cdot 2^{O(k_{\PS}\cdot |Q_\Au |\cdot |\Atoms(\Au)|)})$. 
By construction, it easily follows that a $\Sigma$-labeled complete $k_\PS$-tree $\tpl{\{1,\ldots,k_\PS\}^{*},\Lab_\Sigma}$ 
is accepted by the \NPTA\ $\Pu_{\textit{wf}}$ \emph{iff} there is an environment strategy tree $\Tau$ of $\GS(\PS)$ such that $\tpl{\{1,\ldots,k_\PS\}^{*},\Lab_\Sigma}$ is some  \emph{well-formed} annotated extension  of the $\bot$-completion encoding of $\Tau$.
This concludes the proof of Claim~1.
\end{proof}

\section{\FOUREXPTIME--hardness of ATL* pushdown module-checking}\label{sec:LowerBound}

In this section, we establish the following result.

\begin{thm}\label{theorem:lowerbound}
Pushdown module-checking against \ATLStar\ is
    \FOUREXPTIME--hard even for two-player turn-based $\PMS$. % of fixed size.
\end{thm}

Theorem~\ref{theorem:lowerbound} is proved by a polynomial-time  reduction from the acceptance problem for \THREEXPSPACE--bounded Alternating
 Turing Machines (ATM, for short) with a binary branching degree.
 Formally, such a machine
  is a tuple
$\mathcal{M}=\tpl{\Sigma,Q,Q_{\forall},Q_{\exists},q_0,\delta,F}$,
where $\Sigma$ is the input alphabet  which contains the blank
symbol $\#$, $Q$ is the finite set of states which is partitioned
into $Q=Q_{\forall} \cup  Q_{\exists}$, $Q_{\exists}$ (resp.,
$Q_{\forall}$) is the set of existential (resp., universal)
states, $q_0$ is the initial state, $F\subseteq Q$ is the set  of
accepting states, and the transition function $\delta$ is a
mapping $\delta: Q\times \Sigma\rightarrow (Q \times\Sigma \times
\{\leftarrow,\rightarrow\})^2$. Note that since  $\mathcal{M}$ has a binary branching degree, the transition
function $\delta$ nondeterministically associates to each pair state/input symbol $(q,\sigma)$
two possible moves, where each move is represented by a triple $(q',\sigma',d)$ consisting of a target state
$q'$, the symbol $\sigma'$ to write in the tape cell currently pointed by the reading head, and
a symbol $d\in\{\leftarrow,\rightarrow\}$ encoding the movement of the reading head: $\leftarrow$ (resp., $\rightarrow$) means that
the reading head moves one cell to the left (resp., to the right) of the current cell.

Formally, configurations of $\mathcal{M}$ are words in
$\Sigma^*\cdot(Q\times \Sigma)\cdot\Sigma^*$. A configuration
$C= \eta\cdot(q,\sigma)\cdot\eta'$ denotes that the tape content is
$\eta\cdot\sigma\cdot\eta'$, the current state (resp., current input symbol) is $q$ (resp., $\sigma$), and the reading head
is at position $|\eta|+1$. From a configuration $C$,  the machine $\mathcal{M}$
nondeterministically chooses a triple $(q',\sigma',d)$ in
$\delta(q,\sigma)=
\tpl{(q_l,\sigma_l,d_l),(q_r,\sigma_r,d_r)}$, and then moves
to state $q'$, writes $\sigma'$ in the current tape cell, and its
reading head moves one cell to the left or to the right, according
to $d$. We denote by $succ_l(C)$ and
$succ_r(C)$ the successors of $C$ obtained by choosing respectively
the left and the right triple in
$\tpl{(q_l,\sigma_l,d_l),(q_r,\sigma_r,d_r)}$ (note that the terms `left' and `right' here
should not be confused with the movement of the reading head of the ATM).
The
configuration $C$ is accepting  (resp., universal, resp., existential) if the associated state $q$
is in $F$ (resp., in $Q_\forall$, resp., in $Q_\exists$).

Given an input $\alpha\in\Sigma^{+}$, a (finite) computation tree
of $\mathcal{M}$ over $\alpha$ is a finite tree in which each node is labeled
by a configuration. The root of the tree is labeled by the
initial configuration $(q_0,\alpha(0)) \alpha(1) \ldots \alpha(n-1)$ associated with $\alpha$, where $n=|\alpha|$.
 An \emph{internal} node that is labeled by a universal configuration
$C$  has two
children, corresponding to $succ_l(C)$ and $succ_r(C)$, while an internal
node labeled by an existential configuration $C$  has a single child,
corresponding to either $succ_l(C)$ or $succ_r(C)$.
 The tree is
accepting if each leaf is labeled by an accepting configuration. An input $\alpha\in\Sigma^+$ is \emph{accepted} by
$\mathcal{M}$ if there is an accepting computation tree
of $\mathcal{M}$ over $\alpha$.

If the ATM $\mathcal{M}$ is \THREEXPSPACE--bounded, then there
 is a constant
  $c\geq 1$ such that for each  $\alpha\in \Sigma^+$, the
 space needed by $\mathcal{M}$ on input $\alpha$ is bounded by
  $\Tower(|\alpha|^{c},3)$, where for all $n,h\in\Nat$, $\Tower(n, h)$ denotes a tower of exponentials
of height $h$ and argument $n$ (i.e, $\Tower(n, 0) = n$ and $\Tower(n, h + 1) = 2^{\Tower(n,h)}$).  It is well-known~\cite{CKS81} that
the acceptance problem for \THREEXPSPACE--bounded ATM (with a binary branching degree)
is \FOUREXPTIME-complete. % even if the ATM is assumed to be of fixed size.

 Fix
   a  \THREEXPSPACE--bounded ATM
$\mathcal{M}$ and an input $\alpha\in \Sigma^+$. Let $n=|\alpha|$. W.l.o.g. we  assume that the constant $c$ is $1$ and $n>1$.
 Hence, any reachable configuration of $\mathcal{M}$
over $\alpha$ can be seen as a word in $\Sigma^*\cdot(Q\times
\Sigma)\cdot\Sigma^*$ of length exactly
 $\Tower(n,3)$, and the initial configuration $ (q_0,\alpha(0)) \alpha(1) \ldots \alpha(n-1)$ can be represented as the word
 of  length $\Tower(n,3)$ given by 
 \[
 (q_0,\alpha(0)) \alpha(1) \ldots \alpha(n-1)\cdot (\#)^{t}
 \]
where $t=\Tower(n,3)-n$. Note that
for an  ATM configuration $C=u_1 u_2\ldots u_{\Tower(n,3)}$ and for all $i\in [1, \Tower(n,3)]$ and $\dire\in \{l,r\}$, the
value $u'_i$ of the $i$-th cell of $succ_{\dire}(C)$ is completely determined by the values
$u_{i-1}$, $u_{i}$ and $u_{i+1}$ (taking $u_{i+1}$ for $i=\Tower(n,3)$ and
 $u_{i-1}$ for $i=1$ to be some special symbol, say $\vdash$).  Thus, we denote by
 $\Succ_{\dire}(u_{i-1},u_i,u_{i+1})$  the  value $u'_i$ of the $i$-th cell of $succ_{\dire}(C)$
(note that the function $\Succ_{\dire}$ can be trivially obtained from the transition function of
$\mathcal{M}$).
According to the previous observation, we use the set $\Lambda$ of triples of the form $(u_p,u,u_s)$ where $u\in  \Sigma \cup (Q\times \Sigma)$, and $u_p,u_s\in \Sigma \cup (Q\times \Sigma)\cup \{\vdash\}$.
We prove the following result from which Theorem~\ref{theorem:lowerbound} directly follows.

\begin{thm}\label{theorem:lowerboundReduction} One can construct, in time polynomial in $n$ and the size of $\mathcal{M}$, an open turn-based $\PMS$
$\PS$ and an $\ATLStar$ formula $\varphi$ over the set of agents $\Agents =\{\sys,\env\}$ such that
$\mathcal{M}$ accepts $\alpha$ iff
there is an environment strategy tree in $ \exec(\GS(\PS))$ that satisfies $\varphi$  iff
$\GS(\PS)\not\models^r \neg \varphi$. %Moreover, the size of $\GS(\PS)$ depends only on the size of $\mathcal{M}$.
\end{thm}

The rest of this section is devoted to the proof of Theorem~\ref{theorem:lowerboundReduction}.

\subsection{Encoding of ATM configurations}

We first define an %suitable e
encoding of the ATM configurations by using the following set $\MainP$ of atomic propositions:
\[
\MainP:= \Lambda\cup \{0,1,
\forall,\exists,l,r,acc\} \cup  \{\Start{1},\Start{2},\Start{3},\End{1},\End{2}, \End{3}\}.
\]
In the encoding of an ATM configuration, for each ATM cell, we record the content of the cell,  the location (cell number)
of the cell on the ATM tape, and  the contents of the previous and next cell (if any).  In order to encode the cell number, which is a natural number in $[0,\Tower(n,3)-1]$, for all $1\leq h\leq 3$, we define the notions of \emph{$h$-block} and  \emph{well-formed $h$-block}.
For $h=1,2$, well-formed
$h$-blocks encode integers in $[0,\Tower(n,h)-1]$, while well-formed
\mbox{$3$-blocks}  encode the cells of
ATM configurations.  In particular, for $h=2,3$, a
well-formed $h$-block encoding a  natural number $m\in [0,\Tower(n,h)-1]$ is a sequence of $\Tower(n,h-1)$ well-formed $(h-1)$-blocks,
where the $i^{th}$ $(h-1)$-block encodes both the value and
(recursively)
the position of the $i^{th}$-bit in the binary
representation of $m$.

Formally, for each  $1\leq h\leq 3$, an \emph{$h$-block} $\bl$
is a word  of the form
\[
\{\Start{h}\}\cdot  \bl_0 \ldots \bl_t \cdot \{\tau\} \cdot \{\End{h}\}\mbox{, where}
\]
\begin{itemize}
  \item $t\geq 0$,
  \item  $\tau\in\{0,1\}$
if $h\neq 3$, and $\tau\in \Lambda$ otherwise (we say that $\tau$ is the \emph{content} of $\bl$),
\item if $h=1$, then for all $0\leq i\leq t$, $bl_i\in\{0,1\}$, 
  \item if $h>1$, then  for all $0\leq i\leq t$, $\bl_i$ is an $(h-1)$-block.
\end{itemize}

\noindent Note that the $h$-block $\bl$ is enclosed by the start delimiter $\Start{h}$ and the end delimiter
$\End{h}$. We say that the $h$-block $\bl$ is \emph{well-formed} if the following additional conditions hold:
\begin{itemize}
\item Case $h=1$: $t=n-1$. In this case, the \emph{number}  of $bl$ is the natural number
in $[0,\Tower(n,1)-1]$ whose binary code is given by $bl_0\ldots bl_t$.\vspace{0.2cm}
\item Case $h>1$: $t=\Tower(n,h-1)-1$ and the $(h-1)$-block
$\bl_i$ is well-formed and has \emph{number} $i$ for each $0\leq i\leq t$. 
 If $\bl$ is well-formed, then the \emph{number} of $\bl$ is the natural number
in $[0,\Tower(n,h)-1]$ whose binary code is given by $b_0\ldots b_t$ where $b_i$ is the content of the sub-block $\bl_i$ for all $0\leq i\leq t$.
\end{itemize}

\begin{exa} Let $n=2$. In this case $\Tower(n,2)=16$ and $\Tower(n,1)=4$. Thus, we can encode by well-formed $2$-blocks all the integers in $[0,15]$. For example, let us consider the number 14 whose binary code
$($using $\Tower(n,1)=4$ bits$)$ is given by $0111$ $($assuming that the first bit is the least significant one$)$. For each $b\in\{0,1\}$, the
well-formed $2$-block with content $b$ and number 14 is given by
\[
  \{\Start{2}\}\{\Start{1}\} \{0\}\{0\}\{0\}\{\End{1}\} \{\Start{1}\} \{1\}\{0\}\{1\}\{\End{1}\}  \{\Start{1}\} \{0\}\{1\}\{1\}\{\End{1}\}
   \{\Start{1}\} \{1\}\{1\}\{1\}\{\End{1}\} \{b\} \{\End{2}\}
\]
\noindent Note that the $1$-sub-blocks also encode the position of each bit in the binary code of  14.
Now, let us consider $\tau\in \Lambda$ and $\ell\in [0,2^{16}-1]$, and let $b_0\ldots b_{15}$ be the binary code of $\ell$. Then, the well-formed  $3$-block with content $\tau$ and number $\ell$
is given by the word $\{\Start{3}\} \bl_0,\ldots,\bl_{15}\{\tau\}\{\End{3}\}$, where for each $i\in [0,15]$, $\bl_i$ is the well-formed
$2$-block having content $b_i$ and number $i$.
\end{exa}

%Note that for a $3$-block $\bl$, if the content $\tau$ of $\bl$ is of the form $(u_p,u,u_s)$, then $u$ represents the value  of the encoded ATM cell, while $u_p$ %(resp., $u_s$) represents the value of the previous (resp., next) cell in the ATM configuration.
ATM configurations $C=u_1 u_2\ldots u_k$ (note that here we do not require that $k=\Tower(n,3)$) are then encoded by  words $w_C$  of the form
\[
w_C = \Tag_1 \cdot bl_1 \cdot \ldots \cdot bl_k \cdot \Tag_2\mbox{, where}
\]
\begin{itemize}
  \item  $\Tag_1\in\{\{l\},\{r\}\}$,
  \item  for each $i\in [1,k]$, $bl_i$ is a $3$-block whose content is $(u_{i-1},u_i,u_{i+1})$ (where $u_0=\,\vdash$ and $u_{k+1}=\,\vdash$),
\item  $\Tag_2=\{acc\}$ if $C$ is accepting,  $\Tag_2=\{\exists\}$ if $C$ is  non-accepting and existential, and $\Tag_2=\forall$ otherwise.
\end{itemize}

\noindent  The symbols $l$ and $r$ are used to encode a left and a right ATM successor, respectively. We also use the symbol $l$ to encode the initial configuration.
If $k=\Tower(n,3)$ and for each $i\in [1,k]$, $bl_i$ is a well-formed  $3$-block  having number
$i-1$, then we say that $w_C$ is a \emph{well-formed code} of $C$.
A sequence $w_{C_1}\cdot \ldots \cdot w_{C_p}$ of well-formed ATM configuration codes   is \emph{faithful to the evolution} of $\mathcal{M}$ if for each $1\leq i<p $, either $w_{C_{i+1}}$ starts with the symbol $l$ and $C_{i+1}=succ_l(C_i)$, or $w_{C_{i+1}}$ starts with the symbol $r$ and $C_{i+1}=succ_r(C_i)$.

\paragraph{Definition of $\Prop$ and marked blocks} The set $\Prop$ of atomic propositions is defined as follows:
\[
\Prop =\MainP\cup \{check_1,check_2,check_3,\widehat{check_3}\}
\]
where the atomic propositions in  $\{check_1,check_2,check_3,\widehat{check_3}\}$ are intuitively used to mark the contents of $h$-blocks.
In particular, in the reduction, we also consider \emph{marked} $h$-blocks for each $h=1,2,3$. Formally, for each $h=1,2$, a marked $h$-block $\bl$ 
is defined as an $h$-block but the content of $\bl$ is additionally labeled by proposition $check_h$. A marked $3$-block $\bl$ is defined as a
$3$-block  but the content of $\bl$ is additionally labeled either by proposition $check_3$ (\emph{$check_3$-marked $3$-block}) or by proposition  $\widehat{check_3}$
($\widehat{check_3}$-marked $3$-block).

\subsection{Construction of the open  PMS $\PS$ in Theorem~\ref{theorem:lowerboundReduction}} %% and encoding of accepting computation trees on $\alpha$}
We now describe the behaviour of the open turn-based    $\PMS$ $\PS$  over $\Agents =\{\sys,\env\}$  in Theorem~\ref{theorem:lowerboundReduction}.
Since $\PS$ is turn-based, every configuration is either controlled by the agent $\sys$ (the \emph{system}) or by the 
agent $\env$ (the \emph{environment}). Thus, in the following, for \emph{external} (resp., \emph{internal}) nondeterminism, we mean that at a given configuration, the choices  are resolved by the environment (resp., system) agent, i.e., the configuration is controlled by the environment (resp., the system) agent.
Intuitively, the $\PMS$ generates, for different environment behaviors, all the possible computation trees of $\mathcal{M}$. %Hence, the computation trees  of $\mathcal{M}$ over $\alpha$  correspond to  trees in $ \exec(\GS(\PS))$.
 External nondeterminism is used in order to produce the actual symbols of each ATM configuration code.
 Whenever the $\PMS$
$\PS$ reaches the end of an existential (resp., universal) guessed ATM configuration code $w_C$, it simulates the existential (resp., universal) choice of $\mathcal{M}$ from $C$ by external (resp., internal) nondeterminism, and, in particular, $\PS$ chooses a symbol in $\{l,r\}$ and marks the next guessed ATM configuration with this symbol.
 % (where $l$ denotes a left ATM successor and $r$ denotes a right ATM successor).
This ensures that, once we fix the environment behavior, we really get a tree $T$ where each existential ATM configuration code is followed by (at least) one ATM configuration code marked by a symbol in $\{l,r\}$,  and every universal configuration is followed (in different branches) by two ATM configurations codes, one marked by the symbol $l$ and the other one marked by the symbol $r$.

We have to check that the guessed computation tree $T$ (corresponding to environment choices) corresponds to a legal computation tree of $\mathcal{M}$ over $\alpha$. To that purpose, we have to check several properties about each computation path $\pi$ of $T$, in particular:
\begin{description}
  \item[($\CondOne$)] the ATM configurations codes are well-formed  (i.e., the $\Tower(n,3)$-bit counter is properly updated),
  \item[($\CondTwo$)] $\pi$ is faithful to the evolution of $\mathcal{M}$,
  \item[($\CondThree$)] the first ATM configuration of $\pi$ corresponds to the initial ATM configuration over the input $\alpha$.
\end{description}
\noindent The $\PMS$ $\PS$ cannot guarantee by itself these requirements. Thus, these checks are performed by a suitable $\ATLStar$ formula $\varphi$. However, in order to construct an $\ATLStar$ formula of size polynomial in $n$ and in the size of the ATM $\mathcal{M}$, we need to `isolate' the (arbitrary) selected path $\pi$ from the remaining part of the tree. This is the point where we use the stack of the $\PMS$ $\PS$. As the ATM configurations codes are guessed symbol by symbol, they are pushed onto the stack of the $\PMS$ $\PS$.
This phase is called \emph{push-phase}. 

In the push phase, the $\PMS$ $\PS$ ensures that whenever an $acc$-node $x$ (i.e., a node with label $\{acc\}$) is reached in the unwinding $\Unw(\GS(\PS))$, then the finite path $\pi$ from the root to node $x$ is labeled 
by  a sequence of ATM configuration codes where the last ATM configuration is accepting: we call such finite paths $\pi$ of $\Unw(\GS(\PS))$ \emph{accepting} computation paths. 
Moreover, the stack content associated with node $x$ corresponds to the reverse of the labeling of $\pi$. 
Note that in this phase the unique nodes which are controlled by the system player are the nodes labeled
by the proposition $\forall$, where intuitively the system player simulates the universal choices of the ATM $\mathcal{M}$ from a universal configuration: 
the $\forall$-node has two children, one labeled by $l$ (the first symbol of an ATM $l$-successor) and the other one labeled by $r$ (the first symbol of an ATM $r$-successor).
Thus, $\mathcal{M}$ accepts $\alpha$ \emph{iff} there is an environment strategy tree $\Tau$ of $\GS(\PS)$ such that (i) each path of $\Tau$ from the root visits an $acc$-node  and  (ii)
for each $acc$-node $x$ of $\Tau$, the 
accepting computation path $\pi$ associated with node $x$ satisfies Conditions~$\CondOne$--$\CondThree$. 

\paragraph{Pop Phase.} Whenever an $acc$-node $x$ is reached, the $\PMS$ moves to the so called \emph{pop-phase}. Let $\pi$ be the  accepting computation path associated with $x$ (i.e., the finite path of
$\Unw(\GS(\PS))$ from the root to $x$). Recall that $\PS$ ensures that the labeling of $\pi$ is a sequence of ATM configuration codes where the last ATM configuration is accepting and the stack content of
$x$ corresponds to the reverse of the $\pi$'s labeling.  Let us denote by $\tpl{T_\pi,\Lab_\pi}$ the subtree of the unwinding $\Unw(\GS(\PS))$
 rooted at the last node of $\pi$ (i.e., node $x$).

\begin{figure}[tb]%[htp]
  \centering
  \caption{Subtree of the unwinding $\Unw(\GS(\PS))$ of the open $\PMS$ $\PS$ rooted at an $acc$-node (pop-phase)}\label{FigureLowerBoundFirst}
  \begin{minipage}{.44\textwidth}
\begin{tikzpicture}[scale=0.63]
\draw[draw=none,use as bounding box](-5.0,2.5) rectangle (3.3,-7.5);

\coordinate [label=above:{\footnotesize $acc$}] (RootCheckTree) at (0.0,0);
\fill [black] (RootCheckTree) circle (2pt);
\coordinate [label=left:{\footnotesize $\End{3}$\,}] (FirstEndThree) at (0.0,-0.8);

\path[->, thin,black] (RootCheckTree) edge  (0.0,-0.6);

 \coordinate [label=left:{\footnotesize $\Start{3}$\,}] (FirstStartThree) at (-1.4,-2.2);
 \coordinate (FirstStartUp) at (-1.7,-1.9);
 \coordinate (FirstStartBelow) at (-1.1,-2.5);
 \draw[thin, dotted, black] (FirstEndThree) edge (FirstStartUp);
  \draw[thin, dotted, black] (FirstEndThree) edge (FirstStartBelow);
  \draw[thin, dotted, black] (FirstStartUp) edge (FirstStartBelow);

\fill [black] (FirstStartThree) circle (2pt);
\draw[thick, black] (FirstEndThree) edge (FirstStartThree);

\path[<->, thin,dotted,black] (-3.0,-0.8) edge  (-3.0,-2.2);
\coordinate [label=left:\textcolor{blue}{\footnotesize $3$-block in}] (NoteA) at (-3.0,-1.3);
\coordinate [label=left:\textcolor{blue}{\footnotesize reverse order}] (NoteB) at (-3.0,-1.7);

 \coordinate [label=right:{\footnotesize $\Start{3}$\,}] (FirstStartThreeMark) at (1.4,-2.2);
 \draw[thin, dotted, black] (1.2,-2.6) edge (1.2,-3.2);
  \draw[thin, dotted, black] (1.6,-2.6) edge (1.6,-3.2);

  \coordinate (FirstStartMarkUp) at (1.1,-2.5);
 \coordinate (FirstStartMarkBelow) at (1.7,-1.9);
 \draw[thin, dotted, black] (FirstEndThree) edge (FirstStartMarkUp);
  \draw[thin, dotted, black] (FirstEndThree) edge (FirstStartMarkBelow);
  \draw[thin, dotted, black] (FirstStartMarkUp) edge (FirstStartMarkBelow);

  \coordinate [label=right:{\footnotesize $\CheckB{3}$}] (NoteCheckFirst) at (0.6,-1.3);

\fill [black] (FirstStartThreeMark) circle (2pt);
\draw[thick, black] (FirstEndThree) edge (FirstStartThreeMark);
\node[shape=circle,draw=black,inner sep=2pt,fill=white](A) at (FirstEndThree) {};
\fill [black] (FirstEndThree) circle (2pt);

\coordinate [label=left:{\footnotesize $\End{3}$\,}] (SecondEndThree) at (-1.4,-3.0);
\path[->, thin,black] (FirstStartThree) edge  (-1.4,-2.8);

 \coordinate [label=left:{\footnotesize $\Start{3}$\,}] (SecondStartThree) at (-2.8,-4.4);
 \coordinate (SecondStartUp) at (-3.1,-4.1);
 \coordinate (SecondStartBelow) at (-2.5,-4.7);
 \draw[thin, dotted, black] (SecondEndThree) edge (SecondStartUp);
  \draw[thin, dotted, black] (SecondEndThree) edge (SecondStartBelow);
  \draw[thin, dotted, black] (SecondStartUp) edge (SecondStartBelow);

  \draw[thin, dotted, black] (-3.0,-4.8) edge (-3.0,-5.4);
  \draw[thin, dotted, black] (-2.6,-4.8) edge (-2.6,-5.4);
\fill [black] (SecondStartThree) circle (2pt);
\draw[thick, black] (SecondEndThree) edge (SecondStartThree);

 \coordinate [label=right:{\footnotesize $\Start{3}$\,}] (SecondStartThreeMark) at (0,-4.4);
 \coordinate (SecondStartMarkUp) at (0.3,-4.1);
 \coordinate (SecondStartMarkBelow) at (-0.3,-4.7);
 \draw[thin, dotted, black] (SecondEndThree) edge (SecondStartMarkUp);
  \draw[thin, dotted, black] (SecondEndThree) edge (SecondStartMarkBelow);
  \draw[thin, dotted, black] (SecondStartMarkUp) edge (SecondStartMarkBelow);
  \coordinate [label=right:{\footnotesize $\CheckB{3}$}] (NoteCheckSecond) at (-0.8,-3.5);

  \draw[thin, dotted, black] (-0.2,-4.8) edge (-0.2,-5.4);
  \draw[thin, dotted, black] (0.2,-4.8) edge (0.2,-5.4);
\fill [black] (SecondStartThreeMark) circle (2pt);
\draw[thick, black] (SecondEndThree) edge (SecondStartThreeMark);
\node[shape=circle,draw=black,inner sep=2pt,fill=white](A) at (SecondEndThree) {};
\fill [black] (SecondEndThree) circle (2pt);

\coordinate [label=right:{\footnotesize  \, =  System node}] (SystemNode) at (-2.5,-6.5);
\node[shape=circle,draw=black,inner sep=2pt,fill=white](SN) at (-2.5,-6.5) {};
\fill [black] (SystemNode) circle (2pt);

\coordinate [label=right:{\footnotesize \,  =   Environment node}] (EnvironmentNode) at (-2.5,-7.1);
\fill [black] (EnvironmentNode) circle (2pt);

\end{tikzpicture}
 \end{minipage}\quad
\begin{minipage}{.44\textwidth}
\begin{tikzpicture}[scale=0.63]
\draw[draw=none,use as bounding box](-2.0,4.5) rectangle (3.3,-6.0);

 \coordinate [label=above:{\footnotesize $\End{3}$\,}] (CheckOneEndThree) at (0,3.4);
 \coordinate [label=below right:{\footnotesize $\Start{3}$\,}] (CheckOneStartThree) at (0,2.0);
 \coordinate [label=right:{\footnotesize $\CheckB{3}$}] (NoteCheckFirst) at (0.0,2.8);

  \coordinate   (CheckOneLeft) at (-0.4,2.0);
  \coordinate   (CheckOneRight) at (0.4,2.0);

  \draw[thin, dotted, black] (CheckOneEndThree) edge (-0.8,2.8);

  \draw[thin, dotted, black] (CheckOneEndThree) edge (CheckOneLeft);
  \draw[thin, dotted, black] (CheckOneEndThree) edge (CheckOneRight);
  \draw[thin, dotted, black] (CheckOneLeft) edge (CheckOneRight);
  \draw[thick,black] (CheckOneEndThree) edge  (CheckOneStartThree);

   \fill [black] (CheckOneStartThree) circle (2pt);
   \node[shape=circle,draw=black,inner sep=2pt,fill=white](A) at (CheckOneEndThree) {};
 \fill [black] (CheckOneEndThree) circle (2pt);

\coordinate [label=above:{\footnotesize $\exists$ or $\forall$}] (RootConfig) at (0.0,0);
\coordinate   (RootLeft) at (-0.5,0.0);
  \coordinate   (RootRight) at (0.5,0.0);
  \draw[thin, dotted, black] (CheckOneStartThree) edge (RootConfig);
   \draw[thin, dotted, black] (CheckOneStartThree) edge (RootLeft);
  \draw[thin, dotted, black] (CheckOneStartThree) edge (RootRight);
  \draw[thin, dotted, black] (RootLeft) edge (RootRight);

   \path[<->, thin,dotted,black] (6.0,3.4) edge  (6.0,2.0);
\coordinate [label=above:\textcolor{blue}{\footnotesize $3$-block in}] (NoteA) at (6.0,2.6);
\coordinate [label=above:\textcolor{blue}{\footnotesize reverse order}] (NoteB) at (6.0,2.2);

 \path[<->, thin,dotted,black] (6.0,2.0) edge  (6.0,0.0);
\coordinate [label=above:\textcolor{blue}{\footnotesize no marked $3$-blocks until}] (NoteC) at (6.0,0.8);
\coordinate [label=above:\textcolor{blue}{\footnotesize the next ATM configuration}] (NoteD) at (6.0,0.2);

\path[<->, thin,dotted,black] (6.0,0.0) edge  (6.0,-5.4);
\coordinate [label=above:\textcolor{blue}{\footnotesize ATM configuration code}] (NoteE) at (6.0,-1.1);
\coordinate [label=above:\textcolor{blue}{\footnotesize in reverse order}] (NoteF) at (6.0,-1.5);

\coordinate [label=above:\textcolor{blue}{\footnotesize no marked $3$-blocks in the}] (NoteE) at (6.0,-4.4);
\coordinate [label=above:\textcolor{blue}{\footnotesize next ATM configurations}] (NoteF) at (6.0,-5.0);

\fill [black] (RootConfig) circle (2pt);
\coordinate [label=left:{\footnotesize $\End{3}$\,}] (FirstEndThree) at (0.0,-0.8);

\path[->, thin,black] (RootConfig) edge  (0.0,-0.70);

 \coordinate [label=left:{\footnotesize $\Start{3}$\,}] (FirstStartThree) at (-1.4,-2.2);
 \coordinate (FirstStartUp) at (-1.7,-1.9);
 \coordinate (FirstStartBelow) at (-1.1,-2.5);
 \draw[thin, dotted, black] (FirstEndThree) edge (FirstStartUp);
  \draw[thin, dotted, black] (FirstEndThree) edge (FirstStartBelow);
  \draw[thin, dotted, black] (FirstStartUp) edge (FirstStartBelow);

\fill [black] (FirstStartThree) circle (2pt);
\draw[thick, black] (FirstEndThree) edge (FirstStartThree);

 \coordinate [label=right:{\footnotesize $\Start{3}$\,}] (FirstStartThreeMark) at (1.4,-2.2);
 \draw[thin, dotted, black] (1.2,-2.6) edge (1.2,-3.2);
  \draw[thin, dotted, black] (1.6,-2.6) edge (1.6,-3.2);

  \coordinate (FirstStartMarkUp) at (1.1,-2.5);
 \coordinate (FirstStartMarkBelow) at (1.7,-1.9);
 \draw[thin, dotted, black] (FirstEndThree) edge (FirstStartMarkUp);
  \draw[thin, dotted, black] (FirstEndThree) edge (FirstStartMarkBelow);
  \draw[thin, dotted, black] (FirstStartMarkUp) edge (FirstStartMarkBelow);

  \coordinate [label=right:{\footnotesize $\CheckMark{3}$}] (NoteCheckFirst) at (0.6,-1.3);

\fill [black] (FirstStartThreeMark) circle (2pt);
\draw[thick, black] (FirstEndThree) edge (FirstStartThreeMark);
\fill [black] (FirstEndThree) circle (2pt);

\coordinate [label=left:{\footnotesize $\End{3}$\,}] (SecondEndThree) at (-1.4,-3.0);
\path[->, thin,black] (FirstStartThree) edge  (-1.4,-2.9);

 \coordinate [label=left:{\footnotesize $\Start{3}$\,}] (SecondStartThree) at (-2.8,-4.4);
 \coordinate (SecondStartUp) at (-3.1,-4.1);
 \coordinate (SecondStartBelow) at (-2.5,-4.7);
 \draw[thin, dotted, black] (SecondEndThree) edge (SecondStartUp);
  \draw[thin, dotted, black] (SecondEndThree) edge (SecondStartBelow);
  \draw[thin, dotted, black] (SecondStartUp) edge (SecondStartBelow);

  \draw[thin, dotted, black] (-3.0,-4.8) edge (-3.0,-5.4);
  \draw[thin, dotted, black] (-2.6,-4.8) edge (-2.6,-5.4);
\fill [black] (SecondStartThree) circle (2pt);
\draw[thick, black] (SecondEndThree) edge (SecondStartThree);

 \coordinate [label=right:{\footnotesize $\Start{3}$\,}] (SecondStartThreeMark) at (0,-4.4);
 \coordinate (SecondStartMarkUp) at (0.3,-4.1);
 \coordinate (SecondStartMarkBelow) at (-0.3,-4.7);
 \draw[thin, dotted, black] (SecondEndThree) edge (SecondStartMarkUp);
  \draw[thin, dotted, black] (SecondEndThree) edge (SecondStartMarkBelow);
  \draw[thin, dotted, black] (SecondStartMarkUp) edge (SecondStartMarkBelow);
  \coordinate [label=right:{\footnotesize $\CheckMark{3}$}] (NoteCheckSecond) at (-0.8,-3.5);

  \draw[thin, dotted, black] (-0.2,-4.8) edge (-0.2,-5.4);
  \draw[thin, dotted, black] (0.2,-4.8) edge (0.2,-5.4);
\fill [black] (SecondStartThreeMark) circle (2pt);
\draw[thick, black] (SecondEndThree) edge (SecondStartThreeMark);
\fill [black] (SecondEndThree) circle (2pt);

\end{tikzpicture}
 \end{minipage}
\end{figure}

We now describe the branching behaviour of $\PS$ along $\tpl{T_\pi,\Lab_\pi}$ (pop-phase).
The structure of $\tpl{T_\pi,\Lab_\pi}$ is also illustrated in Figures~\ref{FigureLowerBoundFirst} and~\ref{FigureLowerBoundSecond}.
By using both internal and external nondeterminism, the $\PMS$  pop the (labeling of the) entire computation path $\pi$ from the stack. In this way, the $\PMS$ $\PS$ partitions the sanity checks for $\pi$ into separate branches. The labelings of these branches  correspond  to the \emph{reverse} of the $\pi$'s labeling but they are augmented with additional information by means of the extra atomic propositions $\CheckB{3},\CheckMark{3},\CheckB{2},\CheckB{1}$. In particular, in the pop-phase, the unique nondeterministic or branching nodes (i.e., the nodes with at least two children) are  end nodes, i.e.,  nodes labeled by one of the propositions in $\{\End{1},\End{2},\End{3}\}$. These nodes have, in particular, a binary branching degree. Moreover:

\begin{itemize}
  \item the branching behaviour at the branching $\End{3}$-nodes along $\tpl{T_\pi,\Lab_\pi}$ is subdivided in two sub-phases. In the first sub-phase, the branching $\End{3}$-nodes are controlled by the system player, and  $\PS$ marks by \emph{internal} nondeterminism the \emph{$\Lambda$-content} of exactly one $3$-block $\bl_3$ of $\pi$ with the special proposition $\CheckB{3}$
      (i.e., the content of $\bl_3$ is additionally labeled by proposition $check_3$). This means, in particular, that for each $3$-block $\bl_3$ of $\pi$, there is a path of $\tpl{T_\pi,\Lab_\pi}$ from the root such that
      the unique  $\CheckB{3}$-marked $3$-block corresponds to $\bl_3$. 
       This is illustrated in the left part of Figure~\ref{FigureLowerBoundFirst}. Note that in the first sub-phase a branching $\End{3}$-node
         has two children whose labels are of the form $\{\lambda\}$ and $\{\lambda,check_3\}$, respectively, for some $\lambda\in \Lambda$. %where for each play of $\tpl{T_\pi,\Lab_\pi}$ from the root, there is exactly one
      % $3$-block $\bl_3$ of $\pi$ whose content is marked by the special symbol $\CheckB{3}$. 
      After having marked the content of a $3$-block with $\CheckB{3}$, $\PS$ moves to the second
       sub-phase, where the branching $\End{3}$-nodes are controlled by the environment player. In particular,
      in case the marked $3$-block $\bl_3$ does not belong to  the first configuration code of $\pi$, $\PS$ marks by \emph{external} nondeterminism the \emph{$\Lambda$-content} of exactly  one $3$-block $\bla_3$ with the special proposition
       $\CheckMark{3}$ by ensuring that $\bl_3$ and $\bla_3$ belong  to two consecutive configurations codes along $\pi$. Hence, for all $3$-blocks $\bl_3$ and $\bla_3$
       of $\pi$ such that $\bl_3$ and $\bla_3$ belong to adjacent configurations and $\bla_3$ follows $\bl_3$ along the reverse of the $\pi$'s labeling, there is 
       a path of $\tpl{T_\pi,\Lab_\pi}$ from the root such that
      the unique  $\CheckB{3}$-marked $3$-block corresponds to $\bl_3$ and the unique  $\CheckMark{3}$-marked $3$-block corresponds to $\bla_3$. 
        This is illustrated in the right part of
       Figure~\ref{FigureLowerBoundFirst}.
 \item The branching behaviour at the $\End{2}$-nodes and $\End{1}$-nodes along $\tpl{T_\pi,\Lab_\pi}$, which is illustrated in Figure~\ref{FigureLowerBoundSecond}, is as follows. The $\End{2}$-nodes are controlled by the system player, while the $\End{1}$-nodes are controlled by the environment nodes. In particular, from  each $\End{2}$-node $x_{\End{2}}$ associated with the first symbol of the reverse of some $2$-block
   $\bl_2$, $\PS$ generates by \emph{internal} nondeterminism a tree copy of the reverse of $\bl_2$. More specifically, node $x_{\End{2}}$
     has two children $y$ and $y_{check_2}$ whose labels are $\{b\}$ and $\{b,check_2\}$, respectively, for some $b\in\{0,1\}$ (see Figures~\ref{FigureLowerBoundSecond}(b) and~\ref{FigureLowerBoundSecond}(c)).
     Moreover, the labeled tree (we call \emph{check $2$-block-tree}) obtained from the subtree of  $\tpl{T_\pi,\Lab_\pi}$ rooted a node $x_{\End{2}}$ by pruning node $y$ and its descendants is
  structured as follows (see Figure~\ref{FigureLowerBoundSecond}(c)):
  \begin{itemize}
    \item there is an infinite path $\rho$ from the $\End{2}$-node  $x_{\End{2}}$  whose labeling consists of   
        a marked copy (of the reverse) of $\bl_2$ (the \emph{content} $b$ of $\bl_2$ is additionally labeled by the special proposition $\CheckB{2}$) followed by the suffix $\emptyset^{\omega}$ (see Figures~\ref{FigureLowerBoundSecond}(c));
    \item there are additional branches  chosen by \emph{external} nondeterminism starting at the $\End{1}$-nodes of the infinite path $\rho$. As illustrated in 
    Figure~\ref{FigureLowerBoundSecond}(c), these additional branches represent marked copies of the (reverse of) $1$-sub-blocks $\bl_1$ of $\bl_2$ (the \emph{content} of $\bl_1$ is additionally labeled by the special proposition $\CheckB{1}$).
  \end{itemize}
    Note that for the
     $\End{1}$-nodes of $\tpl{T_\pi,\Lab_\pi}$, only the ones belonging to check $2$-block trees are branching.
\end{itemize}
Note that for each $h=1,2,3$,  in a marked $h$-block $\bl$, only the content (i.e., the symbol preceding the end-symbol) of $\bl$ is marked. 

Hence,  the subtree $\tpl{T_\pi,\Lab_\pi}$ of the unwinding $\Unw(\GS(\PS))$ of $\GS(\PS)$ associated with this pop-phase and the specific accepting computation path $\pi$, satisfies the following: the labeling of each \emph{main} path (i.e., a path of $\tpl{T_\pi,\Lab_\pi}$ starting at the root which does not get trapped into a check $2$-block-tree) corresponds to the reverse of the $\pi$'s labeling (followed by a suffix with label $\emptyset^{\omega}$) with the unique difference that exactly one $3$-block $\bl_3$ is marked by $\CheckB{3}$ and
(in case $\bl_3$ does not belong to the first ATM configuration code of $\pi$)  exactly one $3$-block $\bla_3$ is marked by $\CheckMark{3}$.
The $\PMS$ $\PS$ ensures that $\bl_3$ and $\bla_3$ belong  to two consecutive configurations codes along $\pi$ (where $\bl_3$ precedes $\bla_3$ along the main path) and, independently from the environment choices, all the $3$-blocks $\bl_3$ of $\pi$ are checked (i.e., there is a main path whose $\CheckB{3}$-marked block corresponds to $\bl_3$).

The additional check $2$-block-trees are intuitively used to \emph{isolate} $2$-blocks  for ensuring by an $\ATLStar$ formula $\varphi$ that the ATM configuration codes along the $\pi$'s labeling are well-formed and the $\pi$'s labeling is faithful to the evolution of $\mathcal{M}$. In particular, as detailed in the proof of Lemma~\ref{lemma:constructionOfATLStarFOrmula}, the $\ATLStar$ formula $\varphi$ requires that the given environment strategy tree of $\GS(\PS)$ satisfies the following:
\begin{itemize}
  \item all the environment choices in each $2$-block check-tree are enabled,
  \item  the environment choices from the $\{\End{3}\}$-nodes which are descendants of $acc$-nodes (i.e., $\End{3}$-nodes associated with the pop-phase) and are controlled by the environment player   are \emph{deterministic}. This entails that  
   the subtree rooted at the $\Start{3}$-node of a $\CheckB{3}$-marked $3$-block $\bl_3$ which does not belong to the first ATM configuration code  \emph{contains exactly one} $\CheckMark{3}$-marked $3$-block $\bla_3$.
\end{itemize}

Then, by exploiting the previous two requirements, the $\ATLStar$ formula $\varphi$ existentially quantifies over strategies  of the system player whose outcomes get trapped into a check $2$-block-tree in order to ensure that for the given sequence $\nu$ of ATM configuration codes (associated with an accepting computation path $\pi$ of the push-phase), the following holds:
\begin{itemize}
  \item the configuration codes along $\nu$ are well-formed,
  \item for each $\CheckB{3}$-marked $3$-block $\bl_3$ which does not belong to the first ATM configuration code of $\nu$, the associated  $\CheckMark{3}$-marked $3$-block $\bla_3$ satisfies the following:  $\bl_3$  and $\bla_3$ have the same number and the $\Lambda$-contents of $\bl_3$  and $\bla_3$ are consistent with the transition function of $\mathcal{M}$. Since $\bl_3$ and $\bla_3$ belong to two adjacent configuration codes along $\nu$, the previous conditions ensure that $\nu$ is faithful to the evolution of $\mathcal{M}$. Note that in order to enforce that $\bl_3$ and $\bla_3$ have the same number, for each $2$-sub-block $\bl_2$ of $\bl_3$, the formula $\varphi$ requires the existence of a system strategy $f$ starting at the $\End{2}$-node of $\bl_2$ which gets trapped into the check $2$-block-tree of a $2$-sub-block $\bla_2$ of $\bla_3$ such that the copy of $\bla_2$ in the check $2$-block-tree and $\bl_2$ have the same number and the same content.
      Note that the additional $check_1$-branches of the check $2$-block-tree of $\bla_2$  are used to check by an $\LTL$ formula, asserted at the outcomes of the system strategy $f$, that $\bl_2$ and the copy of $\bla_2$ have the same number.
\end{itemize}

\begin{figure}[tb]%[htp]
  \centering
  \caption{Marked copies of $2$-blocks in the pop-phase of the open $\PMS$ $\PS$}\label{FigureLowerBoundSecond}
  \begin{minipage}{.3\textwidth}
\begin{tikzpicture}[scale=0.63]
\draw[draw=none,use as bounding box](-2.5,2.5) rectangle (3.3,-8.0);

\coordinate [label=right:{\footnotesize  \, =  System node}] (SystemNode) at (-2.5,1.0);
\node[shape=circle,draw=black,inner sep=2pt,fill=white](SN) at (-2.5,1.0) {};
\fill [black] (SystemNode) circle (2pt);

\coordinate [label=right:{\footnotesize \,  =   Environment node}] (EnvironmentNode) at (2.5,1.0);
\fill [black] (EnvironmentNode) circle (2pt);

\coordinate [label=right:{\footnotesize  \, =  owner agent depending on the context}] (BranchingNode) at (9,1.0);
\node[shape=rectangle,draw=black,inner sep=3pt,fill=white](SEN) at (9,1.0) {};
\fill [black] (BranchingNode) circle (2pt);

\coordinate [label=left:{\footnotesize $\End{3}\,$}] (EndThreeNode) at (0.0,0);
\path[->, thin,black] (EndThreeNode) edge  (0,-0.75);

\draw[thin, dotted, black] (EndThreeNode) edge (0.8,-0.8);
\coordinate [label=above right:{\footnotesize   maybe marked}] (L1) at (0.8,-0.8);
\coordinate [label=above right:{\footnotesize   copy of $3$-block}] (L1) at (0.8,-1.3);

\node[shape=rectangle,draw=black,inner sep=3pt,fill=white](A) at (0.0,0) {};
\draw[thin, dotted, black] (0,0.6) edge (EndThreeNode);

\coordinate [label=left:{\footnotesize $\lambda\in\Lambda$}] (Content) at (0,-0.8);

\fill [black] (Content) circle (2pt);
\fill [black] (EndThreeNode) circle (2pt);
\path[->, thin,black] ( Content) edge  (0,-1.4);

\coordinate [label=right:\textcolor{red}{\footnotesize $\phantom{,}\End{2}$}] (Beg) at (0,-1.6);
\coordinate [label=right:\textcolor{red}{\footnotesize $\Start{2}$}] (End) at (0,-3.2);

\node[shape=circle,draw=red,inner sep=2pt,fill=white](B) at (0,-1.6) {};

\draw(Beg.east) edge[thick,dotted, bend right,red] (7.6,0.1);
\draw(End.east) edge[thick,dotted, bend right,red] (7.6,-6.3);

\fill [red] (Beg) circle (2pt);
\fill [red] (0,-3.0) circle (2pt);
\draw[thick, dotted, red] (Beg) edge (0.7,-3.0);
\draw[thick, dotted, red] (Beg) edge (-0.7,-3.0);
\draw[thick, dotted, red] (0.7,-3.0) edge (-0.7,-3.0);

\coordinate [label=right:{\footnotesize $\phantom{,}\End{2}$}] (BegLast) at (0,-4.4);
\coordinate [label=right:{\footnotesize $\Start{2}$}] (EndLast) at (0,-6.0);

\node[shape=circle,draw=black,inner sep=2pt,fill=white](B) at (0,-4.4) {};

\draw[thin, dotted, black] (End) edge (0,-4.2);

\fill [black] (BegLast) circle (2pt);
\fill [black] (0,-5.8) circle (2pt);
\draw[thick, dotted, black] (BegLast) edge (0.7,-5.8);
\draw[thick, dotted, black] (BegLast) edge (-0.7,-5.8);
\draw[thick, dotted, black] (0.7,-5.8) edge (-0.7,-5.8);

\coordinate [label=right:{\footnotesize $\Start{3}$}] (StartThreeNode) at (0,-6.6);

\path[->, thin,black] (0,-5.8) edge  (0,-6.54);

\fill [black] (StartThreeNode) circle (2pt);
\draw[thin, dotted, black] (StartThreeNode) edge (0,-7.3);

\coordinate [label=right:{(a) Tree-encoding of $3$-block}] (Caption) at (-3,-7.9);
\end{tikzpicture}

  \end{minipage}\quad
\begin{minipage}{.31\textwidth}
\begin{tikzpicture}[scale=0.63]
\draw[draw=none,use as bounding box](-3,2.5) rectangle (3.3,-8.0);

\coordinate [label=right:{\footnotesize $\phantom{,}\End{2}$}] (BegCheck) at (0,0.3);
\coordinate [label=above right:\textcolor{red}{\footnotesize $(b,\CheckB{2})$}] (Check) at (1,-0.5);
\path[->, thin,black] (BegCheck) edge  (0,-0.45);
\path[->, thin,red] (BegCheck) edge  (1,-0.45);
\node[shape=circle,draw=black,inner sep=2pt,fill=white](A) at (0,0.3) {};
\fill [black] (BegCheck) circle (2pt);

\fill [black,red] (Check) circle (2pt);

\draw[thick,dotted, red] (1,-0.5) edge (6.5,-0.5);

\coordinate [label=left:{\footnotesize $b\in\{0,1\}$}] (ContentCheck) at (0,-0.5);
\coordinate [label=right:{\footnotesize $\End{1}$}] (SBegCheck) at (0,-1.3);
\coordinate [label=right:{\footnotesize $\Start{1}$}] (SEndCheck) at (0,-2.7);

\fill [black] (ContentCheck) circle (2pt);
%\node[shape=circle,draw=black,inner sep=2pt,fill=white](B) at (0,-1.3) {};
\fill [black] (SBegCheck) circle (2pt);
\fill [black] (SEndCheck) circle (2pt);
\path[->, thin,black] (ContentCheck) edge  (0,-1.25);
\draw[very thick, black] (SBegCheck) edge (SEndCheck);

%\coordinate  (CommLInit) at (-0.2,-2.0);
%\coordinate  (CommL) at (-1.0,-1.9);
\coordinate [label=left:\textcolor{blue}{\footnotesize $1$-block in}] (LineAL) at (0,-1.5);
\coordinate [label=left:\textcolor{blue}{\footnotesize reverse order}] (LineBL) at (0,-2.0);
%\draw(CommL.east) edge[->, bend left,blue] (CommLInit);

\coordinate [label=right:{\footnotesize $\End{1}$}] (SBegCheckLast) at (0,-4.1);
\coordinate [label=right:{\footnotesize $\Start{1}$}] (SEndCheckLast) at (0,-5.5);
\coordinate [label=right:{\footnotesize $\Start{2}$}] (EndCheck) at (0,-6.3);

\draw[thick,dotted, black] (SEndCheck) edge (SBegCheckLast);
\path[->, thin,black] (SEndCheckLast) edge  (0,-6.25);
\fill [black] (SBegCheckLast) circle (2pt);
\fill [black] (SEndCheckLast) circle (2pt);
\fill [black] (EndCheck) circle (2pt);
\draw[very thick, black] (SBegCheckLast) edge (SEndCheckLast);

\coordinate [label=right:{(b) Tree encoding of $2$-block}] (Caption) at (-3,-7.9);

\end{tikzpicture}
 \end{minipage}\quad
  \begin{minipage}{.31\textwidth}
\begin{tikzpicture}[scale=0.63]
\draw[draw=none,use as bounding box](-2,2.5) rectangle (5.3,-8.0);

%\coordinate [label=right:{\footnotesize $\begOne$}] (BegCheck) at (0,0);
%\coordinate [label=right:{\footnotesize $\checkOne$}] (Check) at (0,-0.5);

%\fill [black] (0,-0.8) circle (2pt);

%\path[->, thin,black] (0,-0.8) edge  (0,-1.55);

\coordinate [label=above :{\footnotesize $\{b,\CheckB{2}\}$}] (ContentCheck) at (0,-0.5);
\coordinate  (ContentCheckNote) at (4,-0.5);
\coordinate [label=right:{\footnotesize $\End{1}$}] (SBegCheck) at (0,-1.3);
\coordinate [label=right:{\footnotesize $\Start{1}$}] (SEndCheck) at (0,-2.7);
\coordinate [label=left:{\footnotesize $\Start{1}$}] (SEndCheckMark) at (-1.3,-2.7);

\draw[thick,dotted, black] (SEndCheckMark) edge (-1.3,-3.7);%LAURA
\coordinate [label=left :{\footnotesize $\emptyset^{\omega}$}] (LabEmpty) at (-1.3,-3.5);%LAURA

\path[->, thin,black] (ContentCheck) edge  (0,-1.25);
\fill [black] (ContentCheck) circle (2pt);
\fill [black] (SBegCheck) circle (2pt);

\fill [black] (SEndCheck) circle (2pt);
\fill [black] (SEndCheckMark) circle (2pt);
\draw[very thick, black] (SBegCheck) edge (SEndCheck);
\draw[very thick, black] (SBegCheck) edge (SEndCheckMark);

\coordinate  (CommRInit) at (0.1,-2.0);
\coordinate  (CommR) at (0.8,-2.1);
\coordinate [label=right:\textcolor{blue}{\footnotesize $1$-block in}] (LineAR) at (0.8,-1.7);
\coordinate [label=right:\textcolor{blue}{\footnotesize reverse order}] (LineBR) at (0.8,-2.2);
\draw(CommR.west) edge[->, bend left,blue] (CommRInit);

\coordinate [label=left:{\footnotesize $\CheckB{1}$}] (LineAL) at (-0.5,-1.8);
 \coordinate [label=left:{\footnotesize $\CheckB{1}$}] (LineAL) at (-0.5,-4.6);

\coordinate [label=right:{\footnotesize $\End{1}$}] (SBegCheckLast) at (0,-4.1);
\coordinate [label=right:{\footnotesize $\Start{1}$}] (SEndCheckLast) at (0,-5.5);
\coordinate [label=left:{\footnotesize $\Start{1}$}] (SEndCheckMarkLast) at (-1.3,-5.5);

\draw[thick,dotted, black] (SEndCheckMarkLast) edge (-1.3,-6.5);%LAURA
\coordinate [label=left :{\footnotesize $\emptyset^{\omega}$}] (LabEmptyTwo) at (-1.3,-6.3);%LAURA

\coordinate [label=right:{\footnotesize $\Start{2}$}] (EndCheck) at (0,-6.3);
\coordinate  (EndCheckNote) at (4,-6.3);

\draw[thick,dotted, black] (EndCheck) edge (0,-7.1);%LAURA
\coordinate [label=left :{\footnotesize $\emptyset^{\omega}$}] (LabEmptyThree) at (0,-6.7);%LAURA

\draw[thick,dotted, black] (SEndCheck) edge (SBegCheckLast);
\path[->, thin,black] (SEndCheckLast) edge  (0,-6.25);
\fill [black] (SBegCheckLast) circle (2pt);
\fill [black] (SEndCheckLast) circle (2pt);
\fill [black] (SEndCheckMarkLast) circle (2pt);
\fill [black] (EndCheck) circle (2pt);
\draw[very thick, black] (SBegCheckLast) edge (SEndCheckLast);
\draw[very thick, black] (SBegCheckLast) edge (SEndCheckMarkLast);
\path[<->, thin,dotted,black] (4,0) edge  (EndCheckNote);
 \coordinate [label=left:\textcolor{blue}{\footnotesize $2$-block in}] (NoteA) at (5.0,-4.6);
\coordinate [label=left:\textcolor{blue}{\footnotesize reverse order}] (NoteB) at (5.4,-5.0);

\coordinate [label=right:{(c) Check $2$-block-tree}] (Caption) at (-2,-7.9);

\end{tikzpicture}
  \end{minipage}
%  \vspace{-0.4cm}
\end{figure}

Recall that $\Prop =\MainP\cup \{check_1,check_2,check_3,\widehat{check_3}\}$. We now formally define the $\Prop$-labeled trees associated with the \emph{accepting} environment strategy trees of $\GS(\PS)$, i.e. the environment strategy  trees where each path from the root visits an $\{acc\}$-labeled node.
 In the following, a  $2^{AP}$-labeled tree is \emph{minimal} if the children of each node have distinct labels. A \emph{branching node} of a tree is a node having at least two distinct children.

\paragraph{Tree-codes} A \emph{tree-code} is a \emph{finite  minimal}  $2^{AP}$-labeled tree $\tpl{T,\Lab}$ such that
 \begin{itemize}
   \item for each maximal path $\pi$ from the root, $\Lab(\pi)$ is a sequence of ATM configuration codes;
   \item a node $x$ is labeled by $\{acc\}$ iff $x$ is a leaf;
   \item each node labeled by $\{\forall\}$ has two children, one labeled by $\{l\}$ and one labeled by $\{r\}$.
 \end{itemize}

\noindent Intuitively, tree-codes correspond  to the maximal portions of the \emph{accepting} environment strategy trees of $\GS(\PS)$ where $\PS$ performs push operations (push-phase). We now extend a tree-code $\tpl{T,\Lab}$ with extra nodes in such a way that each leaf $x$ of $\tpl{T,\Lab}$
is expanded in a tree, called \emph{check-tree} (pop-phase).

\paragraph{Check-trees} The definition of check-trees is based on the notion of \emph{check $2$-block-tree} and \emph{simple check-tree}. The structure of
a  check $2$-block-tree for a $2$-block $\bl_2$ is illustrated in Figure~\ref{FigureLowerBoundSecond}(c). Note that the unique
branching nodes are labeled by $\{\End{1}\}$. In the accepting environment strategy trees of $\GS(\PS)$,  these nodes are controlled by the environment.
Formally, a  check $2$-block-tree for a $2$-block $\bl_2$ is a  \emph{minimal}  $2^{AP}$-labeled tree $\tpl{T,\Lab}$ such that: 
\begin{itemize}
    \item there is a path   from the root (\emph{main path})    whose labeling is $\rho\cdot \emptyset^\omega$, where $\rho$
    is the reverse of the marked copy of $\bl_2$ (the content of $\bl_2$ is additionally labeled by proposition $check_2$);
    \item for each $\{\End{1}\}$-labeled node $x$ of the main path, there is an infinite path $\pi_s$ from $x$ (\emph{secondary branch}) such that denoted by 
    $\bl_1$ the $1$-subblock of $\bl_2$ associated with node $x$, the labeling of $\pi_s$ is  $\rho_1\cdot \emptyset^\omega$, where $\rho_1$ 
    is the reverse of the marked copy of $\bl_1$ (the content of $\bl_1$ is additionally labeled by proposition $check_1$); 
    \item each node of $\tpl{T,\Lab}$ is either a node of the main path or a node of some secondary branch.
  \end{itemize}
 A \emph{partial check $2$-block-tree} for $\bl_2$ is obtained
from the check $2$-block-tree for $\bl_2$ by pruning some choices from the $\{\End{1}\}$-branching nodes.

Given a sequence $\nu$ of ATM configuration codes, a \emph{simple check-tree} for $\nu$ is
a  \emph{minimal}  $2^{AP}$-labeled tree $\tpl{T,\Lab}$ such that
 \begin{itemize}
   \item for each path $\pi$ from the root, $\Lab(\pi)$ corresponds to the \emph{reverse} of $\nu$ followed by $\emptyset^{\omega}$ but there is exactly one $3$-block $\bl_3$ of $\nu$ whose content is additionally marked by proposition
       $\CheckB{3}$, and in case $\bl_3$ does not belong to the first configuration code of $\nu$, there is  exactly one $3$-block $\bla_3$ whose content is marked by proposition $\CheckMark{3}$; moreover, $\bla_3$ and $\bl_3$ belong to two consecutive configuration codes, and $\bla_3$ precedes $\bl_3$ along $\nu$;
   \item  for each $3$-block $\bl_3$ of $\nu$, there is a path $\pi$ from the root such that the node associated with the content of $\bl_3$ is additionally labeled  by proposition $\CheckB{3}$ (i.e., all the $3$-blocks of $\nu$ are checked);
   \item each branching node $x$ has label $\{\End{3}\}$ and two children: one labeled by $\{\lambda\}$ and the other one labeled by $\{\lambda,tag\}$  for some $\lambda\in \Lambda$ and $tag\in \{\CheckB{3},\CheckMark{3}\}$. If  $tag=\CheckB{3}$ (resp., $tag=\CheckMark{3}$), we say that $x$ is a $\CheckB{3}$-branching (resp., $\CheckMark{3}$-branching) node.
 \end{itemize}

Finally, a \emph{check-tree} for $\nu$ is a \emph{minimal}  $2^{AP}$-labeled tree $\tpl{T,\Lab}$ which is obtained from some
simple check-tree $\tpl{T',\Lab'}$ for $\nu$ by adding for each node $x$ of $T'$ with label $\{\End{2}\}$ an additional child $y$ and a subtree rooted at $y$
so that the subtree rooted at $x$ obtained by removing all the descendants of $x$ in $T'$ is a partial check $2$-block-tree for the $2$-block associated with node $x$ in $T'$.

Thus, in a check-tree, we have four types of branching nodes:  $check_3$-branching nodes, $\{\End{2}\}$-branching nodes,  $\widehat{check_3}$-branching nodes, and $\{\End{1}\}$-branching nodes. In the accepting environment strategy trees of $\GS(\PS)$, $check_3$-branching nodes and  $\{\End{2}\}$-branching nodes  
are controlled by the system, while $\widehat{check_3}$-branching nodes and $\{\End{1}\}$-branching nodes  are controlled by the environment.

\paragraph{Extended tree-codes} An \emph{extended tree-code} is a minimal $2^{AP}$-labeled tree $\tpl{T_e,\Lab_e}$ such that there is a tree-code $\tpl{T,\Lab}$ so that $\tpl{T_e,\Lab_e}$ is obtained from $\tpl{T,\Lab}$ by replacing  each leaf $x$ (recall that $x$ is labeled by $\{acc\}$)  with a check-tree  for the sequence of labels associated with the path of $\tpl{T,\Lab}$ starting at the root and leading to $x$. By construction and the intuitions given about the $\PMS$ $\PS$, we obtain the following result.

\begin{lem}\label{lemma:constructionOfOPD} One can build, in time polynomial in the size of the ATM $\mathcal{M}$,
an open turn-based $\PMS$ $\PS$ over $\Prop$  and $\Agents =\{\env,\sys\}$ such that the following hold, where an  environment strategy tree is accepting if each path from the root visits
an $\{acc\}$-labeled node:
\begin{itemize}
  \item the set of $2^{\Prop}$-labeled trees $\tpl{T,\Lab}$ associated with the  accepting   environment strategy trees $\tpl{T,\Lab,\Trans}$ in  $\exec(\GS(\PS))$ coincides with the set of extended tree-codes;
  \item for each  accepting  environment strategy tree $\tpl{T,\Lab,\Trans}$ in  $\exec(\GS(\PS))$, the unique nodes controlled by the system
  in a check-subtree of $\tpl{T,\Lab,\Trans}$ are the $check_3$-branching nodes and the $\{\End{2}\}$-branching nodes.
\end{itemize}
\end{lem}
\begin{proof}

Since the $\PMS$ $\PS$ is turn-based, each configuration is either controlled by the environment or by the system agent. Thus, in specifying
the transition function $\Delta$ of $\PS$, we can abstract away from the set of full decisions, and we just specify for each state $p$ and stack symbol
$\gamma$, the set of pairs $(p',\beta)$ such that $(p',\beta)\in \Delta(p,\gamma,\decision)$ for some full decision $\decision$. In particular,
the transition function of $\Delta$ consists of the following types of transitions:
\begin{itemize}
  \item push transitions $p\, \der{\push(\gamma)}\, p'$ meaning that for each stack symbol $\gamma'$, $(p',\gamma\cdot\gamma')\in \Delta(p,\gamma',\decision)$ for some full decision $\decision$;
  \item pop transitions $p\, \der{\push(\gamma)}\, p'$ meaning that   $(p',\varepsilon)\in \Delta(p,\gamma,\decision)$ for some full decision $\decision$;
   \item internal transitions $p\, \der{}\, p'$ that do not use the stack, meaning that for each stack symbol $\gamma$, there is some full decision $\decision$
    such that \emph{either} $\gamma=\gamma_0$ and $(p',\varepsilon)\in \Delta(p,\gamma,\decision)$,  \emph{or} $\gamma\neq \gamma_0$ and $(p',\gamma)\in (p,\gamma,\decision)$.
\end{itemize}
The initial state of $\PS$ is denoted by $in$ and the set of states is given by  
\[
Q_{push}\cup Q_{pop}\cup \{q_\emptyset\}
\]
where $in\in Q_{push}$, $Q_{push}$ is the set of states used in the push phase, $Q_{pop}$ is the set of states used in the pop-phase, and the state 
$q_\emptyset$ is a sink state.  The unique transition from state  $q_\emptyset$ is the internal transition $q_\emptyset\, \der{}\, q_\emptyset$. Moreover, the propositional labeling 
of $q_\emptyset$ is the empty set. Note that each configuration with control state $q_\emptyset$ is deterministic.

\noindent  The stack alphabet $\Gamma$ is defined as follows:
\[
\Gamma := \MainP \cup \{in\} \cup (\Lambda\times \{first,first_{in}\})
\]
where the symbol $in$ is pushed onto the stack in the first step of the push phase, while 
a symbol $(\lambda,t)\in \Lambda\times \{first,first_{in}\}$ is pushed onto the stack on generating
the content $\lambda$ of the first $3$-block of a guessed ATM configuration code $C$. In particular, $t=first_{in}$ means that 
$C$ is the first guessed ATM configuration code, while $t=first$ means it isn't. The  flags in $\{first,first_{in}\}$ are used in the 
pop-phase for ensuring that the generation of the propositions $check_3$ and $\widehat{check_3}$ is consistent with the definition of 
check-tree.   \vspace{0.2cm}
 
\noindent \emph{Push Phase.} The set $Q_{push}$  of states used in the push phase is defined as follows:
\[
\begin{array}{ll}
Q_{push}:= & \{in\}\cup \bigl((\Lambda\cup \{\bot\})\times (Q\cup \{\bot\})\times \{in,\bot\} \times \MainP'\bigr) \vspace{0.2cm}\\
\MainP':=  & \MainP\cup (\{\bl_1,\bl_2\}\times \{0,1\}) \cup (\Lambda\times \{first,first_{in}\})
\end{array}
\]
The intuitive meaning of a state $(\lambda_{\bot},q_{\bot},in_{\bot}, m)\in Q_{push}$ is as follows:
\begin{itemize}
  \item The symbol $\lambda_{\bot}\in \Lambda\cup \{\bot\}$ keeps track of the last $\Lambda$-symbol of the prefix of the current guessed ATM configuration code
  generated so far if such a prefix contains a $\Lambda$-symbol; otherwise,  $\lambda_{\bot}=\bot$.  
  \item The symbol $q_{\bot}\in Q\cup \{\bot\}$ keeps track of the state associated to the current guessed ATM configuration code $C$
  if the prefix of $C$ generated so far contains a $\Lambda$-symbol  $(u_p,u,u_q)$ where $u$ is of the form $(\sigma,q_\bot)$; otherwise,  $q_{\bot}=\bot$.
  \item The symbol $in_{\bot}\in \{in,\bot\}$ keeps track whether the current  guessed ATM configuration code $C$ is the first one to be generated ($in_{\bot}=in$)
  or not ($in_\bot =\bot$). 
  \item The main symbol $m$ has the following meaning: if $m\in \MainP$ (resp., $m=(\lambda,t)\in \Lambda\times \{first,first_{in}\}$), then $m$ (resp., $\lambda$) is the symbol currently generated
  for the current guessed ATM configuration code $C$.  The flag $t\in \{first,first_{in}\}$ means that $\lambda$ is the content of the first $3$-block of $C$ with $t=first_{in}$ 
  iff $C$ is the first guessed ATM configuration code.
  If instead $m=(\bl_1,b)$ (resp., $m=(\bl_2,b)$) for some $b\in\{0,1\}$, then $b$ represents the content of a $1$-block (resp., $2$-block) of $C$.
\end{itemize}
The propositional label of state $(\lambda_{\bot},q_{\bot},in_{\bot},m)$ is $\{m\}$ if $m\in \MainP$, is $\lambda$ if $m=(\lambda,t)\in \Lambda\times \{first,first_{in}\}$, and is $\{b\}$ if $m=(\bl_k,b)$ 
for some $k=1,2$ and $b\in\{0,1\}$. The propositional label of the initial state $in$ is $\{l\}$. Moreover, all the configurations associated with the states in $Q_{push}$ are controlled by the environment
with the exception of the configurations associated with the push states of the form $(\lambda_{\bot},q_{\bot},\forall)$, which are instead controlled by the system. 
Transitions from states in $Q_{push}$ are push transitions. These transitions are defined  as follows, where $acc\in \MainP $ is also used  as control state in $Q_{pop}$ and has propositional labeling
$\{acc\}$. 

\begin{itemize}
\item  Transitions from state   $in$: $ in\, \der{\push(in)}\, (\bot,\bot,in,\Start{3})$.\vspace{0.1cm}
\item  Transitions from states   $(\lambda_\bot,q_\bot,in_\bot,l),(\lambda_\bot,q_\bot,in_\bot,r)\in Q_{push}$:\\
$ (\lambda_\bot,q_\bot,in_\bot,l)\, \der{\push(l)}\, (\lambda_\bot,q_\bot,in_\bot,\Start{3})$ and $(\lambda_\bot,q_\bot,in_\bot,r) \, \der{ \push(r)}\,(\lambda_\bot,q_\bot,in_\bot,\Start{3})$.\vspace{0.1cm}
\item  Transitions from states  $(\lambda_\bot,q_\bot,in_\bot,\Start{k})\in Q_{push}$, where $k=1,2,3$:
  \begin{itemize}
  \item  $ (\lambda_\bot,q_\bot,in_\bot,\Start{k})\, \der{\push(\Start{k})}\, (\lambda_\bot,q_\bot,in_\bot,\Start{k-1})$ for each $k=2,3$.
  \item  $ (\lambda_\bot,q_\bot,in_\bot,\Start{1})\, \der{\push(\Start{1})}\, (\lambda_\bot,q_\bot,in_\bot,b)$ for each $b\in \{0,1\}$.\vspace{0.1cm}
\end{itemize}
   \item  Transitions from states  $(\lambda_\bot,q_\bot,in_\bot,b)\in Q_{push}$, where $b\in \{0,1\}$:
   \begin{itemize}
     \item $(\lambda_\bot,q_\bot,in_\bot,b)\, \der{\push(b)}\, (\lambda_\bot,q_\bot,in_\bot,b')$ for all $b'\in \{0,1\}$; 
     \item $(\lambda_\bot,q_\bot,in_\bot,b)\, \der{\push(b)}\, (\lambda_\bot,q_\bot,in_\bot,(\bl_1,b'))$ for all $b'\in \{0,1\}$.\vspace{0.1cm}
   \end{itemize}
   \item  Transitions from states  $(\lambda_\bot,q_\bot,in_\bot,(\bl_k,b))\in Q_{push}$, where $k=1,2$ and $b\in \{0,1\}$:\\ 
    $ (\lambda_\bot,q_\bot,in_\bot,(\bl_k,b))\, \der{\push(b)}\, (\lambda_\bot,q_\bot,in_\bot,\End{k})$.\vspace{0.1cm}
   \item  Transitions from states  $(\lambda_\bot,q_\bot,in_\bot,\End{1})\in Q_{push}$:
    \begin{itemize}
     \item  $ (\lambda_\bot,q_\bot,in_\bot,\End{1})\, \der{\push(\End{1})}\, (\lambda_\bot,q_\bot,in_\bot,\Start{1})$; 
       \item  $(\lambda_\bot,q_\bot,in_\bot,\End{1})\, \der{\push(\End{1})}\, (\lambda_\bot,q_\bot,in_\bot,(\bl_2,b))$ for all $b\in \{0,1\}$.\vspace{0.1cm}
   \end{itemize}
    \item Transitions from states  $(\bot,q_\bot,in_\bot,\End{2})\in Q_{push}$: 
      \begin{itemize}
     \item   $ (\bot,q_\bot,in_\bot,\End{2})\, \der{\push(\End{2})}\, (\bot,q_\bot,in_\bot,\Start{2})$; 
       \item  $(\bot,q_\bot,in_\bot,\End{2})\, \der{\push(\End{2})}\, (\bot,q'_\bot,in_\bot,(\lambda,t))$   for all $(\lambda,t)\in \Lambda\times \{first,first_{in}\}$ and $q'_\bot\in Q\cup \{\bot\}$ such that the following
   holds, where $\lambda = (u_p,u,u_s)$:\vspace{0.1cm}
   \begin{itemize}
     \item \emph{either} $u\in \Sigma$ and $q'_\bot =\bot$, \emph{or}   $u$ is of the form $(\sigma,q'_\bot)\in \Sigma\times Q$; 
     \item  $u_p=\vdash$;
     \item $t=first_{in}$ if $in_\bot=in$, and $t=first$ otherwise.\vspace{0.1cm}
   \end{itemize}
   \end{itemize}  
    \item Transitions from states  $(\lambda,q_\bot,in_\bot,\End{2})\in Q_{push}$ where $\lambda\in \Lambda$: 
      \begin{itemize}
     \item   $ (\lambda,q_\bot,in_\bot,\End{2})\, \der{\push(\End{2})}\, (\lambda,q_\bot,in_\bot,\Start{2})$; 
       \item  $(\lambda,q_\bot,in_\bot,\End{2})\, \der{\push(\End{2})}\, (\lambda,q'_\bot,in_\bot,\lambda')$   for all $\lambda'\in \Lambda$ and $q'_\bot\in Q\cup \{\bot\}$ such that the following
   holds, where $\lambda = (u_p,u,u_s)$ and  $\lambda' = (u'_p,u',u'_s)$:\vspace{0.1cm}
   \begin{itemize}
     \item \emph{either} $u'\in \Sigma$ and $q'_\bot =q_\bot$, \emph{or}   $u'$ is of the form $(\sigma,q'_\bot)\in \Sigma\times Q$ and $q_\bot = \bot$; 
     \item  $u'_p=u$ and $u'=u_s$.\vspace{0.1cm}
   \end{itemize}
   \end{itemize}
   \item  Transitions from states $(\lambda_\bot,q_\bot,in_\bot,(\lambda',t))\in Q_{push}$, where $(\lambda',t)\in \Lambda\times \{first,first_{in}\}$:  
   \[
      (\lambda_\bot,q_\bot,in_\bot,(\lambda',t))\, \der{\push((\lambda',t))}\, (\lambda',q_\bot,in_\bot,\End{3}) 
   \]       
   \item  Transitions from states $(\lambda_\bot,q_\bot,in_\bot,\lambda')\in Q_{push}$, where $\lambda'\in \Lambda$:  
   \[
      (\lambda_\bot,q_\bot,in_\bot,\lambda')\, \der{\push(\lambda')}\, (\lambda',q_\bot,in_\bot,\End{3}) 
   \] 
\item  Transitions from states  $(\lambda_\bot,q_\bot,in_\bot,\End{3})\in Q_{push}$:
  \begin{itemize}
  \item  if $\lambda_\bot$ is not of the form $(u_p,u,\vdash)$: $ (\lambda_\bot,q_\bot,in_\bot,\End{3})\, \der{\push(\End{3})}\, (\lambda_\bot,q_\bot,in_\bot,\Start{3})$;
  \item  else if $q_\bot\neq \bot$ and $q_\bot$ is universal and non-accepting: 
  \[ (\lambda_\bot,q_\bot,in_\bot,\End{3})\, \der{\push(\End{3})}\, (\lambda_\bot,q_\bot,in_\bot,\forall);\] 
  \item  else if $q_\bot\neq \bot$ and $q_\bot$ is existential and non-accepting: \[ (\lambda_\bot,q_\bot,in_\bot,\End{3})\, \der{\push(\End{3})}\, (\lambda_\bot,q_\bot,in_\bot,\exists);\]  
  \item  else if $q_\bot\neq \bot$ and $q_\bot$ is accepting: $ (\lambda_\bot,q_\bot,in_\bot,\End{3})\, \der{\push(\End{3})}\, acc$;
  \item else:  $ (\lambda_\bot,q_\bot,in_\bot,\End{3})\, \der{\push(\End{3})}\, q_\emptyset$.\vspace{0.1cm}
\end{itemize}
\item  Transitions from states   $(\lambda_\bot,q_\bot,in_\bot,\exists),(\lambda_\bot,q_\bot,in_\bot,\forall)\in Q_{push}$: 
\begin{itemize}
     \item $ (\lambda_\bot,q_\bot,in_\bot,\exists)\, \der{\push(\exists)}\, (\bot,\bot,\bot,dir)$ for all $dir\in\{l,r\}$; 
     \item $(\lambda_\bot,q_\bot,in_\bot,\forall) \, \der{ \push(\forall)}\,(\bot,\bot,\bot,dir)$ for all $dir\in\{l,r\}$.
   \end{itemize}
\end{itemize}\vspace{0.1cm}

\noindent Let $\nu$ be a  sequence of ATM configuration codes of the form 
$\nu=\rho \cdot \{acc\}$. Note that each symbol of $\rho$ is of the form $\{p\}$ where $p\in \MainP$.
Thus, $\rho$ corresponds to a word $\rho'$ over $\MainP$. Let $\rho''$ obtained from $\rho'$ by replacing the first symbol of $\rho'$ with $in$ and by replacing for each
configuration code $C$ along $\rho'$, the content $\lambda$ of the first $3$-block of $C$ with $(\lambda,t)$, where $t=first_{in}$
if $C$ is the first configuration code of $\rho'$, and $t=first$ otherwise. We denote by $\Stack(\nu)$ the stack content 
given by $(\rho'')^R\cdot \gamma_0$, where $(\rho'')^R$ is the reverse of $\rho''$.
By construction, the following two claims hold.
\vspace{0.2cm}
 
\noindent \emph{Claim 1.} Let $\Tau$ be an accepting environment strategy tree of $\GS(\PS)$. Then, the finite labeled tree obtained from $\Tau$
by pruning all the subtrees of $\Tau$ rooted at the children of $acc$-nodes is a tree-code. Moreover, for each $acc$-node $x$ of $\Tau$, let
$\pi$ be the finite path from the root to node $x$. Then, the stack content of node $x$ is   $\Stack(\nu)$ where $\nu$ is the labeling of $\pi$.  \vspace{0.2cm}

\noindent \emph{Claim 2.} Let $\tpl{T,\Lab}$ be a tree-code. Then there exists  an accepting environment strategy tree $\Tau$ of $\GS(\PS)$ such that  the finite labeled tree obtained from $\Tau$
by pruning all the subtrees of $\Tau$ rooted at the children of $acc$-nodes is isomorphic to $\tpl{T,\Lab}$.   \vspace{0.2cm}

\noindent \emph{Pop Phase.} The set $Q_{pop}$  of states used in the pop phase is defined as follows:
\[
 Q_{pop} := Q_{pop}^1 \cup Q_{pop}^2
\]
where $Q_{pop}^1$ is used to generate the nodes of the \emph{main} paths of a check-tree (i.e., the paths that do not get trapped into check 2-block trees), while
the states in $Q_{pop}^1$ are used to generate check 2-block trees. We first consider the set $Q_{pop}^2$ which is defined as follows:
\[
Q_{pop}^1 := \{acc\}\cup (\{check_3\}\times \Lambda)\cup (\{check_3^0, check_3^1,check_3^2,\widehat{check_3}\}\times  \MainP)  
\]
Intuitively, for a state $(t,m)\in Q_{pop}^1$, $m$ represents the last symbol which has been popped from the stack. Moreover, the flag 
$t\in \{check_3^0, check_3^1,check_3^2,\widehat{check_3}\}$ has the following meaning:
 \begin{itemize}
  \item  $t=check_3^0$ iff the proposition $check_3$ has not been generated so far.
      \item  $t=check_3$ iff $m\in \Lambda$ (i.e, $m$ is the content of a $3$-block) and proposition $check_3$ is currently generated.
  \item  $t=check_3^1$ iff the proposition $check_3$ has  been generated and is associated with a $3$-block of the current ATM configuration code.
      \item  $t=check_3^2$ iff the proposition $check_3$ has  been generated and is associated with a $3$-block of an ATM configuration code preceding the current one. 
  \item  $t=\widehat{check_3}$ iff the proposition $\widehat{check_3}$ has  been generated. 
\end{itemize}
For each $(t,m)\in Q_{pop}^1$, the propositional labeling of state $(t,m)$ is $\{t,m\}$ if $(t,m)\in \{check_3,\widehat{check_3}\}\times \Lambda$, and $\{m\}$ otherwise.
 All the configurations associated with the states in $Q_{pop}^1$ are controlled by the environment
with the exception of the configurations associated with the  states of the form $(check_3^0,\End{3})$, which are instead controlled by the system. 
Transitions from states in $Q_{pop}^1$ are pop transitions. They are defined as follows  
  where $(check_2,b)\in Q_{pop}^2$ for each $b\in\{0,1\}$.
\begin{itemize} 
\item  Transitions from state   $acc$:
  \begin{itemize}
  \item  $acc\, \der{\pop(\End{3})}\, (check_3^0,\End{3})$.
  \item  $acc\, \der{\pop(\gamma)}\, q_\emptyset$  for all $\gamma\neq \End{3}$.\vspace{0.1cm}
\end{itemize}
\item  Transitions from states  $(check_3,\lambda)\in Q_{pop}^1$ (note that $\lambda\in\Lambda$):
  \begin{itemize}
  \item  $(check_3,\lambda)\, \der{\pop(\End{2})}\, (check_3^1,\End{2})$.
  \item  $(check_3,\lambda)\, \der{\pop(\gamma)}\, q_\emptyset$  for all $\gamma\neq \End{2}$.\vspace{0.1cm}
\end{itemize}
\item Transitions from states $(t,\End{2})\in Q_{pop}^1$:
  \begin{itemize}
  \item  $(t,\End{2})\, \der{\pop(\gamma)}\, q_\emptyset$  for all $\gamma\notin \{0,1\}$.
  \item  $(t,\End{2})\, \der{\pop(b)}\, (t,b)$ and $(t,\End{2})\, \der{\pop(b)}\, (check_2,b)$  for all $b\in \{0,1\} $.\vspace{0.1cm}
\end{itemize}
  \item Transitions from states $(t,\End{3})\in Q_{pop}^1$:
  \begin{itemize}
  \item  $(t,\End{3})\, \der{\pop(\gamma)}\, q_\emptyset$  for all $\gamma\notin \Lambda\cup (\Lambda\times \{first,first_{in}\})$.
  \item  $(t,\End{3})\, \der{\pop(\lambda)}\, (t,\lambda)$  for all $t\in \{check_3^1,\widehat{check_3}\}$ and $\lambda\in \Lambda$.
   \item  $(t,\End{3})\, \der{\pop((\lambda,first))}\, (t,\lambda)$  for all $t\in \{check_3^1,\widehat{check_3}\}$ and $(\lambda,t)\in \Lambda\times \{first,first_{in}\}$.
   \item  $ (check_3^0,\End{3})\, \der{\pop(\lambda)}\, (check_3^0,\lambda)$ and
  $ (check_3^0,\End{3})\, \der{\pop(\lambda)}\, (check_3,\lambda)$ for all $\lambda\in \Lambda$.
   \item  $ (check_3^0,\End{3})\, \der{\pop((\lambda,first))}\, (check_3^0,\lambda)$ and
  $ (check_3^0,\End{3})\, \der{\pop((\lambda,first))}\, (check_3,\lambda)$ for all $\lambda\in \Lambda$.
  \item     $ (check_3^0,\End{3})\, \der{\pop((\lambda,first_{in}))}\, (check_3,\lambda)$ for all $\lambda\in \Lambda$.
    \item  $ (check_3^2,\End{3})\, \der{\pop(\lambda)}\, (check_3^2,\lambda)$ and
  $ (check_3^2,\End{3})\, \der{\pop(\lambda)}\, (\widehat{check_3},\lambda)$ for all $\lambda\in \Lambda$.  
   \item  
  $ (check_3^2,\End{3})\, \der{\pop((\lambda,t))}\, (\widehat{check_3},\lambda)$ for all $(\lambda,t)\in \Lambda\times \{first,first_{in}\}$.  \vspace{0.1cm}
\end{itemize}
\item Transitions from states $(t,m) \in Q_{pop}^1\setminus  (\{check_3\}\times \Lambda)$ with $m\notin\{\End{2},\End{3}\}$:
  \begin{itemize}
  \item  $(t,m)\, \der{\pop(\gamma)}\, q_\emptyset$  for all $\gamma\notin \MainP$.
  \item  $(t,m)\, \der{\pop(\gamma)}\, (t,m)$, where  $\gamma\in \MainP$, and  either $m\notin \{\exists,\forall\}$ 
  or $t\neq check_3^1$.
  \item  $(check_3^1,m)\, \der{\pop(\gamma)}\, (check_3^2,m)$  for all $\gamma\in\MainP$ and $m\in\{\exists,\forall\}$. 
\end{itemize}
\end{itemize}\vspace{0.2cm}

The states in $Q_{pop}^2$ are used to generate the non-root nodes of   check 2-block trees. The set $Q_{pop}^2$   is defined as follows:
\[
Q_{pop}^2 := (\{check_1,check_2\}\times \{0,1\})\cup (\{check_1^0\}\times \{\Start{1},\End{1},\Start{2}, 0,1\})   \cup (\{check_1^1\}\times \{\Start{1},0,1\})  
\]
Intuitively, for a state $(t,m)\in Q_{pop}^2$, $m$ represents the last symbol which has been popped from the stack. Moreover, the flag 
$t\in \{check_2, check_1,check_1^0,check_1^0\}$ has the following meaning:
 \begin{itemize}
  \item  $check_2$ is associated with the unique root's child $x$ of  a check 2-block tree $\tpl{T,\Lab}$. The node $x$ is labeled with the marked content
  $(check_2,b)$ of the $2$-block encoded by $\tpl{T,\Lab}$.
  \item  $t=check_1^0$ is associated with the nodes of the main path of a check 2-block tree $\tpl{T,\Lab}$ whose labels are in$\{\Start{1},\End{1},\Start{2}, 0,1\}$.
  \item  $t=check_1^1$ (resp., $t=check_1$) is related  to the nodes of the secondary branches  of a check 2-block tree $\tpl{T,\Lab}$ whose labels are in $\{\Start{1},0,1\}$  
  (resp., whose labels are of form $(check_1,b)$, i.e., the marked contents of $1$-blocks).
\end{itemize}
For each $(t,m)\in Q_{pop}^2$, the propositional labeling of state $(t,m)$ is $\{t,m\}$ if $t\in \{check_1,check_2\}$, and $\{m\}$ otherwise.
 All the configurations associated with the states in $Q_{pop}^2$ are controlled by the system
with the exception of the configurations associated with the  state  $(check_1^0,\End{1})$, which are instead controlled by the environment. 
Transitions from states in $Q_{pop}^2$ are pop transitions. They are defined as follows. Note that all the configurations associated with states
in  $Q_{pop}^2\setminus \{(check_1^0,\End{1})\}$ are deterministic. 
\begin{itemize}
\item  Transitions from states  $(check_2,b)\in Q_{pop}^2$ (note that $b\in \{0,1\}$):
  \begin{itemize}
  \item  $ (check_2,b)\, \der{\pop(\End{1})}\, (check_1^0,\End{1})$.
  \item  $ (check_2,b)\, \der{\pop(\gamma)}\, q_\emptyset$ for all  $\gamma\neq \End{1}$.\vspace{0.1cm}
\end{itemize}
\item  Transitions from states  $(check_1^0,m)\in Q_{pop}^2$ (note that $m\in  \{\Start{1},\End{1},\Start{2}, 0,1\}$):
  \begin{itemize}
  \item  $ (check_1^0,m)\, \der{\pop(\gamma)}\, (check_1^0,\gamma)$   for all $m\in \{\Start{1},0,1\}$ and $\gamma\in \{\Start{1},\End{1},\Start{2},0,1\}$. 
     \item  $ (check_1^0,\End{1})\, \der{\pop(b)}\, (check_1^0,b)$ and $(check_1^0,\End{1})\, \der{\pop(b)}\, (check_1,b)$ for all $b\in \{0,1\}$. 
\item  $ (check_1^0,m)\, \der{\pop(\gamma)}\, q_\emptyset$ if either $m=\Start{2}$ or $\gamma \notin  \{\Start{1},\End{1},\Start{2},0,1\} $.\vspace{0.1cm}
\end{itemize}
\item  Transitions from states  $(check_1,b)\in Q_{pop}^2$ (note that $b\in \{0,1\}$):
  \begin{itemize}
  \item  $ (check_1,b)\, \der{\pop(b')}\, (check_1^1,b')$  for all $b'\in \{0,1\}$. 
  \item   $ (check_1,b)\, \der{\pop(\gamma)}\, q_\emptyset$ for all  $\gamma\notin \{0,1\}$. \vspace{0.1cm}
\end{itemize}
\item  Transitions from states  $(check_1^1,m)\in Q_{pop}^2$ (note that $m\in  \{\Start{1},  0,1\}$):
  \begin{itemize}
\item  $ (check_1^1,b)\, \der{\pop(\gamma)}\, (check_1^1,\gamma)$ for all $b \in \{0,1\}$ and $\gamma\in \{\Start{1},0,1\}$. 
  \item  $ (check_1^1,m)\, \der{\pop(\gamma)}\, q_\emptyset$ if either $m\notin \{0,1\}$ or $\gamma\notin \{\Start{1},0,1\}$. 
\end{itemize}
\end{itemize}\vspace{0.2cm}

 By construction, the following  claim holds, 
\vspace{0.2cm}

\noindent \emph{Claim 3.} Let $\nu$ be a sequence of ATM configuration codes of the form $\nu=\rho\cdot \{acc\}$ and $\Tau$ be the $\CGT$ obtained by unwinding   $\GS(\PS)$
 from configuration $(acc,\Stack(\nu))$. Then, the environments strategy trees of $\Tau$ correspond to the check trees associated with $\nu$.   \vspace{0.2cm}
 
 By Claims 1--3, it follows that  the set of $2^{\Prop}$-labeled trees $\tpl{T,\Lab}$ associated with the  accepting  environment strategy trees $\tpl{T,\Lab,\Trans}$ in  $\exec(\GS(\PS))$ coincides with the set of extended tree-codes. Moreover,  the unique nodes controlled by the system
  in a check-subtree of $\tpl{T,\Lab,\Trans}$ are the $check_3$-branching nodes and the $\{\End{2}\}$-branching nodes. This concludes the proof of Lemma~\ref{lemma:constructionOfOPD}.
\end{proof}

\subsection{Construction of the   ATL*  formula $\varphi$ in Theorem~\ref{theorem:lowerboundReduction}}
We now illustrate in detail the construction of the  $\ATLStar$ formula $\varphi$ in Theorem~\ref{theorem:lowerboundReduction}.
To this end, we need an additional definition and a preliminary result.
 
\begin{defi}[Well-formed Check-trees]\label{def:WFCheckTrees}
A check-tree $\tpl{T,\Lab}$ for a sequence $\nu$ of ATM configuration codes is \emph{well-formed} if
\begin{itemize}
   \item $\tpl{T,\Lab}$ satisfies the \emph{goodness property}, which means that:
   \begin{itemize}
     \item there are no $\widehat{check_3}$-branching nodes,\footnote{Recall that a $\widehat{check_3}$-branching node is a $\End{3}$-node having two children, one labeled by $\{\widehat{check_3},\lambda\}$ and one which is not marked and is labeled by $\{\lambda\}$ for some $\lambda\in\Lambda$.} i.e., the unique branching $\End{3}$-nodes are the $check_3$-branching nodes.\footnote{Recall that the $check_3$-branching nodes in the check-subtrees of the accepting environment strategy trees of $\GS(\PS)$ are controlled by the system player.}  This entails that the subtree rooted at the $\{\Start{3}\}$-node of a
   $check_3$-marked $3$-block contains at most one  $\widehat{check_3}$-marked $3$-block.
   \item Each $\{\End{1}\}$-node in a partial check $2$-block-tree has two children (i.e., all the choices in the $\{\End{1}\}$-branching nodes are enabled).\footnote{Recall that the $\{\End{1}\}$-nodes 
   in the check-subtrees of the accepting environment strategy trees of $\GS(\PS)$ are controlled by the environment.}
   \end{itemize}
   \item  The ATM configuration codes in $\nu$ are well-formed;
   \item $\nu$ starts with the code of the initial %ATM
   configuration for $\alpha$;
   \item \emph{fairness condition:} $\nu$ is faithful to the evolution of $\mathcal{M}$ and for each path visiting a (well-formed) $\CheckB{3}$-marked $3$-block   $\bl_3$ and a (well-formed) $\CheckMark{3}$-marked $3$-block $\bla_3$, $\bl_3$ and $\bla_3$ have the same number.
 \end{itemize}
\end{defi}

Next we show the following preliminary result.

\begin{lem}\label{lemma:preliminaryOfATLStarFOrmula} One can construct in time polynomial in $n$ and $|\Prop|$, three $\CTLStar$ formulas $\varphi_{\good}$,  $\varphi_{\init}$ and $\varphi_{3\-\bl}$ over $\Prop$ satisfying the following for each check-tree $\tpl{T_c,\Lab_c}$, where $\nu$ is the sequence of ATM configuration codes associated with $\tpl{T_c,\Lab_c}$:
   \begin{itemize}
  \item $\tpl{T_c,\Lab_c}$ satisfies $\varphi_{\good}$ \emph{iff}  $\tpl{T_c,\Lab_c}$ satisfies the goodness property in Definition~\ref{def:WFCheckTrees};
  \item $\tpl{T_c,\Lab_c}$ satisfies $\varphi_{\init}$  \emph{\emph{iff}} the first configuration code of $\nu$ is associated with an ATM configuration of the form
  $(q_0,\alpha(0)) \alpha(1) \ldots
\alpha(n-1)\cdot (\#)^{k}$ for some $k\geq 0$;% (initialization);
  \item if $\tpl{T_c,\Lab_c}$ is good, then $\tpl{T_c,\Lab_c}$ satisfies $\varphi_{3\-\bl}$ \emph{iff} the $3$-blocks along $\nu$ are well-formed.
\end{itemize}
\end{lem}
\begin{proof}
Fix a check-tree $\tpl{T_c,\Lab_c}$  and let $\nu$ be the sequence of ATM configuration codes associated with $\tpl{T_c,\Lab_c}$.

\noindent The $\CTLStar$ formula $\varphi_{\good}$ ensuring the goodness property in Definition~\ref{def:WFCheckTrees} is defined as follows:
\[
\begin{array}{ll}
\varphi_{\good}  := &  \A\Always\, \bigl(\End{3}  \rightarrow \neg(\E\Next \CheckMark{3} \wedge  \E\Next \neg\CheckMark{3})\bigr)\, \,\wedge\,\,
\\
&  \A\Always\, \bigl(\CheckB{2} \rightarrow \A\Always(\End{1} \rightarrow (\E\Next\CheckB{1}\wedge \E\Next\neg\CheckB{1}))\bigr)
\end{array}
\]
where
\begin{itemize}
  \item  the first conjunct ensures that there are no $\widehat{check_3}$-branching nodes, i.e., no $\End{3}$-node of the check-tree has both a child marked by  $\widehat{check_3}$ and a child which is not marked by $\widehat{check_3}$;
  \item the second conjunct asserts that  each $\End{1}$-node associated with a marked $2$-block has exactly two children.
  Recall that each $\End{1}$-node associated with a marked $2$-block has at most two children, one which is not marked and the other one which is marked
  by  ${check_1}$.  
\end{itemize}
The definition of the  $\CTLStar$ formula $\varphi_{\init}$ is involved but standard.
\[
\begin{array}{l}
\varphi_{\init}  :=   \E\Eventually\,\Bigl(\, (acc\vee \exists\vee \forall)\wedge  ((\neg l\wedge \neg r)\,\until\, (l\wedge\neg\E\Next \displaystyle{\bigvee_{p\in\Prop}}\,p))\, \wedge
\\
 (\End{3} \rightarrow \Next \psi_{\#})\,\until\, (\End{3} \wedge \Next(\psi_n \wedge (\neg \End{3}\, \until\, (\End{3} \wedge \Next  (\psi_{n-1}\wedge\ldots (\neg \End{3}\,\until\,( \End{3}\wedge \Next(\psi_1\wedge \Next\Always \neg \End{3})))\ldots))))) \Bigr)
\end{array}
\]
where  $\psi_{\#}:= \displaystyle{\bigvee_{(u_p,\#,u_s)\in\Lambda}}(u_p,\#,u_s)$,  $\psi_1:= \displaystyle{\bigvee_{(u_p,(q_0,\alpha(0)),u_s)\in\Lambda}}(u_p,(q_0,\alpha(0)),u_s)$, and for all  $2\leq i\leq n$,
$\psi_i:= \displaystyle{\bigvee_{(u_p, \alpha(i-1),u_s)\in\Lambda}}(u_p, \alpha(i-1),u_s)$.

\noindent Recall that the labelings of the paths along the check-tree $\tpl{T_c,\Lab_c}$ are associated to the reverse of $\nu$ and
the first symbol (resp., the last symbol) of a configuration code is of the form $\{p\}$ where
$p\in \{l,r\}$ (resp., $p\in \{acc,\exists,\forall\}$). Moreover, each path of the check-tree
 $\tpl{T_c,\Lab_c}$ has a suffix labeled by $\emptyset^{\omega}$.
 Thus, the previous formula asserts that the last configuration code along the reverse of $\nu$
 (corresponding to the first configuration code of $\nu$) has the form
$(q_0,\alpha(0)) \alpha(1) \ldots
\alpha(n-1)\cdot (\#)^{k}$ for some $k\geq 0$.

\paragraph{Construction of the \CTLStar\ formula $\varphi_{3\-\bl}$} Assuming that the check-tree  $\tpl{T_c,\Lab_c}$ is good, the $\CTLStar$ formula $\varphi_{3\-\bl}$ requires that
the $3$-blocks along $\nu$ are well-formed (hence, the $n$-bit and $2^{n}$-bit counters in a $3$-block are properly updated).
\[
\varphi_{3\-\bl} := \varphi_{2\-\bl}\wedge \varphi_{2,\first} \wedge \varphi_{2,\last} \wedge \varphi_{2,\inc}
\]
The conjunct $\varphi_{2\-\bl}$ checks  that the $2$-blocks are well-formed. Again we recall that the labelings of the paths along the check-tree $\tpl{T_c,\Lab_c}$ are associated to the reverse of $\nu$.
 \[
\begin{array}{ll}
\varphi_{2\-\bl}  := &  \A\Always \Bigl( \End{1}  \rightarrow   (\Next^{n+2}\Start{1}\wedge
\displaystyle{\bigwedge_{i=1}^{n+1}\bigvee_{b\in\{0,1\}}}\Next^{i} b)\Bigr)\,\wedge\,
\A\Always \Bigl( (\End{1}\wedge \Next^{n+3}\Start{2})   \rightarrow
\displaystyle{\bigwedge_{i=2}^{n+1}}\Next^{i} 0\Bigr)\,\wedge\,
\\
&  \A\Always \Bigl( (\neg \Start{1}\wedge \Next\End{1})   \rightarrow
\displaystyle{\bigwedge_{i=3}^{n+2}}\Next^{i} 1\Bigr)\,\wedge\, \A\Always \Bigl( (\End{1}\wedge \Next^{n+3}\End{1})   \rightarrow

\\
&\displaystyle{\bigvee_{i=2}^{n+1}} \Bigl[ (\Next^{i}1\wedge  \Next^{n+3+i}0 )\wedge \displaystyle{\bigwedge_{j=2}^{i-1}\bigvee_{b\in\{0,1\}}}(\Next^{j}b\wedge \Next^{n+3+j}b )\wedge  \displaystyle{\bigwedge_{j=i+1}^{n+1}}(\Next^{j}0\wedge  \Next^{n+3+j}1 )\Bigr]\Bigr)
\end{array}
\]
where:
\begin{itemize}
  \item the first conjunct in the definition  of $\varphi_{2\-\bl}$ ensures well-formedness of $1$-blocks. Recall that the reverse of a
  well-formed $1$-block is of the form $\{\End{1}\}\{b\} \{b_1\}\ldots \{b_n \}\{\Start{1}\}$, where $b,b_1,\ldots,b_n\in \{0,1\}$
  and $b$ is the content of the $1$-block.
  \item The second  conjunct  ensures that
the first  $1$-block $bl_1$  of a $2$-block  has number $0$, i.e., the reverse of $bl_1$ has  the form
 $\{\End{1}\}\{b\} \{0\}\ldots \{0 \}\{\Start{1}\}$ for some $b\in\{0,1\}$.
  \item The  third conjunct  ensures that
the last $1$-block $bl_1$  of a $2$-block  has number $2^{n}-1$, i.e., the reverse of $bl_1$ has the form
 $\{\End{1}\}\{b\} \{1\}\ldots \{1 \}\{\Start{1}\}$ for some $b\in\{0,1\}$.
 \item Finally, the last conjunct ensures that  for two adjacent $1$-blocks $\bl_1$ and $\bla_1$ along a $2$-block, $\bl_1$ and $\bla_1$ have consecutive  numbers.
\end{itemize}

The second conjunct $\varphi_{2,\first}$ in the definition of $\varphi_{3\-\bl}$  ensures that the first $2$-block $\bl_2$ of a $3$-block along $\nu$ has number $0$, i.e., the content of each $1$-sub-block of $\bl_2$ is $0$.
\[
\varphi_{2,\first}  :=   \A\Always \Bigl( \bigl[\End{2}\wedge \Next(\neg\End{2}\,\until\,\Start{3})\bigr] \, \longrightarrow \,
\Next\bigl[(\neg\End{2}\wedge (\End{1} \rightarrow \Next 0))\,\until\,\Start{3}\bigr]\Bigr)
\]
The second conjunct $\varphi_{2,\last}$ guarantees that the last $2$-block $\bl_2$ of a $3$-block has number $2^{2^{n}}-1$, i.e., the content of each $1$-sub-block of $\bl_2$ is $1$.
\[
\varphi_{2,\last}  :=   \A\Always \Bigl( \bigl[\neg\Start{2}\wedge \Next \End{2}\wedge \Eventually\Start{2}\bigr] \, \longrightarrow \,
\Next\bigl[(\neg\Start{2}\wedge (\End{1} \rightarrow \Next 1))\,\until\, \Start{2}\bigr]\Bigr)
\]
Finally, the last conjunct $\varphi_{2,\inc} $ in the definition of $\varphi_{3\-\bl}$ guarantees that
for all adjacent $2$-blocks $\bl_2$ and $\bla_2$ of a $3$-block along $\nu$, $\bl_2$ and $\bla_2$  have consecutive numbers. For this, assuming that $\bla_2$ follows $\bl_2$ along the reverse of $\nu$, we need to check that there is a $1$-sub-block $\overline{\bl}_1$ of $\bl_2$ whose content is $1$ and the following holds:
 \begin{itemize}
  \item  the $1$-sub-block of $\bla_2$ with the same number as $\overline{\bl}_1$ has content $0$;
  \item Let $\bl_1$ be a $1$-sub-block  of $\bl_2$ distinct from $\overline{\bl}_1$, and $\bla_1$  be the $1$-sub-block of
  $\bla_2$   having the same number as $\bl_1$. Then,   $\bl_1$ and $\bla_1$  have the same content if $\bl_1$ precedes $\overline{\bl}_1$ along the reverse of  $\bl_2$;
  otherwise, the content of $\bl_1$ is $0$ and the content of $\bla_1$  is $1$.
\end{itemize}
In order to check these conditions, we exploit the branches of a check $2$-block-tree $\tpl{T',\Lab'}$ in $\tpl{T_c,\Lab_c}$ associated with (a copy of) $\bla_2$
which lead to $check_1$-marked copies of the $1$-sub-blocks of $\bla_2$ (see Figure~\ref{FigureLowerBoundSecond}(c)). 
 Note that these branches
consist  (of the reverse) of a $check_1$-marked  $1$-sub-block  of $\bla_2$ followed by the suffix $\emptyset^{\omega}$.
Moreover, since $\tpl{T_c,\Lab_c}$ is good
all the choices in the $\End{1}$-nodes of $\tpl{T',\Lab'}$ are enabled (i.e, for each $1$-sub-block  $\bla_1$ of $\bla_2$, there is a branch for the $check_1$-marked copy of $\bla_1$).
Then, the  formula $\varphi_{2,\inc}$ is defined as follows.
\[
\begin{array}{ll}
\varphi_{2,\inc} := &  \A\Always \Bigl( (\End{2} \wedge \Next (\neg\Start{3}\,\until\, \End{2})) \, \longrightarrow \, \Next\Bigl[\bigl\{ \neg \End{2}\wedge (\End{1}\rightarrow \displaystyle{\bigvee_{b\in\{0,1\}}} \theta(b,b)) \bigr\}
\\
& \quad\quad \,\until\, \bigl\{\theta(1,0)\wedge \End{1} \wedge \Next ((\neg \End{2}\wedge (\End{1}\rightarrow \theta(0,1))) \,\until\, \End{2})
   \bigr\}\Bigr]\,\,\Bigr)
\end{array}
\]
\noindent  where for all $b,b'\in\{0,1\}$,  the auxiliary subformula $\theta(b,b')$ in the definition of $\varphi_{2,\inc} $
requires that for the current $1$-sub-block $\bl_1$ of $\bl_2$ and for the path from $\bl_1$ which leads to the $check_1$-marked copy $\bla_1$ of the $1$-sub-block
of $\bla_2$  having the same number as $\bl_1$, the following holds:
the content of $\bl_1$ is $b$ and the content of $\bla_1$ is $b'$.

\[
\theta(b,b'):= \Next b\wedge \E\Bigl(\bigl[\neg \End{2}\,\until\, (\End{2} \wedge \Next (\CheckB{2}\wedge \Eventually(\CheckB{1}\wedge b')))\bigr]\wedge \displaystyle{\bigwedge_{i=1}^{n}\bigvee_{c\in\{0,1\}}}\bigl[\Next^{i+1}c\wedge \Eventually (\CheckB{1}\wedge \Next^{i}c)\bigr] \Bigr)
\]
This concludes the proof of Lemma~\ref{lemma:preliminaryOfATLStarFOrmula}.
\end{proof}

An extended tree-code $\tpl{T_e,\Lab_e}$ is   \emph{well-formed}  if each check-tree in $\tpl{T_e,\Lab_e}$ is
  well-formed.
Evidently, there is a  well-formed extended tree-code if and only if there is an accepting computation tree of $\mathcal{M}$ over $\alpha$.
By exploiting Lemma~\ref{lemma:preliminaryOfATLStarFOrmula}, we now  establish the following result  that together
with Lemma~\ref{lemma:constructionOfOPD} provides a proof of Theorem~\ref{theorem:lowerboundReduction}.

\begin{lem}\label{lemma:constructionOfATLStarFOrmula} One can construct in time polynomial in $n$ and $|\Prop|$, an $\ATLStar$  formula $\varphi$ over $\Prop$ and
 $\Agents =\{\env,\sys\}$
such that for each environment strategy tree  $\Tau=\tpl{T,\Lab,\Trans}$ in  $\exec(\GS(\PS))$, $\Tau$ is a model of $\varphi$ iff $\tpl{T,\Lab}$ is a well-formed extended tree-code.
\end{lem}
\begin{proof}  By Lemma~\ref{lemma:constructionOfOPD}, the set of $2^{\Prop}$-labeled trees  associated with the  \emph{accepting}   environment strategy trees of  $ \GS(\PS)$ coincides with the set of extended tree-codes. Let $\varphi_{\good}$, $\varphi_{\init}$, and $\varphi_{3\-\bl}$ be the $\CTLStar$ formulas of Lemma~\ref{lemma:preliminaryOfATLStarFOrmula} having
size polynomial in $n$ and $|\Prop|$. Note that since the paths quantifiers of $\CTLStar$ correspond to the strategic quantifiers $\Exists{\emptyset}$ and $\Exists{\Agents}$, each $\CTLStar$ formula 
can be seen as an $\ATLStar$ formula.  
Then,  
the \ATLStar\ formula $\varphi$ is given by
\[
\varphi:= \A\Eventually\,acc \wedge \A\Always(acc \rightarrow (\varphi_{\good} \wedge \varphi_{\init}\wedge \varphi_{3\-\bl}  \wedge \varphi_{\conf}\wedge \varphi_{\fair}))
\]
where for an environment  strategy tree  $\Tau=\tpl{T,\Lab,\Trans}$ of the $\PMS$ $\PS$ of Lemma~\ref{lemma:constructionOfOPD}, the first conjunct ensures that $\Tau$ is accepting (recall that $\Tau$ is accepting iff each path from the root visits an $\{acc\}$-labeled node), while the subformulas $\varphi_{\good}$, $\varphi_{\init}$, $\varphi_{3\-\bl}$, $\varphi_{\conf}$, and $\varphi_{\fair}$ ensure that each check-tree $\tpl{T_c,\Lab_c}$ of $\Tau$ is well-formed. Hence, an environment strategy tree  $\Tau=\tpl{T,\Lab,\Trans}$ of  $ \GS(\PS)$  satisfies $\varphi$ iff $\tpl{T,\Lab}$ is a well-formed extended tree-code. 

Fix a check-tree $\tpl{T_c,\Lab_c}$ of  an accepting  environment  strategy tree of the $\PMS$ $\PS$, and let
  $\nu$ be   the sequence of ATM configuration codes associated with $\tpl{T_c,\Lab_c}$. By Lemma~\ref{lemma:preliminaryOfATLStarFOrmula}, we have that:
\begin{itemize}
  \item  $\tpl{T_c,\Lab_c}$ satisfies $\varphi_{\good}$   iff it   satisfies the goodness property in Definition~\ref{def:WFCheckTrees};
  \item $\varphi_{\init}$  guarantees that the first configuration code of $\nu$ is associated with an ATM configuration of the form
  $(q_0,\alpha(0)) \alpha(1) \ldots
\alpha(n-1)\cdot (\#)^{k}$ for some $k\geq 0$;% (initialization);
  \item $\varphi_{3\-\bl}$   enforces well-formedness of $3$-blocks along $\nu$.
\end{itemize}
We now consider the conjuncts $\varphi_{\conf}$ and $\varphi_{\fair}$  of $\varphi$ which ensure the following properties for the given check-tree  $\tpl{T_c,\Lab_c}$:
\begin{itemize}
    \item   $\varphi_{\conf}$   requires that the   ATM configuration codes  along $\nu$ are well-formed;
  \item   $\varphi_{\fair}$  ensures that $\nu$ satisfies the fairness condition in Definition~\ref{def:WFCheckTrees}. 
\end{itemize}
 By means of the formulas $\varphi_{\good}$ and $\varphi_{3\-\bl}$, we can assume that
the check-tree $\tpl{T_c,\Lab_c}$ is good and all the $3$-blocks along $\nu$ are well-formed.
For defining the  $\ATLStar$ formulas   $\varphi_{\conf}$  and $\varphi_{\fair}$,
we exploit the following pattern: starting from an $\{\End{2}\}$-node $x_{\bl_2}$ related to a $2$-block $\bl_2$ of the good check-tree  $\tpl{T_c,\Lab_c}$,
we need \emph{to isolate} another $2$-block $\bla_2$ following $\bl_2$ along the reverse of $\nu$ and checking, in particular, that
$\bl_2$ and $\bla_2$ have the same number. Moreover, for the case of the formula $\varphi_{\conf}$, we require that the $3$-block of $\bla_2$ is adjacent
to the $3$-block of $\bl_2$ within the same ATM configuration code, while for the case of the formula $\varphi_{\fair}$, we require
that the $3$-block of $\bl_2$ (resp., $\bla_2$) is $\CheckB{3}$-marked (resp., $\CheckMark{3}$-marked) in the considered path of $\tpl{T_c,\Lab_c}$.

Recall that in a good check-tree, the unique nodes controlled by the system  are the $\CheckB{3}$-branching nodes and the $\{\End{2}\}$-nodes, and each unmarked $2$-block is associated with a check $2$-block-tree ($2$-$\BCT$ for short). In particular, in a $2$-$\BCT$, all the nodes, but the root (which is an $\{\End{2}\}$-node), are controlled by the environment. Moreover, each strategy of the system selects exactly one child for each node controlled by the system.
Hence, there is a strategy $f_{\bl_2}$ of the player system such that
 \begin{itemize}
  \item (*) each play consistent with the strategy $f_{\bl_2}$ starting from the $\{\End{2}\}$-node $x_{\bl_2}$ ``gets trapped" in the $2$-$\BCT$ of $\bla_2$, and
  \item (**)
 each path starting from the node $x_{\bl_2}$ and leading to some marked $1$-block of the $2$-$\BCT$ for  $\bla_2$ is consistent with the strategy
  $f_{\bl_2}$.
\end{itemize}
Thus, in order \emph{to isolate} a $2$-block $\bla_2$, an $\ATLStar$ formula ``guesses'' the strategy $f_{\bl_2}$ and check that conditions (*) and (**) are fulfilled by simply requiring that each outcome from the current node $x_{\bl_2}$ visits a node marked by proposition $\CheckB{2}$. Additionally, by exploiting
the branches of the $2$-$\BCT$ leading to marked $1$-blocks, we can check by a formula of size polynomial in $n$ and the size of $\mathcal{M}$ that $\bl_2$ and $\bla_2$ have the same number.
We now proceed with the technical details about the construction of the $\ATLStar$ formulas  $\varphi_{\conf}$  and $\varphi_{\fair}$.

\paragraph{Construction of the \ATLStar\ formula $\varphi_{\conf}$}  The $\ATLStar$ formula $\varphi_{\conf}$ is defined as follows.
\[
\varphi_{\conf} :=  \varphi_{3,\first} \wedge \varphi_{3,\last} \wedge \varphi_{3,\inc}
\]
The  conjunct $\varphi_{3,\first}$ requires  that the first $3$-block $\bl_3$ of an ATM configuration code along $\nu$ has number $0$, i.e., the content of each $2$-sub-block of $\bl_3$ is $0$.
\[
\varphi_{3,\first}  :=   \A\Always \Bigl( \bigl[\End{3}\wedge \Next(\neg\End{3}\,\until\,(l\vee r))\bigr] \, \longrightarrow \,
\Next\bigl[(\neg\End{3}\wedge (\End{2} \rightarrow \Next 0))\,\until\,(l\vee r)\bigr]\Bigr)
\]
The second conjunct $\varphi_{3,\last}$ guarantees that the last $3$-block $\bl_3$ of an ATM configuration code has number $\Tower(n,3)-1$, i.e., the content of each $2$-sub-block of $\bl_3$ is $1$).
\[
\varphi_{3,\last}  :=   \A\Always \Bigl( \bigl[\neg\Start{3}\wedge \Next \End{3}\wedge \Eventually\Start{3}\bigr] \, \longrightarrow \,
\Next\bigl[(\neg\Start{3}\wedge (\End{2} \rightarrow \Next 1))\,\until\, \Start{3}\bigr]\Bigr)
\]
 The last conjunct $\varphi_{3,\inc} $ in the definition of $\varphi_{\conf}$ checks that
for all \emph{adjacent} $3$-blocks $\bl_3$ and $\bla_3$ of an ATM configuration code along $\nu$, $\bl_3$ and $\bla_3$  have consecutive numbers. For this, assuming that $\bla_3$ follows $\bl_3$ along the reverse of $\nu$, we need to check that there is a $2$-sub-block $\overline{\bl}_2$ of $\bl_3$ whose content is $1$ and the following holds:
 \begin{itemize}
  \item  the $2$-sub-block of $\bla_3$ with the same number as $\overline{\bl}_2$ has content $0$;
  \item Let $\bl_2$ be a $2$-sub-block  of $\bl_3$ distinct from $\overline{\bl}_2$, and $\bla_2$  be the $2$-sub-block of
  $\bla_3$   having the same number as $\bl_2$. Then,   $\bl_2$ and $\bla_2$  have the same content if $\bl_2$ precedes $\overline{\bl}_2$ along the reverse of  $\bl_3$;
  otherwise, the content of $\bl_2$ is $0$ and the content of $\bla_2$  is $1$.
\end{itemize}
Formula $\varphi_{3,\inc}$ is then defined as follows.
\[
\begin{array}{ll}
\varphi_{3,\inc} := &  \A\Always \Bigl( (\End{3} \wedge  \Next((\neg l\wedge \neg r)\,\until\, \End{3}) ) \, \longrightarrow \, \Next\Bigl[\bigl\{ \neg \End{3}\wedge (\End{2}\rightarrow \displaystyle{\bigvee_{b\in\{0,1\}}} \eta(b,b)) \bigr\}
\\
& \quad\quad \,\until\, \bigl\{\eta(1,0)\wedge \End{2} \wedge \Next ((\neg \End{3}\wedge (\End{2}\rightarrow \eta(0,1))) \,\until\, \End{3})
   \bigr\}\Bigr]\,\,\Bigr)
\end{array}
\]
where for all $b,b'\in\{0,1\}$, we exploit  the auxiliary formula $\eta(b,b')$ to
require from the current $\End{2}$-node $x$ of the current $2$-sub-block $\bl_2$ of $\bl_3$ that the content of $\bl_2$ is $b$ and
the $2$-sub-block $\bla_2$ of $\bla_3$ having the same number as $\bl_2$ has content $b'$. In order to ensure the last condition, the
formula  $\eta(b,b')$ asserts the existence of
 a strategy $f_x$ of the player system
such that the following two conditions hold:
 \begin{enumerate}
  \item  each outcome of $f_x$ from the node $x$  visits a node marked by $\CheckB{2}$ whose parent ($\End{2}$-node) belongs to a  $2$-block  of $\bla_3$. This ensures that all the outcomes ``get trapped"
  in the \emph{same} check $2$-block-tree associated with some $2$-block $\bla_2$ of $\bla_3$. Moreover,   $\bla_2$ has content $b'$.
  \item  For each outcome $\pi'$ of $f_x$ from $x$ which leads to a marked $1$-sub-block $\bla_1$ (hence, a marked copy of a $1$-sub-block of $\bla_2$), denoting by $\bl_1$ the $1$-sub-block
  of $\bl_2$ having the same number as $\bla_1$, it holds that $\bl_1$ and $\bla_1$ have the same content. This ensures that $\bl_2$ and $\bla_2$ have the same number.
\end{enumerate}
The first (resp., second) condition is implemented by the first (resp., second) conjunct in the argument of the strategic quantifier $\Exists{\sys}$ in the definition of $\eta(b,b')$ below.
\[
\begin{array}{ll}
\eta(b,b'):= &\Next b\wedge \Exists{\sys}\Bigl(\bigl[\neg \End{3}\,\until\, (\End{3} \wedge \Next (\neg \End{3} \,\until \, (\CheckB{2}\wedge b')))\bigr]\,\wedge\,\\
& \phantom{\Next b\wedge \Exists{\sys}\Bigl(} \bigl[\Eventually\CheckB{1}\rightarrow \Next((\neg \End{2}\wedge (\End{1}\rightarrow \Next\eta_{1}))\,\until\,\Start{2})\bigr] \Bigr)
\end{array}
\]
\[
\eta_1 :=  \Bigl(\displaystyle{\bigwedge_{i=1}^{i=n}\bigvee_{b\in \{0,1\}}}((\Next^{i}\, b)\wedge \Eventually(\CheckB{1}\wedge \Next^{i} b))\Bigr) \, \longrightarrow \,  \displaystyle{\bigvee_{b\in \{0,1\}}}(b\,\wedge\, \Eventually(\CheckB{1}\wedge  b))
\]
Note that for each outcome $\pi'$ of strategy $f_x$ which leads to a marked $1$-sub-block $\bla_1$  of $\bla_2$, the subformula
$\eta_1$ of $\eta(b,b')$ is asserted at the content node of each $1$-sub-block $\bl_1$ of $\bl_2$. Thus, $\eta_1$ requires that whenever
 $\bl_1$ and $\bla_1$ have the same number, then $\bl_1$ and $\bla_1$ have the same content as well.

\paragraph{Construction of the \ATLStar\ formula $\varphi_{\fair}$} We can assume that
the check-tree $\tpl{T_c,\Lab_c}$ is good and all the ATM configuration codes along $\nu$ are well-formed.
Since $\tpl{T_c,\Lab_c}$ satisfies the goodness property, for each $\CheckB{3}$-marked $3$-block $\bl_3$
which does not belong to the first configuration code of $\nu$, there is \emph{exactly one}
$\CheckMark{3}$-marked $3$-block $\bla_3$ in
the subtree of $\tpl{T_c,\Lab_c}$ rooted at the $\Start{3}$-node of $\bl_3$. Moreover, $\bl_3$ and $\bla_3$ belong to two adjacent configuration codes along $\nu$.
  Thus,
 by construction, in order to ensure that $\nu$
is faithful to the evolution of $\mathcal{M}$, it suffices to require that for each (well-formed) $\CheckB{3}$-marked $3$-block $\bl_3$ in
$\tpl{T_c,\Lab_c}$ which does not belong to the first configuration code of $\nu$, the associated  (well-formed) $\CheckMark{3}$-marked $3$-block $\bla_3$ satisfies the following conditions, where
 $(u_p,u,u_s)$ (resp., $(u'_p,u',u'_s)$) is the content of
  $\bl_3$ (resp., $\bla_3$)
\begin{itemize}
  \item $\bl_3$ and $\bla_3$ have the same number,
  \item $u= \Succ_l(u'_p,u',u'_s)$ if $l$ marks the ATM configuration code of $\bl_3$, and $u= \Succ_r(u'_p,u',u'_s)$ otherwise.
\end{itemize}
\noindent Thus, formula $\varphi_{\fair}$ is defined as follows:
\[
\begin{array}{ll}
\varphi_{\fair} := & \displaystyle{\bigwedge_{\dire\in\{l,r\}}} \A\Always \Bigl( \Bigl[\CheckB{3} \wedge [(\neg l \wedge \neg r)\,\until\,(\dir\wedge \Next(\exists \vee\forall))]\Bigr]  \longrightarrow    \Bigl[ \bigl((\neg\End{3}\wedge (\End{2}\rightarrow \psi_=))\,\until\, \Start{3}\bigr)
\\
& \,\wedge\, \displaystyle{\bigvee_{(u_p,u,u_s),(u'_p,u',u'_s)\in\Lambda:\, u= \Succ_\dire(u'_p,u',u'_s) }} \bigl((u_p,u,u_s) \wedge \E\Eventually (\CheckMark{3}\wedge (u'_p,u',u'_s)) \bigr)\Bigr]\,\Bigr)
\end{array}
\]
\noindent where  the auxiliary formula $\psi_=$  in the definition of $\varphi_{\fair} $
requires from the current $\End{2}$-node $x$ of the current $2$-sub-block $\bl_2$ of $\bl_3$ that the $2$-sub-block $\bla_2$ of $\bla_3$ having the same number as $\bl_2$ has the same content as $\bl_2$ too. In order to ensure the last condition, the
formula  $\psi_=$ asserts the existence of
 a strategy $f_x$ of the player system
such that the following holds:
 \begin{enumerate}
  \item  each outcome of $f_x$ from the node $x$  visits a node marked by $\CheckB{2}$ whose parent ($\End{2}$-node) belongs to a $\CheckMark{3}$-marked $3$-block.  This ensures that all the outcomes ``get trapped"
  in the \emph{same} $2$-block check-tree associated with some $2$-block $\bla_2$ of $\bla_3$. Moreover,  $\bl_2$ and $\bla_2$ have the same content.
  \item  For each outcome $\pi'$ of $f_x$ from $x$ which leads to a marked $1$-sub-block $\bla_1$ (hence, a marked copy of a $1$-sub-block of $\bla_2$), denoting by $\bl_1$ the $1$-sub-block
  of $\bl_2$ having the same number as $\bla_1$, it holds that $\bl_1$ and $\bla_1$ have the same content. This ensures that $\bl_2$ and $\bla_2$ have the same number.
\end{enumerate}
\noindent  Thus, formula $\psi_=$ is defined as follows.
\[
\begin{array}{ll}
\psi_= := & \Exists{\sys}\Bigl(\displaystyle{\bigvee_{b\in\{0,1\}}}\bigl[\Next b\wedge \Eventually\{\CheckMark{3}\wedge (\neg\End{3}\,\until\,(b\wedge\CheckB{2}))\}\bigr]\,\wedge\,\\
& \phantom{\Exists{\sys}\Bigl(} \bigl[\Eventually\CheckB{1}\rightarrow \Next((\neg \End{2}\wedge (\End{1}\rightarrow \Next\eta_{1}))\,\until\,\Start{2})\bigr] \Bigr)
\end{array}
\]
where $\eta_1$ corresponds to the homonymous subformula of the auxiliary formula $\eta(b,b')$ used in the definition of $\varphi_{3,\inc}$. This concludes the proof of Lemma~\ref{lemma:constructionOfATLStarFOrmula}.
\end{proof}

%-------------------------------------------------------------------------------
\section{Conclusion}\label{sec:conc}
%-------------------------------------------------------------------------------

Module checking is a useful game-theoretic framework to deal with branching-time specifications. The setting is simple and powerful as it allows to capture the essence of the adversarial interaction between an open system (possibly consisting of several independent components) and its unpredictable environment. The work on module checking has brought an important contribution to the strategic reasoning field, both in computer science and AI~\cite{AlurHK02}. It is known~\cite{JamrogaM14} that  \CTL/\CTLStar\ module checking %has come to the fore as it has been shown that it 
is incomparable with \ATL/\ATLStar\ model checking. %~\cite{JamrogaM14}. 
In particular the former can keep track of all moves made in the past, while the latter cannot. This is a severe limitation in \ATL/\ATLStar\ and has been studied under the name of irrevocability of strategies in~\cite{AGJ07}. Remarkably, this feature can be handled with more sophisticated logics such as \emph{Strategy Logics}~\cite{CHP10,MMPV14}, \emph{\ATL\ with strategy contexts}~\cite{LaroussinieM15}, and \emph{quantified} \CTL\ \cite{LaroussinieM14}. However, for such logics, the relative model checking question for finite-state multi-agent systems
(modelled by finite-state concurrent game structures) turns out to be non-elementarily decidable.

In this paper,  we have addressed %and carefully investigated the computational complexity
  %of
  the module-checking problem of multi-agent pushdown systems (\PMS) against \ATL\ and \ATLStar\
  specifications.
$\PMS$ endow finite-state multi-agent systems with an additional expressive power, the possibility of using
 	a stack to store unbounded information. The stack is the standard low level mechanism which allows to structure agents in modules and to implement recursive calls and returns of modules. Hence, the considered  framework is suitable for formally reasoning on the behaviour of software agents with (recursive) procedural modularity.
As a main contribution, we have established the exact computational complexity of pushdown module-checking against $\ATL$ and $\ATLStar$. While for \ATL, the considered problem is \TWOEXPTIME-complete, which is the
same complexity as pushdown module-checking   for \CTL,  for   \ATLStar,  pushdown module-checking   turns out to be \FOUREXPTIME-complete, hence exponentially harder than  both \CTLStar\ pushdown module-checking and \ATLStar\ model-checking of \PMS. As future work, we aim to investigate the considered problems
  in the setting of \emph{imperfect
information under memoryless strategies}. We recall that this
setting is decidable in the finite-state case~\cite{AlurHK02}.
However, moving to pushdown systems one has to distinguish
whether the missing information relies in the control states, in the
pushdown store, or both. We recall that in pushdown module-checking only the former case is decidable for specifications
given in $\CTL$ and $\CTLStar$~\cite{Aminof13pushdown-jc}.

Another interesting question to investigate is the exact computational complexity of pushdown module checking against the fragment $\ATLPlus$ of \ATLStar, 
where each temporal modality is immediately preceded either by a strategic quantifier or by a Boolean connective. Our results just imply that pushdown module checking against $\ATLPlus$ lies somewhere between \TWOEXPTIME\ and \FOUREXPTIME. 
\bibliographystyle{alphaurl}
\bibliography{bib2}

\end{document}